\documentclass[%
groupedaddress,
 amsmath,amssymb,
 aps,
]{revtex4-1}

\usepackage{graphicx}
\usepackage{dcolumn}
\usepackage{bm}

\begin{document}

\preprint{APS/123-QED}

\title{Capillary Interactions on Fluid Interfaces: Opportunities for Directed Assembly}

\author{Nima Sharifi-Mood}
\author{Iris B. Liu}%
\author{Kathleen~J.~Stebe}
 \email{kstebe@seas.upenn.edu}
\affiliation{Department of Chemical and Biomolecular Engineering, University of Pennsylvania, Philadelphia, PA, 19104, USA\\}%


\begin{abstract}
A particle placed in soft matter distorts its host and creates an energy landscape. This can occur, for example, for particles in liquid crystals, for particles on lipid bilayers or for particles trapped at fluid interfaces. Such energies can be used to direct particles to assemble with remarkable degrees of control over orientation and structure. These notes explore that concept for capillary interactions, beginning with particle trapping at fluid interfaces, addressing pair interactions on planar interfaces and culminating with curvature capillary migration. Particular care is given to the solution of the associated boundary value problems to determine the energies of interaction. Experimental exploration of these interactions on planar and curved interfaces is described. Theory and experiment are compared.  These interactions provide a rich toolkit for directed assembly of materials, and, owing to their close analogy to related systems, pave the way to new explorations in materials science.
\end{abstract}

\pacs{Valid PACS appear here}
\maketitle

Why is assembly of colloids at fluid interfaces interesting? Colloidal particles are often directed to assemble by applying external electromagnetic fields to steer them into well-defined structures at given locations \cite{Whitesides}.  We have been developing alternative strategies based on energy landscapes that emerge when a colloid is placed within soft matter.  Particles attached to fluid interfaces are an important example.\\ 
 
Particles become trapped at interfaces because they eliminate a patch of fluid interface, with a significant decrease in energy \cite{Pieranski}. Once attached, they can distort the interface to minimize wetting energies.  The distortions have an associated energy,  the product of interfacial area and the surface tension.  Why do particles attract by capillarity? They attract to minimize the area in their distortions, approaching each other rise to rise, fall to fall, lowering the interfacial area \cite{Kralchevsky,Stamou}. Closer to contact, rearrangement of the wetting configurations and associated solid-fluid wetting energies can play a role. The interactions prove tremendously rich. Neighboring particles interact and assemble.  Non-spherical particles can adopt preferred orientations \cite{Arjun,Botto}.  The mechanics of the assemblies that form, be they flexible or rigid, depend on the details of the particle shape even for fixed wetting energies \cite{Lorenzo}.\\

Different aspects of particle shapes and wetting boundaries matter at different separation distances. Particles that are far apart all obey a universal interaction, with various details of the shape and wetting becoming progressively more important as they approach-are they elongated? Rough? Does the contact line remain fixed or rearrange?  All of these issues play a role.\\

In evaluating the area of the interface between interacting particles in the far field, one particle interacts with the interface curvature imposed by the other.  This observation allows another exciting approach: Why be limited to curvature fields created by neighboring particles?  Why not mold the interface curvature to define fields that drive particles along pre-determined paths to well-defined locations \cite{Marcello}?\\

Understanding these interactions allows us to develop schemes to make reconfigurable materials, since interfaces and their associated capillary energy landscapes can be readily reconfigured.  We can gain a better insight into these systems by thinking about the equations that govern the shape of the fluid interface, the Young-Laplace equation and the wetting boundary conditions at the particle surface.  Since we will focus on small particles, we can limit this discussion to the small slopes limit.  For confined fluids in sub-millimeter container, mean curvatures (see below for defination) are typically constant.  Constant or zero mean curvature surfaces can prove extremely rich and relevant in experiment.  For the zero mean curvature case, the interface shape is governed by the Laplacian!  For this, we can invoke valuable analogies to electrostatics, which, when treated with care, can provide rich guidance.   Owing to the linearity of the equations, we solve certain canonical cases in closed form to guide our thinking. These notes are structured to take us through recent work, emphasizing canonical solutions and their implications.\\

There are other examples of colloidal interactions in soft matter systems. Liquid crystals are one important host medium. Particles immersed in liquid crystals distort the director field to elicit an elastic energy response.  Preferred paths and locations for assembly can be defined by molding the director field and its associated defect structures \cite{Cavallaro-LC}.  Particles adhered to lipid bilayer are another system in which such fields can be generated and exploited.  Adhered particles deform the host membrane, eliciting interactions mediated by membrane bending energies and tension \cite{Safinya,Pincus,Lipowsky,Granick}.  There are certainly other systems which we have yet to explore.  In the following notes, we do not provide a thorough literature review, but  do highlight certain work especially relevant to the discussions.  
\section{The interactions of microparticles in confined systems:  It is all about the boundaries}
We are particularly interested in systems at the scale of hundreds of microns and smaller.  For example, we are interested in particles with radius $a$ small enough that the Bond number is small, i.e. $Bo = {{\rho g{a^2}} \mathord{\left/
 {\vphantom {{\rho g{a^2}} \gamma }} \right.
 \kern-\nulldelimiterspace} \gamma } \ll 1$ and hence gravity can be neglected, here $\gamma$ is the surface tension, $\rho$ is the fluid density, and $g$ is the acceleration due to gravity.\\
 
We are also interested in molding fluid interfaces in small volumes, with linear dimensions small enough so that gravity can be neglected, but large enough that particle interactions with boundaries are not important. In these cases, all of the cues to mold the interface arise from the boundaries either of the particles or the containers used to hold the fluid.\\

Particles are the sources of the capillary energies.  The energies depend on the particle wetting conditions.  When particles distort an interface, the distortions decay over particle length scales, and ultimately define the range and form of the interactions.  ``Point charge'' treatments fail to capture these effect and lead to incorrect formulations of the leading order energy terms.  So, throughout this text, we will focus on the local interface shapes around the particles.  In systems with multiple length scales, we use matched asymptotic analysis \cite{Nayfeh} to bound the range of validity of our functional forms.\\

We discuss wetting boundary conditions for vapor-liquid interfaces, although the discussion is equally valid for any two juxtaposed immiscible liquids.  At any contact line, where fluid $1$ (e.g. the liquid $L$), fluid $2$ (e.g. the vapor $V$), and solid ($S$) meet, the wetting configuration can either obey equilibrium wetting conditions characterized by the equilibrium contact angle $\theta_0$ and the wetting energies of the solids and fluids $\gamma_{SL}$ and $\gamma_{SV}$ according to Young equation \cite{Young},
 \begin{eqnarray}
 \cos \theta_0=\frac{\gamma_{SL}-\gamma_{SV}}{\gamma_{LV}},
 \end{eqnarray}
  or contact lines can be kinetically trapped at pinning sites including corners, sharp edges, or sites of patchy wetting \cite{Stamou,Manoharan,Furst1}.   We will discuss both, emphasizing when they give remarkably different outcomes.

 \section{The trapping of isolated particles on planar interfaces}   
\subsection{Particle with an equilibrium contact angle}
\begin{figure} 
\centering
\includegraphics[width=0.8 \textwidth]{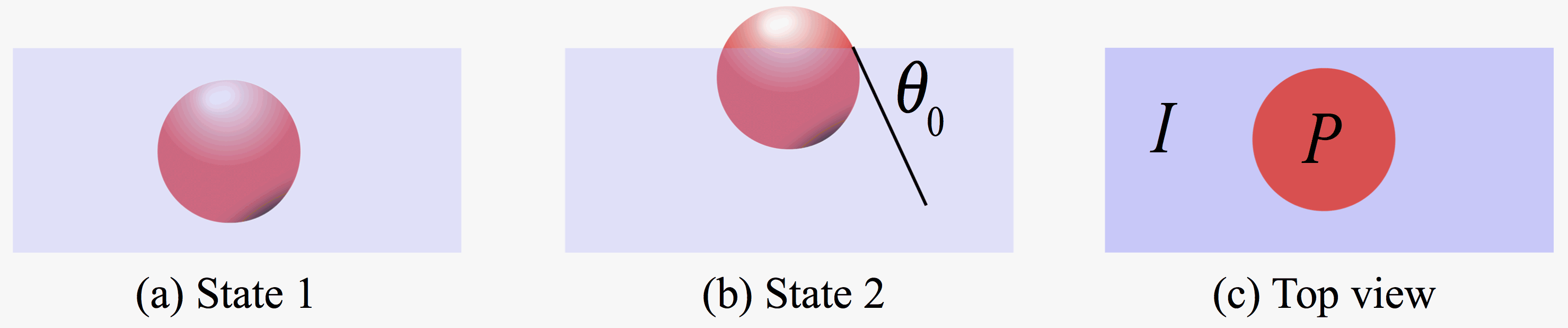}	     	    
 \caption{Schematic representations of (a) a sphere immersed in the liquid phase, State $1$ (b) a sphere at the liquid-vapor interface, State $2$, and (c) top view of a sphere trapped at the interface and particle $P$ and interface $I$ domains. } 
\label{fig_1}
 \end{figure}
 
First consider an isolated, spherical particle placed on an otherwise planar interface (see ref. \cite{Pieranski} for detail). We can evaluate the change in surface energy upon attachment of a sphere that is initially completely immersed in either of the sub-phase fluids (State $1$) to an interface (State $2$, in which the particle is partially immersed in each phase and there is a hole in the interface).  The integration domain can be divided into $I$, external to  $P$, the projection of the circular hole made by the particle in the $x$ - $y$ plane (see fig.~\ref{fig_1} for detail).\\

The free energy change upon particle attachment to the interface can be easily found.   The energy in State $1$ is simply the energy of the intact interface and the wetting energy of the solid in the liquid: 
\begin{eqnarray}
{E_1} = {\gamma _{SL}}4\pi {a^2} + {\gamma _{LV}}\mathop{{\int\!\!\!\!\!\int}\mkern-21mu \bigcirc}\limits_{I + P} 
 {{\rm{d}}x{\rm{d}}y},
\end{eqnarray}
where $I+P$ denotes the entire domain of the interface. The energy in State $2$ is the wetting energy of the solid in each phase and the area of the interface, which is reduced by the particle protruding through it:
\begin{eqnarray}
{E_2} = {\gamma _{SL}}{A_{SL,2}} + {\gamma _{SV}}{A_{SV,2}} + {\gamma _{LV}}\mathop{{\int\!\!\!\!\!\int}\mkern-21mu \bigcirc}\limits_{I} 
 {{\rm{d}}x{\rm{d}}y} ,
\end{eqnarray}
where $A_{SL,2}$ and $A_{SV,2}$ are the wetted areas of the solid in each phase after the particle is adsorbed. The energy difference is 
\begin{eqnarray}
\Delta E = {E_2} - {E_1} = ({\gamma _{SV}} - {\gamma _{SL}})\Delta {A_{SV}} + {\gamma _{LV}}\Delta {A_{LV}}.
\end{eqnarray}
After some manipulation, the changes in the liquid-solid and liquid-vapor areas can be found: 
\begin{align}
- \Delta {A_{LS}} &= \Delta {A_{VS}} = 2\pi {a^2}(1 - \cos {\theta _0})^2,\\
\Delta {A_{LV}} &=  - \mathop{{\int\!\!\!\!\!\int}\mkern-21mu \bigcirc}\limits_P 
 {{\rm{d}}x{\rm{d}}y}  =  - \pi {a^2}{\sin ^2}{\theta _0}.
\end{align}
The change in energy is:
\begin{eqnarray}
\Delta E = {E_2} - {E_1} =  - {\gamma _{LV}}\pi {a^2}{(1 - \left| {\cos {\theta _0}} \right|)^2}.
\end{eqnarray}

The absolute value  in the above expression is found by considering a particle initially immersed in the upper fluid; see ref.~\cite{Pieranski} for details. This is the trapping energy for a particle at a fluid interface.  By attaching, the particle reduces the area of the liquid vapor interface, decreasing the free energy.   This effect is modulated by the particle wetting properties.  For any particle that is not perfectly wet, being at the interface lowers the energy of the system.  The energy changes can be remarkably large.  For example, for air-water interfaces, the surface tension is $72~{{{{mN}}} \mathord{\left/{\vphantom {{{{mN}}} {{m}}}} \right.\kern-\nulldelimiterspace} {{m}}}$ or $18~{{{{{k}}_{{B}}}{{T}}} \mathord{\left/
 {\vphantom {{{{{k}}_{{B}}}{\rm{T}}} {{{n}}{{{m}}^{{2}}}}}} \right.
 \kern-\nulldelimiterspace} {{{n}}{{{m}}^{{2}}}}}$. Microparticles can be trapped with energies of $10^5-10^6~k_BT$.
 
\subsection{Particle with a pinned, undulated contact line}
In this system, the shape of the contact line can be decomposed into Fourier modes \cite{Stamou}. The shape of the interface obeys the Laplace equation $\nabla^2h=0$, and can be described by a multipole expansion excited by each Fourier mode at the contact line: 
\begin{eqnarray}
h(r,\phi ) = {a_0} + {b_0}\ln r + \sum\limits_{m = 1}^\infty  {({a_m}{r^m} + {b_m}{r^{ - m}})\cos m\phi  + ({c_m}{r^m} + {d_m}{r^{ - m}})\sin m\phi }.
\end{eqnarray}

In the absence of external forces and  torques, the monopole and dipole must be zero.\\

Let the interface shape around a spherical particle be
\begin{figure} 
\centering
\includegraphics[width=0.8 \textwidth]{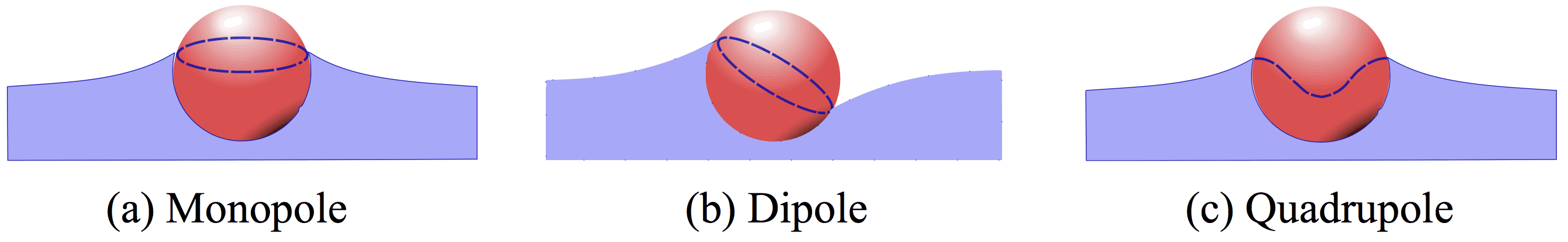}	     	    
 \caption{Schematic representation of the interface deformation of (a) a monopole, (b) a dipole and (c) a quadrupole. } 
\label{fig_2}
 \end{figure}
\begin{eqnarray}
h =  {b_0}\ln r,
\end{eqnarray}
where $r$ is the radial distance from the center of the particle in polar coordinates, and $b_0$ is constant. The shape of the interface in this case is shown in fig.~\ref{fig_2}(a). The capillary force exerted by the interface on any particle can be written as
\begin{eqnarray}
{\bf{F}} = \oint {{\gamma _{LV}}{\bf{m}}~{\rm{d}}s},\label{force}
\end{eqnarray}
where ${\rm{d}}s=a{\rm{d}}\phi$ is the arc length along the contact line, and $\bf{m}$ is defined by
\begin{eqnarray}
{\bf{m}} = {{\bf{e}}_t} \times {{\bf{e}}_n},
\end{eqnarray}
where ${\bf{e}}_n$ is the local unit vector normal to the interface and ${{\bf{e}}_t}$ is the unit vector tangent to the contact line.  This vector ${{\bf{e}}_t} \times {{\bf{e}}_n}$ is tangent to the interface, normal to the contour, and points outward from the boundary.\\ 

The corresponding ${\bf{e}}_n$ and ${{\bf{e}}_t}$ can be calculated in a cylindrical coordinate with origin at the particle center of mass as 
\begin{align}
&{{\bf{e}}_t} = {{\bf{e}}_\phi },\\
&{{\bf{e}}_n} =  \pm \frac{{{{\bf{e}}_z} - \frac{{\partial h}}{{\partial r}}{{\bf{e}}_r} - \frac{1}{r}\frac{{\partial h}}{{\partial \phi }}{{\bf{e}}_\phi }}}{{\sqrt {1 + {{(\frac{{\partial h}}{{\partial r}})}^2} + {{(\frac{1}{r}\frac{{\partial h}}{{\partial \phi }})}^2}} }}\sim \pm \left( {{{\bf{e}}_z} - \frac{{\partial h}}{{\partial r}}{{\bf{e}}_r} - \frac{1}{r}\frac{{\partial h}}{{\partial \phi }}{{\bf{e}}_\phi }} \right)\label{en}.
\end{align}
In the above expression, we have utilized the small gradient approximation, i.e. $|\nabla h| \ll 1$. Taking appropriate derivatives and evaluating at $r=a$,
\begin{eqnarray}
{{\bf{e}}_n} =  \pm {\left. {\left( {{{\bf{e}}_z} - \frac{{{b_0}}}{r}{{\bf{e}}_r}} \right)} \right|_{r = a}}.
\end{eqnarray}
We choose ${{\bf{e}}_n}$ to point from the liquid to the vapor and therefore consider the positive case
\begin{eqnarray}
{{\bf{e}}_n} = {{\bf{e}}_z} - \frac{{{b_0}}}{a}{{\bf{e}}_r}.
\end{eqnarray}
To calculate ${\bf{m}}$, substitute the expressions for ${{\bf{e}}_n}$ and ${{\bf{e}}_t}$ and find the cross product:
\begin{eqnarray}
{\bf{m}} = {{\bf{e}}_t} \times {{\bf{e}}_n} = \left| {\begin{array}{*{20}{c}}
{{{\bf{e}}_r}}&{{{\bf{e}}_\phi }}&{{{\bf{e}}_z}}\\
0&1&0\\
- {{{b_0}} \mathord{\left/
 {\vphantom {{{b_0}} a}} \right.
 \kern-\nulldelimiterspace} a}&0&1
\end{array}} \right| = {{\bf{e}}_r} + \frac{{{b_0}}}{a}{{\bf{e}}_z}.
\end{eqnarray}
Note that ${{\bf{e}}_r}$ can be written in Cartesian coordinates as 
\begin{eqnarray}
{{\bf{e}}_r} = \cos \phi ~{{\bf{e}}_x} + \sin \phi ~{{\bf{e}}_y},
\end{eqnarray}
and hence the force can be evaluated as:
\begin{align}
{\bf{F}} &= \oint {{\gamma _{LV}}{\bf{m}}{\rm{d}}s}  = {\gamma _{LV}}\int_0^{2\pi } {\left( {{{\bf{e}}_r} + \frac{{{b_0}}}{a}{{\bf{e}}_z}} \right)a{\rm{d}}\phi } \nonumber\\
&= {\gamma _{LV}}\int_0^{2\pi } {\left( {\cos \phi {{\bf{e}}_x} + \sin \phi {{\bf{e}}_y} + \frac{{{b_0}}}{a}{{\bf{e}}_z}} \right)a{\rm{d}}\phi }\nonumber\\
&= 2\pi {\gamma _{LV}}{b_0}{{\bf{e}}_z}.
\end{align}
In the absence of an external force,  ${\bf{F}}=0$ and ${b_0}=0$. All higher order multipoles give zero force as is apparent from the symmetries of their distortions.\\ 

A particle with a dipolar contact line is pulled down on one side and up on the other and thus  experiences a force couple [see fig.~\ref{fig_2}(b)]. Absent any external torque, this configuration cannot be supported.  We show this by evaluating the torque associated with the dipole, and equating it to zero. For a particle trapped at an interface creating a dipolar deformation, the interface profile can be written as
\begin{eqnarray}
h = \frac{{{b_1}a}}{r}\cos \phi.
\end{eqnarray}
The tangent to the interface pointing outward from the particle is
\begin{eqnarray}
{\bf{m}} = {{\bf{e}}_r} - \frac{{{b_1}}}{a}\cos \phi~{{\bf{e}}_z}.
\end{eqnarray}
By changing ${{\bf{e}}_r}$ back to Cartesian coordinates using 
\begin{eqnarray}
{\bf{m}} = \left( {\cos \phi~{{\bf{e}}_x} + \sin \phi~{{\bf{e}}_y}} \right) - \frac{{{b_1}}}{a}\cos \phi~{{\bf{e}}_z},
\end{eqnarray}
the torque can be calculated
 \begin{eqnarray}
{\bf{T}} = \oint {{\gamma _{LV}}~{\bf{R}} \times {\bf{m}}~{\rm{d}}s}  ={{\gamma _{LV}}} \int_0^{2\pi } {{\bf{R}} \times ({{\bf{e}}_t} \times {{\bf{e}}_n})}~a~{\rm{d}}\phi,
\end{eqnarray}
where ${\bf{R}}=a~{\bf{e}}_R$ and ${\bf{e}}_R$ is the unit vector in the radial direction in spherical coordinates. Hence, the cross product is 
\begin{eqnarray}
{\bf{R}} \times {\bf{m}} = \left| {\begin{array}{*{20}{c}}
{{{\bf{e}}_r}}&{{{\bf{e}}_\phi }}&{{{\bf{e}}_z}}\\
{a\sin \theta }&0&{a\cos \theta }\\
1&0&{ - \frac{{{b_1}}}{a}\cos \phi }
\end{array}} \right|,
\end{eqnarray}
where $\theta$ is a polar angle. Since ${b_1} \ll a$, we can consider only the leading order up to ${{{b_1}} \mathord{\left/
 {\vphantom {{{b_1}} a}} \right.
 \kern-\nulldelimiterspace} a}$ and rewrite $\cos\theta$ and $\sin\theta$ in the form
\begin{align}
\cos \theta  &= \sin \left( {\frac{\pi }{2} - \theta } \right) = \frac{{{b_1}}}{a}\cos \phi,\nonumber \\
\sin \theta  &= \sqrt {1 - {{\left( {\frac{{{b_1}}}{a}\cos \phi } \right)}^2}}  \sim 1.
\end{align}
Then, ${\bf{R}} \times {\bf{m}}$ can be simplified to
\begin{align}
&{\bf{R}} \times {\bf{m}} = \left| {\begin{array}{*{20}{c}}
{{{\bf{e}}_r}}&{{{\bf{e}}_\phi }}&{{{\bf{e}}_z}}\nonumber\\
a&0&{{b_1}\cos \phi }\\
1&0&{ - \frac{{{b_1}}}{a}\cos \phi }
\end{array}} \right|\\
&= {{\bf{e}}_\phi }({b_1}\cos \phi  + {b_1}\cos \phi ) = 2{b_1}\cos \phi~{{\bf{e}}_\phi },
\end{align}
and the torque can be rewritten in the form
\begin{align}
&{\bf{T}} = \oint {{\gamma _{LV}}{\bf{R}} \times {\bf{m}}{\rm{d}}s}  = \int_0^{2\pi } {{\gamma _{LV}}{\bf{R}} \times } ({{\bf{e}}_t} \times {{\bf{e}}_n}){\rm{d}}s\nonumber\\
&= {\gamma _{LV}}\int_0^{2\pi } {2{b_1}\cos \phi \,{{\bf{e}}_\phi }} \,a{\rm{d}}\phi. 
\end{align}
By substituting ${{\bf{e}}_\phi } =  - \sin \phi~ {{\bf{e}}_x} + \cos \phi~ {{\bf{e}}_y}$, we have:
\begin{align}
{\bf{T}} &= \oint {{\gamma _{LV}}{\bf{R}} \times {\bf{m}}{\rm{d}}s}  = \int_0^{2\pi } {{\gamma _{LV}}{\bf{R}} \times } \left( {{{\bf{e}}_t} \times {{\bf{e}}_n}} \right){\rm{d}}s\nonumber\\
&= {\gamma _{LV}}\int_0^{2\pi } {\,\left[ {\left( { - 2a{b_1}\cos \phi \sin \phi~{{\bf{e}}_x}} \right) + \left( {2a{b_1}\cos \phi \cos \phi~{{\bf{e}}_y}} \right)} \right]{\rm{d}}\phi }\nonumber\\
&= 2\pi a{\gamma _{LV}}{b_1}~ {{\bf{e}}_y}.
\end{align}
In the absence of any external torque, ${\bf{T}}=0$ and ${b_1}=0$.\\

Having discarded the monopole and the dipole, the quadrupole is therefore the first surviving mode in the multipole  expansion \cite{Stamou}. This is important, since this describes the long-range distortion from any particle with any undulated contact line. We can calculate the trapping energy for a particle on the interface with such a distortion. Let $h_{qp}$ be the amplitude of the quadrupolar mode around the particle and calculate the interface height,
\begin{figure} 
\centering
\includegraphics[width=0.3 \textwidth]{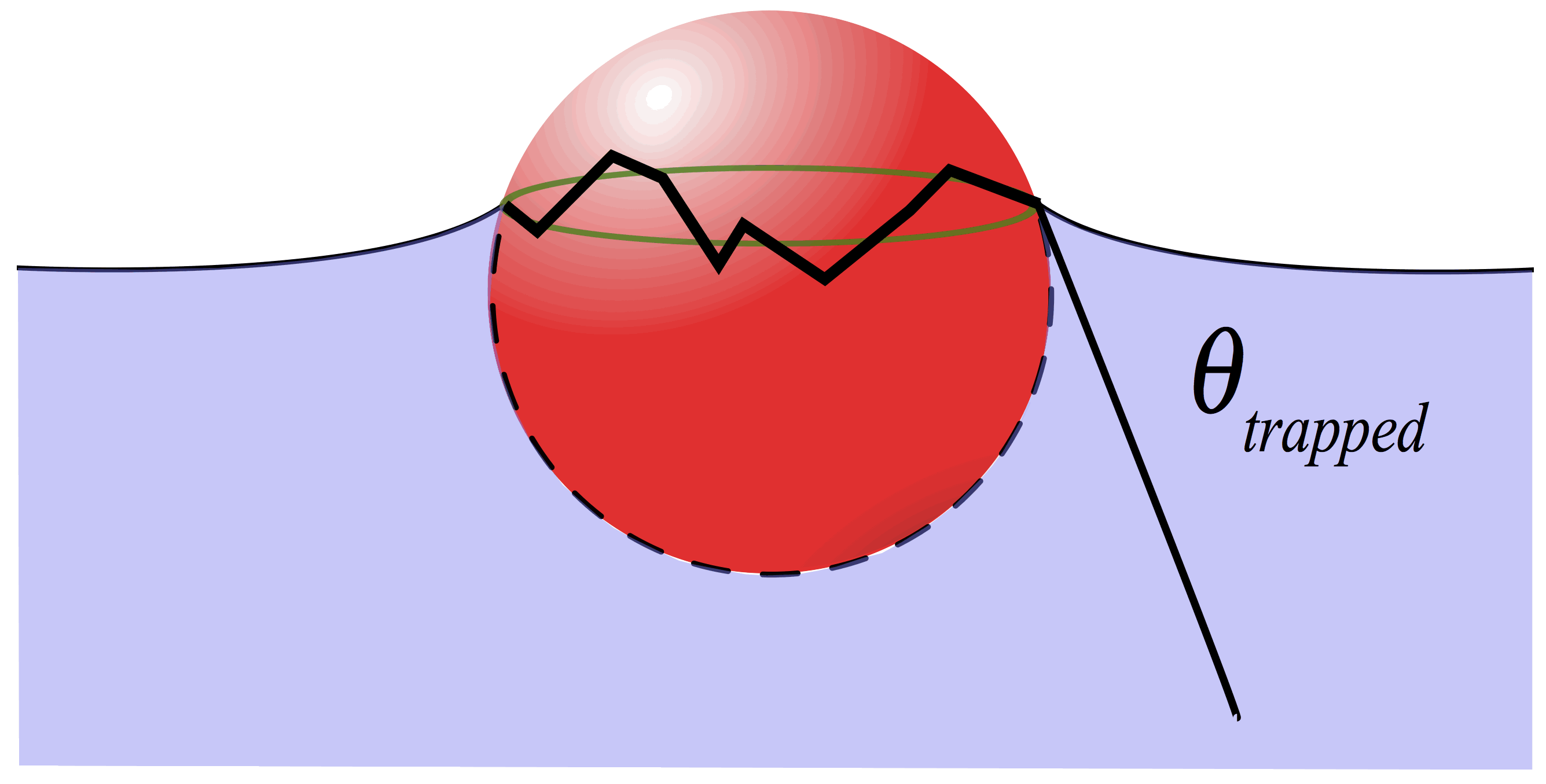}	     	    
 \caption{Schematic representations of a particle with a pinned, undulated contact line trapped at a flat interface.} 
\label{fig_3}
 \end{figure} 
\begin{eqnarray}
h(r,\phi ) = {h_{qp}}\frac{{{a^2}}}{{{r^2}}}\cos 2\phi.
\end{eqnarray}
The shape of the quadrupolar deformation of the interface is shown in fig.~\ref{fig_2}(c). The energy in State $2$ can be cast as: 
\begin{eqnarray}
{E_2} = {\gamma _{SL}}{A_{SL,2}} + {\gamma _{SV}}{A_{SV,2}} + {\gamma _{LV}}\mathop{{\int\!\!\!\!\!\int}\mkern-21mu \bigcirc}\limits_I {{\rm{d}}A_{LV}},
\end{eqnarray}
\begin{align}
&\mathop{{\int\!\!\!\!\!\int}\mkern-21mu \bigcirc}\limits_I 
 {{\rm{d}}{A_{LV}}}  = \mathop{{\int\!\!\!\!\!\int}\mkern-21mu \bigcirc}\limits_I 
 {\left({1 + \frac{{\nabla h \cdot \nabla h}}{2}}\right){\rm{d}}x{\rm{d}}y}. 
\end{align}
The expression for ${\rm{d}}{A_{LV}}$ includes not only the hole made by the particle in the interface, but the interface area increase by the distortion around the particle. Each mode excited has an associated distortion energy.  We focus on the quadrupolar mode because of its special importance. Upon evaluation of these terms, the trapping energy becomes:
\begin{align}
&\Delta E_{planar} = E_2-E_1= - {\gamma _{LV}}\pi {a^2}(1 - \left| {\cos {\theta _{trapped}}} \right|)^{2} + {E_{dist,{h_{qp}}}},\label{p1}\\
&{E_{dist,{h_{qp}}}} = {\gamma _{LV}}\pi h_{qp}^2,\label{p2}
\end{align}
The first term is similar to the equilibrium case except that the angle characterizing the degree of immersion in the trapped state $\theta_{trapped}$ replaces $\theta_{0}$ (see fig.~\ref{fig_3}). The second term is the energy associated with the excess area of the distortion around the particle.\\

\subsection{Non-spherical particles} 
\begin{figure} 
\centering
\includegraphics[width=0.75 \textwidth]{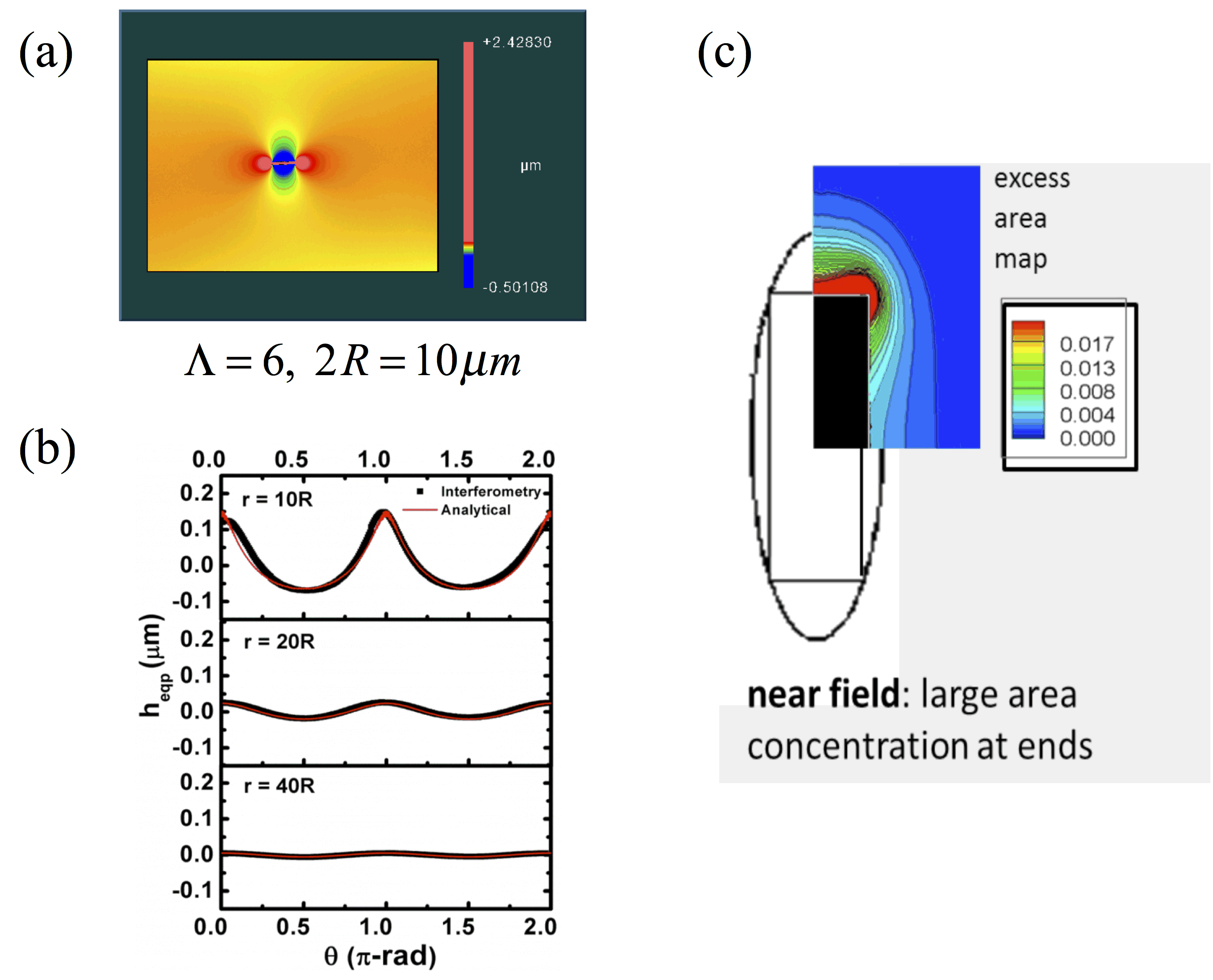}	     	    
 \caption{(a) Interferogram of the interface shape around a partially-wet cylindrical microparticle (60 $\mu m$ long, 10 $\mu m$ wide) at an air-water interface. (b) Interface shape at circles various distances from the particle. Far from the particle, the deformation is a polar quadrupole.  Closer to the particle, the interface is well described by a quadrupole in elliptical coordinates. (c) Simulated excess area around a partially wet cylinderical microparticle. There is significant excess area near the corners, which influences the strength of the capillary bond and orientation during assembly.} 
\label{fig_4}
 \end{figure}
Non-spherical particles also become trapped in planar interfaces.  See, for example, fig.~\ref{fig_4}(a) for the interface shape around a cylindrical microparticle obtained by interferometry.  The main contributions to the trapping energy remain the same, i.e. the hole in the interface, modulated by the particle wetting energies and the energy owing to the excess area of the distortion.  The evaluation of each of these terms is now far more complex, as the minimum free energy wetting configuration of complex shaped particles cannot typically be derived analytically (see ref.~\cite{Lewandowski2010,Botto,Lorenzo} for details).  Can we simplify this problem?\\

We know that in the far field, the particle-induced deformation always decays to a quadrupolar distortion.  Closer to the particle, other expansions can become appropriate, for example, for elongated particles, distortions are described well by quadrupolar modes in elliptical coordinates  within a few radii of contact with the particle \cite{Lewandowski2010,Lorenzo}.  In the very near-field, however, only simulations will capture details like area distortions near sharp edges and corners [see figs.~\ref{fig_4}(b) and (c)].\\

Other features can also play a role, including particle roughness on scales much smaller than the particle length scale \cite{Lucassen}. We can approximate this contact line feature as a wavy contour on a planar edge. The undulations can be  decomposed into Fourier modes. The distortions made by such wavy features die out over distances similar to the wavelength of the undulations and will change the energy landscape around the particle.\\

To conclude this section, particles become trapped at planar fluid interfaces. If they are perfectly smooth spheres obeying equilibrium wetting conditions, they become trapped while letting the interface around them remain unperturbed. If they have pinned contact lines, patchy wetting or non-spherical shapes, they distort the interface around them.  Distortions due to various particle features are observed at different distances from the particle.  All particles make quadrupolar distortions in  the far field.  In the moderate to near-field, features like particle elongation become apparent; closer still, waviness, roughness and sharp edges play a role \cite{Lu}.\\ 
  
\section{Isolated particles trapped on curved interfaces}
What if the interface is curved?  How does the energy trapping the particle in the interface change?  Here we discuss this for a particle with pinned, undulated contact lines with associated quadrupolar modes of amplitude $h_{qp}$\cite{Disk}. We can locally expand any host interface near the particle center in terms of the mean curvature ${H_0} = {{({c_1} + {c_2})} \mathord{\left/
 {\vphantom {{({c_1} + {c_2})} 2}} \right.
 \kern-\nulldelimiterspace} 2} = {{({1 \mathord{\left/
 {\vphantom {1 {{R_1}}}} \right.
 \kern-\nulldelimiterspace} {{R_1}}} + {1 \mathord{\left/
 {\vphantom {1 {{R_2}}}} \right.
 \kern-\nulldelimiterspace} {{R_2}}})} \mathord{\left/
 {\vphantom {{({1 \mathord{\left/
 {\vphantom {1 {{R_1}}}} \right.
 \kern-\nulldelimiterspace} {{R_1}}} + {1 \mathord{\left/
 {\vphantom {1 {{R_2}}}} \right.
 \kern-\nulldelimiterspace} {{R_2}}})} 2}} \right.
 \kern-\nulldelimiterspace} 2}$ and the deviatoric curvature $\Delta {c_0} = {c_1} - {c_2} = {1 \mathord{\left/
 {\vphantom {1 {{R_1}}}} \right.
 \kern-\nulldelimiterspace} {{R_1}}} - {1 \mathord{\left/
 {\vphantom {1 {{R_2}}}} \right.
 \kern-\nulldelimiterspace} {{R_2}}}$, where $c_1$ and $c_2$ are the principal curvatures of the interface defined in terms of the principal radii of curvature $R_1$ and $R_2$ evaluated at the particle center of mass. In  the absence of the particle, assuming $ac\ll 1$ and $\left| {\nabla h} \right| \ll 1$,  where $c$ is either of the principal curvatures, the interface shape is
\begin{eqnarray}
{h_{host}} = \frac{{{{r^2}H_0}}}{2} + \frac{{\Delta {c_0}}}{4}{r^2}\cos 2\phi.
\end{eqnarray}
When a particle attaches to such an interface, the trapping energy becomes:
\begin{eqnarray}
\Delta E = {E_2} - {E_1} = {\gamma _{SL}}\mathop{{\int\!\!\!\!\!\int}\mkern-21mu \bigcirc} 
 {{\rm{d}}{A_{SL}}}  + {\gamma _{SV}}\mathop{{\int\!\!\!\!\!\int}\mkern-21mu \bigcirc} 
 {{\rm{d}}{A_{SV}}}  + {\gamma _{LV}}\mathop{{\int\!\!\!\!\!\int}\mkern-21mu \bigcirc} 
 {{\rm{d}}{A_{LV}}}  -  ~{p\Delta V}.
\end{eqnarray}
\begin{figure} 
\centering
\includegraphics[width=0.75 \textwidth]{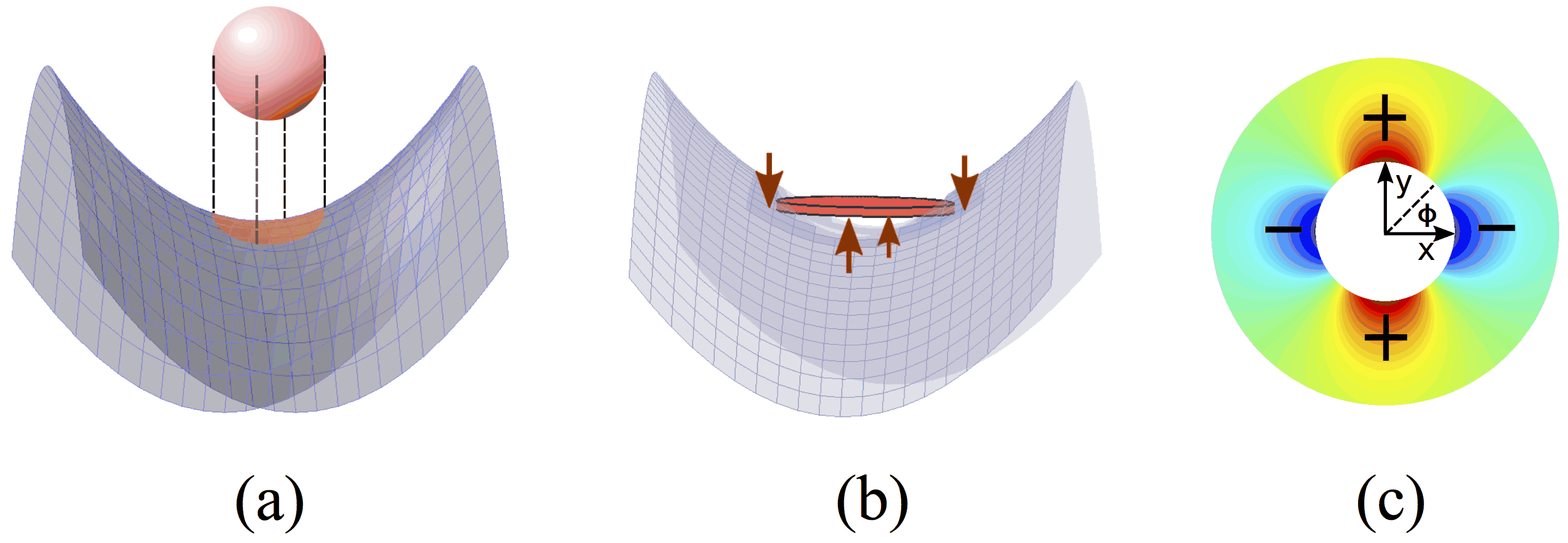}	     	    
 \caption{Schematic representations of (a) the area of the hole under the particle, (b) induced disturbance due to pinned contact line, and (c) induced quadrupolar deformation at the particle contact line (top view). } 
\label{fig_5}
 \end{figure}
 \\
There are several contributions: (i) The wetting energies, which, given  the symmetries of the Fourier modes, are unchanged from the planar case. (ii) The changes in area of the interface. This quantity must be treated with care. To impose the boundary condition at the contact line, the particle must distort the host interface, working against the host interface curvature at the contact line.  In addition, the hole made by the particle in the interface depends on the curvature field [see fig.~\ref{fig_5}(a) for a schematic]. (iii) Finally, for finite mean curvature, the particle performs pressure-volume work against the pressure jump.  To determine these contributions, we solve for the disturbance made to the interface shape by the particle $\eta=h-h_{host}$, which requires solution of a simple boundary value problem,
\begin{align}
&{\nabla ^2}h = 0
\end{align}
with the boundary conditions
\begin{align}
&h(r = a) = {h_{qp}}\cos 2\phi ,\\
&h(r \to \infty ) = {h_{host}},
\end{align}
where
\begin{align}
&h = {h_{host}} + \eta,
\end{align}
and
\begin{align}
&\eta  = \left({{h_{qp}} - \frac{{{a^2}\Delta {c_0}}}{4}}\right)\frac{{{a^2}}}{{{r^2}}}\cos 2\phi ,
\end{align}
The disturbance can be broken into two parts: The particle-imposed distortion $\eta_{qp}=h_{qp}\frac{a^2}{r^2}\cos2\phi$   and the induced disturbance that fights the deviatoric curvature, also called a reflected mode ${\eta _{ind}} =  - \frac{{{a^2}\Delta {c_0}}}{4}\frac{{{a^2}}}{{{r^2}}}\cos 2\phi$ [figs.~\ref{fig_5}(b) and (c)]. Additionally, the particle shifts vertically to situate itself in the interface with finite  mean curvature $\omega_0={{{a^2}{H_0}} \mathord{\left/
 {\vphantom {{{a^2}{H_0}} 2}} \right.
 \kern-\nulldelimiterspace} 2}$. This adjustment of the particle height does work against the pressure drop owing to mean curvature,  
\begin{eqnarray}
\Delta p\mathop{{\int\!\!\!\!\!\int}\mkern-21mu \bigcirc}\limits_P 
 {\left({{h_{host}} - \frac{{{H_0}{a^2}}}{2}}\right){\rm{d}}A}  =  - {\gamma _{LV}\pi}{a^2}\frac{ H_0^2{a^2}}{2}.
\end{eqnarray}
The associated excess area of the interface is: 
\begin{align}
&\Delta {A_{LV,particle}} - \Delta {A_{LV,curved}} =  - \mathop{{\int\!\!\!\!\!\int}\mkern-21mu \bigcirc}\limits_P 
 {\frac{{\nabla {h_{host}} \cdot \nabla {h_{host}}}}{2}{\rm{d}}x{\rm{d}}y}  + \mathop{{\int\!\!\!\!\!\int}\mkern-21mu \bigcirc}\limits_I 
 {\frac{{\nabla {\eta _{ind}} \cdot \nabla {\eta _{ind}}}}{2}{\rm{d}}x{\rm{d}}y}\nonumber \\
&+ \mathop{{\int\!\!\!\!\!\int}\mkern-21mu \bigcirc}\limits_I 
 {{{\nabla {\eta _{ind}} \cdot \nabla {\eta _{qp}}}}{\rm{d}}x{\rm{d}}y}  + \mathop{{\int\!\!\!\!\!\int}\mkern-21mu \bigcirc}\limits_I 
 {\frac{{\nabla {\eta _{qp}} \cdot \nabla {\eta _{qp}}}}{2}{\rm{d}}x{\rm{d}}y}+ \mathop{{\int\!\!\!\!\!\int}\mkern-21mu \bigcirc}\limits_I 
 {{{\nabla {\eta} \cdot \nabla {h_{host}}}}{\rm{d}}x{\rm{d}}y},\label{int-1}
\end{align}
where $\Delta {A_{LV,particle}}$ and $\Delta {A_{LV,curved}}$ are the areas of the interfaces in the presence and absence of the particle on curved interface respectively. The first integral is the curvature contribution to the hole in the interface
 \begin{eqnarray}
 - \mathop{{\int\!\!\!\!\!\int}\mkern-21mu \bigcirc}\limits_P 
 {(\frac{{\nabla {h_{host}} \cdot \nabla {h_{host}}}}{2}){\rm{d}}x{\rm{d}}y}  =  - \pi {a^2}\left({\frac{{{a^2}H_0^2}}{4} + \frac{{{a^2}\Delta c_0^2}}{{16}}}\right).\label{host-11}
 \end{eqnarray}
 The second integral in eq.~(\ref{int-1}) is the curvature contribution to the induced disturbance around the particle.  Using the divergence theorem, this term must be equal and opposite to the deviatoric curvature contribution of the hole under the particle [see eq.~(\ref{host-11})].  The third term  is the interaction of two disturbance terms, and is finite. 
\begin{eqnarray}
\mathop{{\int\!\!\!\!\!\int}\mkern-21mu \bigcirc}\limits_I 
 {{\nabla {\eta _{qp}} \cdot \nabla {\eta _{ind}}}{\rm{d}}x{\rm{d}}y}  =  - \frac{\pi }{2}\Delta {c_0}{a^2}{h_{qp}}.
\end{eqnarray}
The fourth term is ${E_{dist,{h_{qp}}}}$and the fifth term is identically zero. Gathering this together, we can show that  energy change upon attaching to the curved  interface is: 
\begin{eqnarray}
\Delta E = \Delta {E_{planar}} - {\gamma _{LV}}\pi {a^2}\left({\frac{{3{a^2}H_0^2}}{4} + \frac{{{h_{qp}}\Delta {c_0}}}{2}}\right),\label{energy-221}
\end{eqnarray}
where $\Delta {E_{planar}}$ is defined in  eqs.~(\ref{p1}) and (\ref{p2}). For particles on curved interfaces, the trapping energy is reduced for finite mean and deviatoric curvatures. 
In this discussion, we assumed that the particle-sourced quadrupole was aligned with the saddle shape of the interface.  If this were not the case, there would be a capillary torque exerted on the particle which would cause the particle to align \cite{Eric}. We will revisit the implications of these relationships shortly.
\section{Pair interaction on planar interfaces}
\begin{figure} 
\centering
\includegraphics[width=0.6 \textwidth]{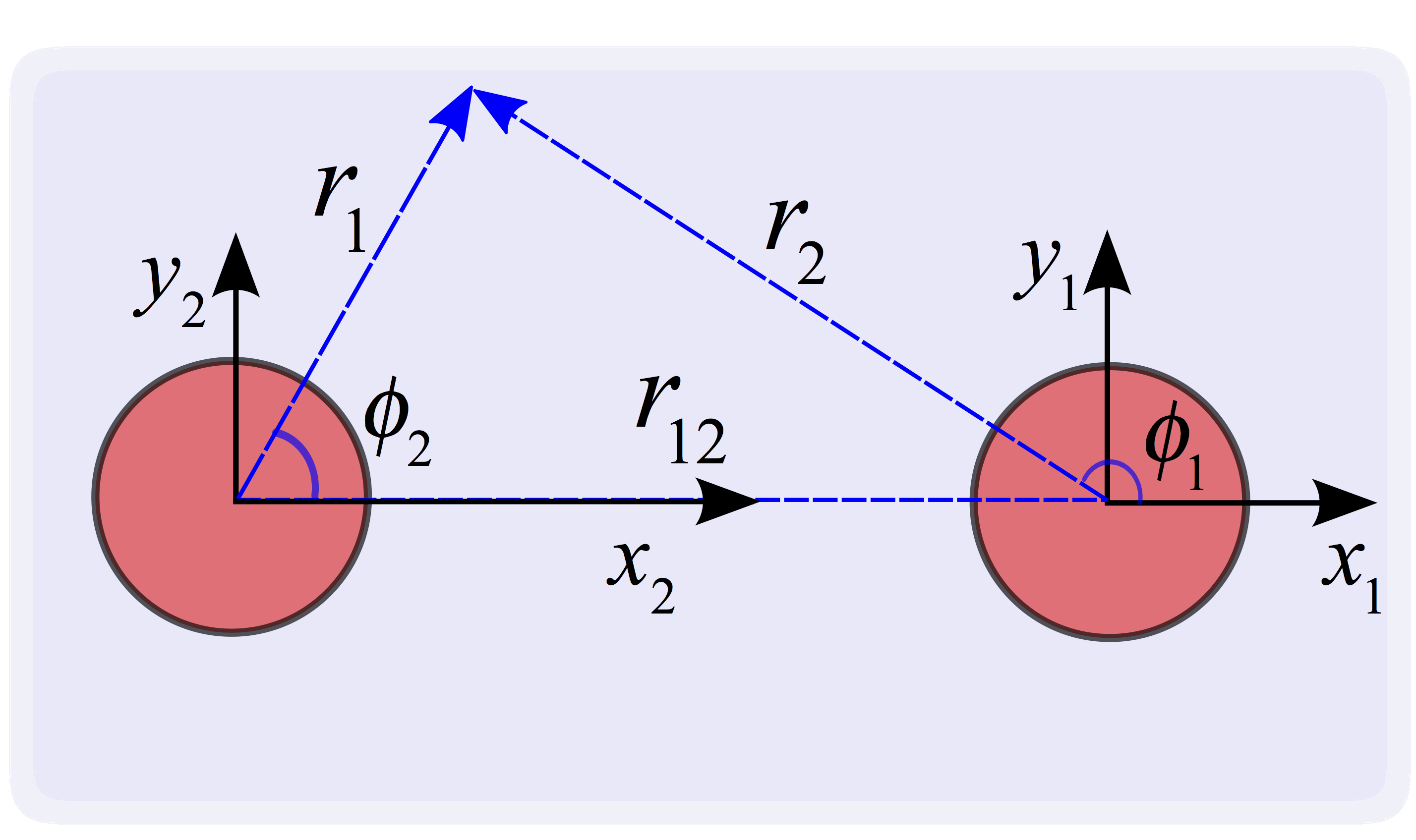}	     	    
 \caption{Schematic representation of two particles on a flat interface (top view). } 
\label{fig_6}
 \end{figure}
 
We calculate the capillary interaction energy of two colloidal particles with pinned contact lines on an otherwise planar interface as shown in fig.~\ref{fig_6}. Here, we first provide the far field solution utilizing the method of reflections and thereafter we capture the exact solution in bipolar coordinate.

\subsection{Method of reflections}\label{reflection}
Suppose two particles $1$ and $2$ of radii $a$ are separated by distance $r_{12}$ so that  ${a \mathord{\left/
 {\vphantom {a {{r_{12}}}}} \right.
 \kern-\nulldelimiterspace} {{r_{12}}}}\ll 1$. The particles have pinned contact lines with associated quadrupolar modes. The shape of the interface around each colloid in the absence of the other can be expressed in terms of polar coordinates located at the centers of the particles
\begin{align}
&h_1 =  \frac{{{h_{{qp_1}}}{a^2}}}{{r_1^2}}\cos 2({\phi _1}- {\alpha}_1)
\end{align}
and
\begin{align}
&h_2 = \frac{{{h_{{qp_2}}}{a^2}}}{{r_2^2}}\cos 2({\phi _2}-\alpha_2),
\end{align}
where $h_{{qp}_j}$ is the amplitude of quadrupolar roughness of the $j$th particle's contact line $\partial S_{p_j}$ and $\alpha_j$ denotes the phase angle of the $j$th particle with respect to the line connecting particle centers.

We can expand the distortion made by particle $2$ in the region of particle $1$ in a Taylor series:
 \begin{eqnarray}
 {h_2} = {\left. {{h_2}} \right|_{{{\bf{r}}_{12}}}} + {\bf{r}}_1 \cdot {\left. {\nabla {h_2}} \right|_{{{\bf{r}}_{12}}}} + {\bf{r}}_1 \cdot {\left. {\frac{{\nabla \nabla {h_2}}}{2}} \right|_{{{\bf{r}}_{12}}}} \cdot {\bf{r}}_1 + ...,\label{taylor}
 \end{eqnarray}
 where ${\bf{r}}_1$ is the position vector with respect to the coordinate centered at particle $1$ and ${\bf{r}}_{12}$ is the position vector connecting center of particle $1$ to the particle $2$.  Evaluating these terms, we find 
 \begin{align}
{\left. {{h_2}} \right|_{{{\bf{r}}_{12}}}} &= \frac{{{h_{{{qp}_2}}}{a^2}}}{{{r_{12} ^2}}} - \frac{{2{h_{{{qp}_2}}}{a^2}}}{{{r_{12} ^3}}}[\cos2\alpha_2-\sin2\alpha_2]{r_1}\cos {\phi _1} \\
&+ \frac{{3{h_{{{qp}_2}}}{a^2}}}{{{r_{12} ^4}}}r_1^2\cos 2({\phi _1}+\alpha_2).\nonumber \label{expansion}
 \end{align}
 
In the above expression, the first term is a constant which does not contribute to the interface area (nor the energy). The second term is a dipolar deformation and cannot persist, i.e.  particle 1  rotates to eliminate this term. The third term is the curvature field created by particle $2$ in the vicinity of particle $1$.  We can solve the boundary value problem to find the shape of the interface in the plane tangent to the interface around particle 1 owing to this curvature field:
 \begin{align}
&{\nabla ^2}{h_1} = 0,
\end{align}
with boundary conditions
 \begin{align}
&{h_1}({r_1} = a) = {h_{{{qp}_1}}}\cos 2({\phi _1} - {\alpha _1})
\end{align}
and
 \begin{align}
&{h_1}({r_1} \to \infty ) = \frac{{3{h_{{{qp}_2}}}{a^2}}}{{r_{12}^4}}r_1^2\cos 2({\phi _1} + {\alpha _2}),
\end{align}
where
 \begin{align}
&{h_1} = \frac{{3{h_{{{qp}_2}}}{a^2}}}{{r_{12}^4}}r_1^2\cos 2({\phi _1} + {\alpha _2}) + {\eta _1}
\end{align}
and
 \begin{align}
&{\eta _1} = {h_{{{qp}_1}}}\frac{{{a^2}}}{{r_1^2}}\cos 2({\phi _1} - {\alpha _1}) - \frac{{3{h_{{{qp}_2}}}{a^2}}}{{r_{12}^4}}\frac{{{a^4}}}{{r_1^2}}\cos 2({\phi _1} + {\alpha _2}).
\end{align}

The disturbance includes a particle sourced term and an induced or reflected term ``undoing'' the curvature created by particle $2$. In the method of reflections which we employ here, we ignore the details of particle $2$, which is now treated as a far field distortion source.  We consider the energy difference between two states. State II is the trapping energy of particle $1$ in the disturbance field created by particle $2$. State I is the trapping energy of particle $1$ without a neighbor.  Thus, the energy of particle 1 interacting with its neighbor becomes: 
\begin{eqnarray}
\Delta {E_1}=E_{II}-E_{I} =  - \gamma_{LV} \pi {a^2}\frac{{6{h_{{{qp}_1}}}{h_{{{qp}_2}}}{a^2}}}{{r_{12}^4}}\cos 2({\alpha _1} + {\alpha _2}).\label{pair-22}
\end{eqnarray}
A similar treatment for particle $2$ yields identical capillary migration energy and therefore we define the total interaction energy between the two particles via a superposition to yield:
\begin{eqnarray}
\Delta E = \Delta {E_1} + \Delta {E_2}=- \gamma_{LV} \pi {a^2}\frac{{12{h_{{qp_1}}}{h_{{qp_2}}}{a^2}}}{{{r_{12} ^4}}}\cos 2({\alpha _1} + {\alpha _2}).
\end{eqnarray} 

\subsection{Exact solution in bipolar coordinate}
In this section we provide an exact solution of the interface height  in the presence of two particles with pinned contact lines in bipolar coordinate \cite{Arfken,Danov_bipolar}. We give the details so students can do the calculation as an exercise.  The height of interface is governed by the Laplace equation:
 \begin{eqnarray}
 {\nabla ^2}h = 0,\label{laplace-el}
 \end{eqnarray}
 
We introduce bipolar coordinates ($u$,$v$), which can be related to Cartesian coordinates by:
\begin{figure} 
\centering
\includegraphics[width=0.5 \textwidth]{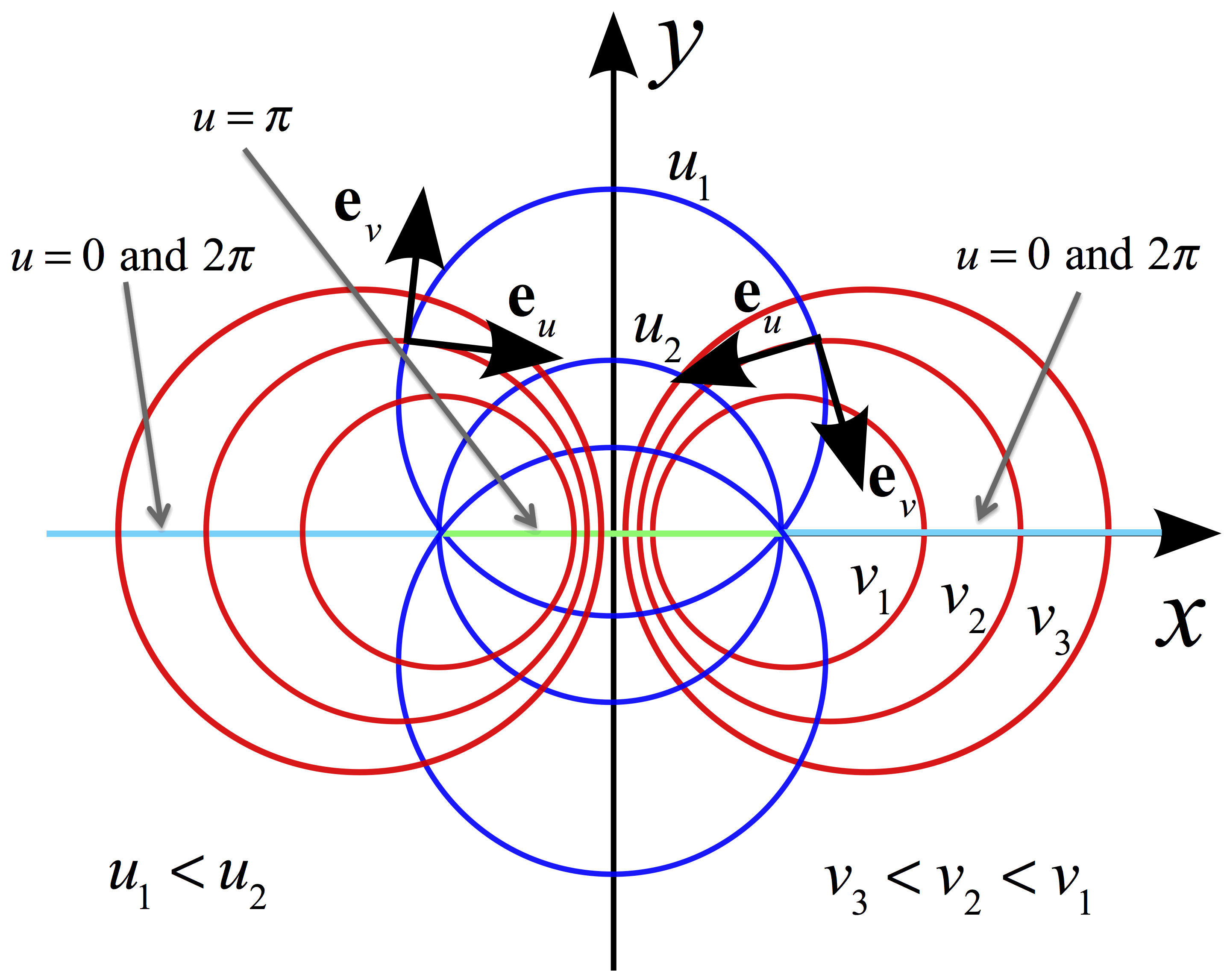}	     	    
 \caption{Sketch of bipolar coordinates ($u$,$v$).} 
\label{fig_7}
 \end{figure}
 \begin{eqnarray}
x = \frac{{\zeta \sinh v}}{{\cosh v - \cos u}},\label{x-el}
 \end{eqnarray}
 \begin{eqnarray}
y = \frac{{\zeta\sin u}}{{\cosh v - \cos u}},\label{y-el}
\end{eqnarray}
where $v$ is a real number, $u$ varies  between $0$ to $2\pi$, and $2\zeta$ is the focal distance.  Families of coordinate curves,  shown in fig.~\ref{fig_7}, correspond to:
 \begin{enumerate}
 \item
Circular circles located at $y=\zeta\cot u$, $u$ = $const.$, $0 \le u < 2\pi $. \\ \indent
 \item
Circular circles located at $x=\zeta\coth v$, $-\infty \le v< \infty $. \\ \indent
\end{enumerate} 
$\zeta$ can be related to the radii of the particles according to
\begin{eqnarray}
{\zeta ^2} = \frac{1}{4}\left( {r_{12}^2 - 4{a^2}} \right).
\end{eqnarray}

The contact lines for the adjacent particles are contours of constant  $v$, specifically, $v=-s_1$ and $v=s_2$ where
\begin{align}
&{s_1} = {s_2} = {\cosh ^{ - 1}}\left( {\frac{{{r_{12}}}}{{2a}}} \right)
\end{align}
and
\begin{align}
&{\cosh ^{ - 1}}u = \ln \left[ {u + \sqrt {{u^2} - 1} } \right],
\end{align}
and the distance between center of each particle to the origin can be evaluated as,
\begin{eqnarray}
{L_1} = {L_2} = \frac{{{b^2} + 4a\left( {b + a} \right)}}{{2\left( {b + 2a} \right)}},
\end{eqnarray}
where $b$ is the minimum separation defined as,
\begin{eqnarray}
b = r_{12} - 2a.
\end{eqnarray}
The Laplacian operator can be written in terms of these coordinates as,
\begin{eqnarray}
{\nabla ^2} = \frac{{{{\left( {\cosh v - \cos u} \right)}^2}}}{{{\zeta ^2}}}\left( {\frac{{{\partial ^2}}}{{\partial {v^2}}} + \frac{{{\partial ^2}}}{{\partial {u^2}}}} \right),
\end{eqnarray}

The pinning boundary condition on the contact line requires that
\begin{eqnarray}
h(v =  - {s_1}) = {h_{q{p_1}}}\cos 2\left( {{\phi _1} - {\alpha _1}} \right) - \frac{{2{h_{q{p_2}}}{a^3}}}{{r_{12}^3}}\left( {\cos 2{\alpha _2} - \sin 2{\alpha _2}} \right)\cos {\phi _1}\label{pinned_c1}
\end{eqnarray}
and
\begin{eqnarray}
h(v = {s_2}) = {h_{q{p_2}}}\cos 2\left( {{\phi _2} - {\alpha _2}} \right) + \frac{{2{h_{q{p_1}}}{a^3}}}{{r_{12}^3}}\left( {\cos 2{\alpha _1} - \sin 2{\alpha _1}} \right)\cos {\phi _2},\label{pinned_c2}
\end{eqnarray}
where the second term on the right hand side of the above equations is the dipole term, which we subtract because the particle has no body torque, and lies in the plane tangent to the interface.  Recalling that the interface height tends to zero far from the particles, we can immediately construct the general solution.
\begin{align}
h(v,u) =&{A_0}v + {B_0} + \sum\limits_{m = 1}^\infty  {\left( {{C_m}\cos mu + {D_m}\sin mu} \right)\frac{{\sinh \left( {m\left( {{s_1} + v} \right)} \right)}}{{\sinh \left( {m\left( {{s_1} + {s_2}} \right)} \right)}}}  + \\ \nonumber
&\left( {{E_m}\cos mu + {F_m}\sin mu} \right)\frac{{\sinh \left( {m\left( {{s_2} - v} \right)} \right)}}{{\sinh \left( {m\left( {{s_1} + {s_2}} \right)} \right)}}.
\end{align}

\begin{figure} 
\centering
\includegraphics[width=0.45 \textwidth]{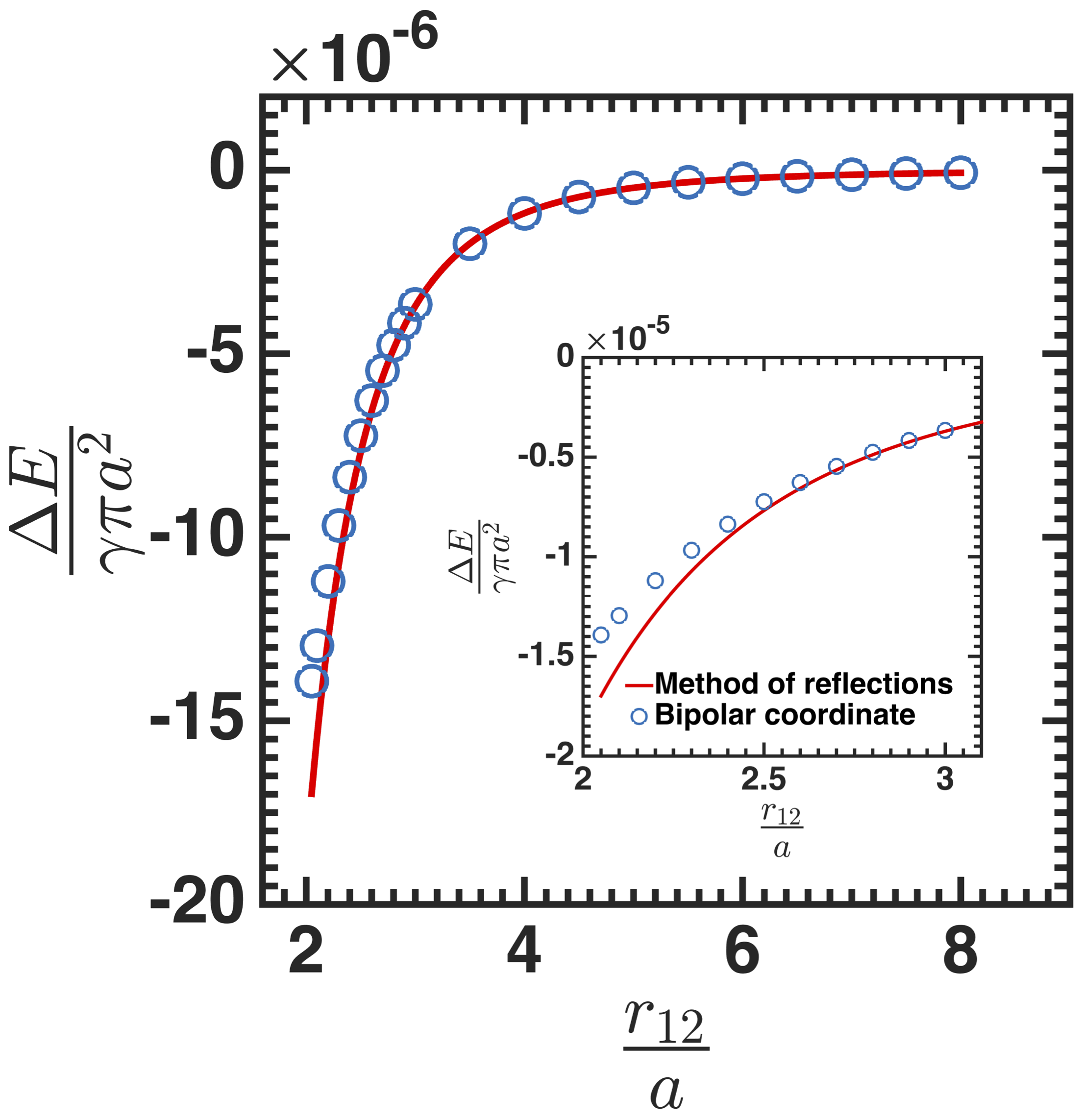}	     	    
 \caption{Non-dimensional energy obtained from method of reflections (solid line) and exact solution obtained in a bipolar coordinate (open symbols) as a function of center-to-center separation distance. The inset shows the near-field region where method of reflections is not accurate.} 
\label{fig_8}
 \end{figure}

In bipolar coordinates the region far from the particles ($r \to \infty$ in polar coordinates) can be represented as $u \to 0$ and $v \to 0$. The unknowns $A_0$, $B_0$, $C_m$, $D_m$, $E_m$ and $F_m$ can be found by applying the boundary conditions on particle edges in eqs.~(\ref{pinned_c1}) and (\ref{pinned_c2}). The  coefficients can be found explicitly by utilizing orthogonality from which we ultimately have,
\begin{align}
&{A_0} = \frac{{{J_0} - {I_0}}}{{{s_2} + {s_1}}},\\
&{B_0} = \frac{{{I_0}{s_2} + {J_0}{s_1}}}{{{s_2} + {s_1}}},\\
&{C_m} = \frac{1}{\pi}\int_0^{2\pi } {\left[ {{h_{q{p_2}}}\cos 2({\phi _2} - {\alpha _2}) + \frac{{2{h_{q{p_1}}}{a^3}}}{{r_{12}^3}}(\cos 2{\alpha _1} - \sin 2{\alpha _1})\cos {\phi _2}} \right] \cos mu~{\rm{d}}u}, \\
&{D_m} =\frac{1}{\pi} \int_0^{2\pi } {\left[ {{h_{q{p_2}}}\cos 2({\phi _2} - {\alpha _2}) + \frac{{2{h_{q{p_1}}}{a^3}}}{{r_{12}^3}}(\cos 2{\alpha _1} - \sin 2{\alpha _1})\cos {\phi _2}} \right]\sin mu~{\rm{d}}u}, \\
&{E_m} = \frac{1}{\pi}\int_0^{2\pi }{\left[ {{h_{q{p_1}}}\cos 2({\phi _1} - {\alpha _1}) - \frac{{2{h_{q{p_2}}}{a^3}}}{{r_{12}^3}}(\cos 2{\alpha _2} - \sin 2{\alpha _2})\cos {\phi _1}} \right]\cos mu~{\rm{d}}u}
\end{align} 
and
\begin{align}
&{F_m} = \frac{1}{\pi}\int_0^{2\pi }{\left[ {{h_{q{p_1}}}\cos 2({\phi _1} - {\alpha _1}) - \frac{{2{h_{q{p_2}}}{a^3}}}{{r_{12}^3}}(\cos 2{\alpha _2} - \sin 2{\alpha _2})\cos {\phi _1}} \right]\sin mu~{\rm{d}}u}, 
\end{align}
where
\begin{align}
&{I_0} = \frac{1}{2\pi}\int_0^{2\pi } {\left[ {{h_{q{p_1}}}\cos 2({\phi _1} - {\alpha _1}) + \frac{{2{h_{q{p_1}}}{a^3}}}{{r_{12}^3}}(\cos 2{\alpha _1} - \sin 2{\alpha _1})\cos {\phi _2}} \right] {\rm{d}}u}, \\
&{J_0} = \frac{1}{2\pi}\int_0^{2\pi }{\left[ {{h_{q{p_2}}}\cos 2({\phi _2} - {\alpha _2}) - \frac{{2{h_{q{p_2}}}{a^3}}}{{r_{12}^3}}(\cos 2{\alpha _2} - \sin 2{\alpha _2})\cos {\phi _1}} \right] {\rm{d}}u}.
\end{align}
The excess energy of the system can be defined by
\begin{eqnarray}
\Delta E = \frac{\gamma }{2}\mathop{{\int\!\!\!\!\!\int}\mkern-21mu \bigcirc}\limits_{I} 
{\left[ {{{(\nabla h)}^2} - {{(\nabla {h_1})}^2} - {{(\nabla {h_2})}^2}} \right]{\rm{d}}A} ,
\end{eqnarray}
where we have:
\begin{eqnarray}
\mathop{{\int\!\!\!\!\!\int}\mkern-21mu \bigcirc}\limits_{I} 
 {(\nabla {h_j})^2{\rm{d}}A}  = \pi h_{{qp_j}}^2,~~(j = 1,2).
\end{eqnarray}
On the other hand, we have
\begin{eqnarray}
{(\nabla h)^2} = \nabla h \cdot \nabla h = \nabla  \cdot (h\nabla h) - h{\nabla ^2}h = \nabla  \cdot (h\nabla h),
\end{eqnarray}
so that
\begin{align}
&\mathop{{\int\!\!\!\!\!\int}\mkern-21mu \bigcirc}\limits_{I} 
 {{{(\nabla h)}^2}{\rm{d}}A}  = \mathop{{\int\!\!\!\!\!\int}\mkern-21mu \bigcirc}\limits_{I} 
 {\nabla  \cdot (h\nabla h){\rm{d}}A}  = \nonumber\\
&- \oint\limits_{{P_1}} {{{\left. {h{\partial _v}h} \right|}_{v =  - {s_1}}}{\rm{d}}s}  + \oint\limits_{{P_2}} {{{\left. {h{\partial _v}h} \right|}_{v = {s_2}}}{\rm{d}}s}  + \oint\limits_{r \to \infty } {{{\left. {h{\partial _r}h} \right|}_{r \to \infty }}{\rm{d}}s} 
\end{align}
where $s_1$ or $s_2$ are the location of particles contact lines in bipolar coordinate and the contour integral far away and enclosed the particles are zero because $h$ and ${\partial _r}h$ are both decaying functions and therefore the integral has no contribution. Furthermore we have,
\begin{eqnarray}
{\partial _v} = \frac{{\cosh v - \cos u}}{\zeta}\frac{\partial }{{\partial v}}
\end{eqnarray}
and
\begin{eqnarray}
{\rm{d}}s = \frac{\zeta}{{\cosh v - \cos u}}{\rm{d}}u,
\end{eqnarray}
and therefore,
\begin{align}
\oint\limits_{{S_{P_1}} \cup {S_{P_2}}} {h({\bf{n}} \cdot \nabla h){\rm{d}}s}  =  - \int_0^{2\pi } {{{\left. {h\frac{{\partial h}}{{\partial v}}} \right|}_{v =  - {s_1}}}{\rm{d}}u}  + \int_0^{2\pi } {{{\left. {h\frac{{\partial h}}{{\partial v}}} \right|}_{v = {s_2}}}{\rm{d}}u}. 
\end{align}
Finally the excess energy of the system can be written as
\begin{eqnarray}
\Delta E = \gamma \left[ {\int_0^{2\pi } {{{\left. {\frac{h}{2}\frac{{\partial h}}{{\partial v}}} \right|}_{v = {s_2}}}{\rm{d}}u}  - \int_0^{2\pi } {{{\left. {\frac{h}{2}\frac{{\partial h}}{{\partial v}}} \right|}_{v =  - {s_1}}}{\rm{d}}u}  - \pi (h_{{qp_1}}^2 + h_{{qp_2}}^2)} \right],
\end{eqnarray}
where we evaluate the two integrals numerically and the partial derivative can be obtained via
\begin{align}
\frac{{\partial h}}{{\partial v}} &= {A_0} + \sum\limits_{m = 1}^\infty  {m\left({{C_m}\cos mu + {D_m}\sin mu}\right)\frac{{\cosh (m({s_1} + v))}}{{\sinh (m({s_1} + {s_2}))}}} \\
&- m\left({{E_m}\cos mu + {F_m}\sin mu}\right)\frac{{\cosh (m({s_2} - v))}}{{\sinh (m({s_1} + {s_2}))}}.\nonumber,
\end{align}
Figure~\ref{fig_8} shows the non-dimensional capillary energy of the two particles with aligned quadrupoles, i.e. $\alpha_1=\alpha_2=0^{\circ}$, on a flat interface as a function of the non-dimensional center-to-center separation distance found using the method of reflections. We have also calculated the exact capillary energy of two particles trapped at a planar fluid interface utilizing the Laplacian solution in a bipolar coordinate. The two solutions are in excellent agreement when the particles are far from each other. However, very near to contact, there is a slight deviation between the two solutions which indicates the importance of higher order reflections that we neglected in our treatment based on the method of reflections.  While the treatment in terms of bipolar coordinates yields the exact solution, the treatment in terms of the method of reflections gives insights into the physics behind the interactions.\\   
 
The key features of the interaction energy between particles with pinned quadrupolar distortions are:
(i) Particles attract if they are in mirror-symmetric orientations. (ii) Particles not in mirror symmetric orientations rotate owing to a local torque to attain such alignments. This treatment differs from that in the literature \cite{Stamou} in that it respects the boundary conditions on the particle surface and the far field boundary conditions imposed by particle.  However, our findings on planar interfaces are very close to the original literature \cite{Stamou}. This approach can be extended to particles in an arbitrary curved interface as described in the next section; in insisting that the boundary conditions on the particle are obeyed in this latter problem, we find interaction energies which differ significantly from the literature. 
\section{Capillary curvature energy}\label{curved-interface}
In this section, we revisit the capillary energy of a particle on a curved fluid interface. We have already addressed the key points- twice!  First, we considered the trapping energy for a particle in a curved interface, and arrived at eq.~(\ref{energy-221}). In that discussion, we treated the curvature fields as spatially constant; in this section, we will relax that assumption.  Second, in  the  discussion of the interaction energy of two particles,  particle $2$ creates a (spatially varying) local deviatoric curvature field near particle 1: 
\begin{eqnarray}
|\Delta {c_0}| =  \frac{12h_{{qp}_2}{a^2}}{r_{12}^4}
\end{eqnarray}
In this expression, we have set $\alpha_2=0$, as $\Delta {c_0}$ is defined along the principal axes.  This allow eq.~(\ref{pair-22}) to reduce to eq.~(\ref{energy-221}) for $\alpha_1=0$.  
In this example, the curvature source is the neighboring particle. In principle, though, any means of pinning or distorting  the interface far from the particle can create a curvature field, which can be locally expanded in the small slope limit in terms of local deviatoric and mean curvatures. We are particularly interested in deviatoric curvature fields, as mean curvature fields are typically constant in confined fluid elements with length scales less than the capillary length and deviatoric curvature fields can vary strongly. For constant mean curvature, the curvature capillary energy is
\begin{eqnarray}
\Delta E = \Delta {E_{0}} - {\gamma _{LV}}\pi {a^2} \frac{{{h_{qp}}\Delta {c_0}}}{2}.\label{energy-22}
\end{eqnarray}

In the more general case of finite  $\alpha_1$, the particle-sourced quadrupole is not aligned along the principal axes of the curvature field emanating from the curvature source,
\begin{eqnarray}
\Delta E = \Delta {E_0} - {\gamma _{LV}}\pi {a^2}\frac{{{h_{qp}}\Delta {c_0}}}{2}\cos 2{\alpha _1}.
\label{energy-23}
\end{eqnarray} 
 
The spatial variation of the deviatoric curvature field $\Delta {c_0}$ drives particle   migration and the local torque for finite $\alpha_1$ drives particles to orient in the plane of the interface so as to align their quadrupolar rise along the rise axis of the host deviatoric saddle surface \cite{Eric}. 
\section{Local expansion of the curvature field in terms of matched asymptotics}\label{asymp}
\begin{figure} 
\centering
\includegraphics[width=0.8 \textwidth]{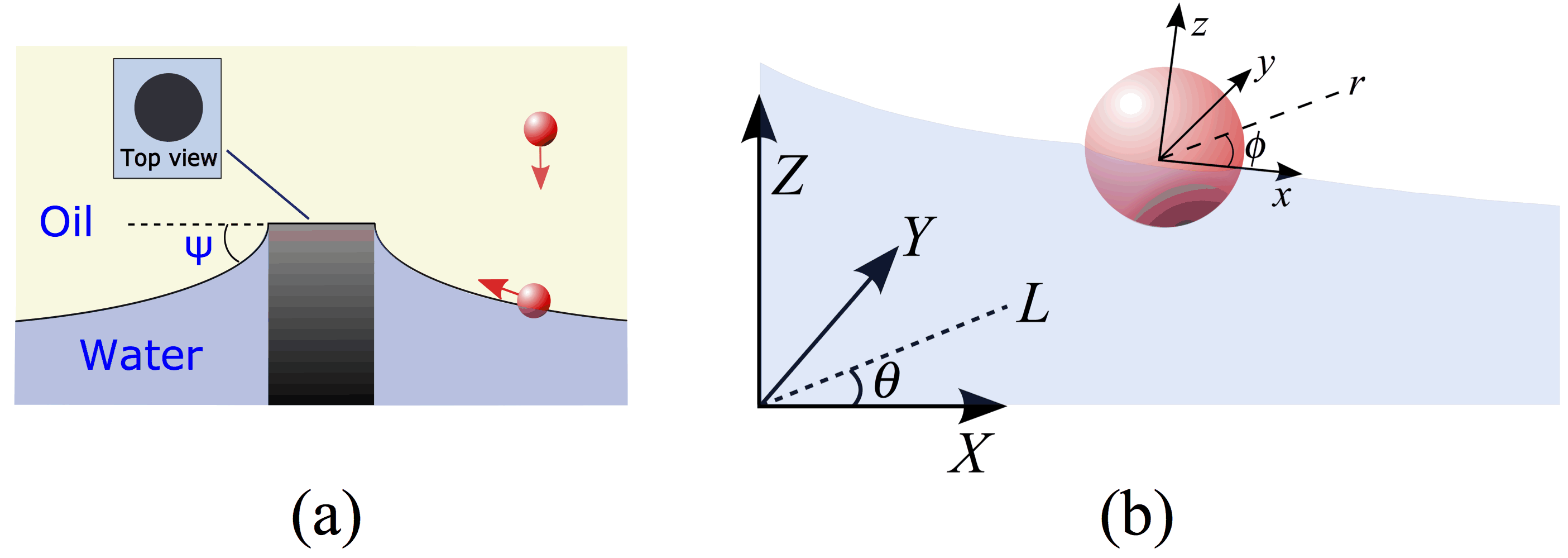}	     	    
 \caption{Schematic representations of (a) particle migration at an interface around a circular micropost and (b) a particle trapped on a curved interface and inner and outer coordinates. } 
\label{fig_9}
 \end{figure}
For a particle that is small compared to the prevailing curvature field (i.e. a particle that is much smaller than the capillary length and far from pinning boundaries), particle-sourced distortions decay over distances comparable to the particle radius. This scenario lends itself to analysis by a matched asymptotic expansion analysis \cite{Nayfeh}. We discuss  the interface shape when a particle of radius $a$ with an undulated pinned contact line in the form of quadrupole is placed on a host interface. In experiments, we form a curved oil-water interface around a circular micropost  [see the schematic in fig.~\ref{fig_9}(a)]4. The interface pins to the edge of the
post, and has a height $H_m$ at the post's edge. The post is centered within a confining ring far away. By adjusting the volume of water, the interface slope at the post's edge is adjusted so that the angle $\psi \sim 15^{\circ}-18^{\circ}$. The interface height within several post radii of the micropost is well approximated by
\begin{eqnarray}
{h_{host}} = {H_m} - {R_m}\tan \psi \ln {\rm{ }}\left( {\frac{L}{{{R_m}}}} \right),\label{host-1}
\end{eqnarray}
where $L$ is the distance from the micropost center. This interface has zero mean curvature $H_0$, and finite deviatoric curvature $\Delta c_0 $ that varies with $L$. The disturbance made by the particle decays monotonically over distances comparable to $a$. Moreover, the particles are small compared to the characteristic length of the host interface $R_c \sim{1 \mathord{\left/
 {\vphantom {1 {\Delta {c_0}}}} \right.
 \kern-\nulldelimiterspace} {\Delta {c_0}}}$ and hence the interface domain can be divided into two distinct regions, an inner region close to the particle where the relevant length scale is $a$ and an outer region further away where the relevant length scale is $R_c$. For $\lambda=a\Delta c_0 \ll 1 $, the problem can be solved systematically by a matched asymptotic expansion scheme \cite{Nayfeh}. \\
First, we non-dimensionalize the equations.  We rewrite eq.~(\ref{host-1}) to have:
\begin{eqnarray}
{{\hat h}^{outer}} = \frac{{{H_m}}}{{{R_c}}} - \frac{{{R_m}\tan \psi }}{{{R_c}}}\ln \left({\frac{{{R_c}\hat L}}{{{R_m}}}}\right).
\end{eqnarray}
This expression is given in terms of  the outer coordinate with the origin at the center of the circular micropost ($\hat X,\hat Y,\hat Z$). Here, $\hat Z={\hat h}^{outer}$ and the expression is nondimensionalized with respect to the host interface radius of curvature $R_c$.\\

In the inner region, we adopt the inner coordinate $(\tilde{x},\tilde{y},\tilde{z})$ defined with respect to the tangent plane. This plane has a slope of $\epsilon=- {{{R_m}\tan \psi } \mathord{\left/
 {\vphantom {{{R_m}\tan \psi } {{L_0}}}} \right.
 \kern-\nulldelimiterspace} {{L_0}}}$ with respect to the outer coordinate [see fig.~\ref{fig_9}(b)], and $L_0$ is the location of particle center of mass with respect to the center of the micropost. This coordinate is made dimensionless with the inner region length scale $a$. The height of the interface in the presence of the particle is  $\tilde z = {{\tilde h}^{inner}}$, where we adopt the Monge representation of the interface. The interface shape in the inner region is  governed via the Young-Laplace equation which, in the limit of small gradients and infinitesimal Bond numbers, reduces to 
\begin{eqnarray}
{{\tilde \nabla }^2}{{\tilde h}^{inner}} = 0.
\end{eqnarray}
We then expand the dimensionless height of the interface in the inner region as a power series to yield
\begin{eqnarray}
{{\tilde h}^{inner}} = \tilde{h}_0 + \tilde{h}_1\lambda  + \tilde{h}_2{\lambda ^2} + O({\lambda ^3}).\label{perturb}
\end{eqnarray} 
The pinning boundary condition at the three phase contact line can be expressed 
\begin{eqnarray}
{{\tilde h}^{inner}}(\tilde r = 1) = \frac{{{h_{qp}}}}{a}\cos 2\phi.
\end{eqnarray}

 The far field boundary condition for the inner solution is obtained by matching the inner solution with the outer solution via Van Dyke's matching procedure \cite{Nayfeh}. This is done systematically by first writing the outer solution in terms of inner variable and then holding the inner variable fixed and expanding for small parameter $\lambda$. This provides the far field boundary condition for the inner region.  Formally, this matching is written
\begin{eqnarray}
\mathop {\lim }\limits_{\tilde r \to \infty } \lambda {{\tilde h}^{inner}}(\tilde r,\phi ) = \mathop {\lim }\limits_{\scriptstyle\lambda  \to 0\hfill\atop
\tilde {r}~{\rm{fixed}}\hfill} {{\hat h }^{outer}}(\tilde r,\phi ).\label{matching}
\end{eqnarray}
Note that, once again, in the above expression $\hat h^{outer}$ is made dimensionless with the outer region length scales, and ${{\tilde h}^{inner}}$ is made dimensionless with the inner region length scales. \\

 We demonstrate the execution of this matching procedure for our particular host interface without loss of generality.  The same procedure can be applied to any host interface, provided ${a \mathord{\left/
 {\vphantom {a {{R_c}}}} \right.
 \kern-\nulldelimiterspace} {{R_c}}} \ll 1$. To perform this matching, we transform  coordinates. The inner coordinate is tangent  with respect to the outer coordinate with slope $\epsilon$. The relationship between the two coordinates providing small slopes $|\epsilon|  \ll 1$ (in dimensional form) can be written as
\begin{align}
&X = {L_0} + x + O(\epsilon ),\nonumber\\
&Y = y,\label{trans}\\
&Z = {Z_0} + z + O(\epsilon ),\nonumber
\end{align}
where $Z_0$ is the interface height at the particle center of mass with respect to the outer coordinate. Using this information, the right hand side of the matching condition expression in eq.~(\ref{matching}) can be written
\begin{eqnarray}
\hat h^{outer} = \frac{{{H_m}}}{{{R_c}}} - \frac{{{R_m}\tan \psi }}{{{R_c}}}\ln \left({\frac{{\sqrt {{{(x + {L_0})}^2} + {y^2}} }}{{{R_m}}}}\right) - \frac{{{H_0}}}{{{R_c}}}.
\end{eqnarray}

By nondimensionalizing and expanding the argument of the $\log$ term for small $\lambda$, we can show,
\begin{eqnarray}
\mathop {\lim }\limits_{\scriptstyle\lambda  \to 0\hfill\atop
\tilde{r}~{\rm{fixed}}\hfill} {{
\hat h }^{outer}}(\tilde r,\phi ) = \frac{{{\lambda ^2}}}{4}{{\tilde r}^2}\cos 2\phi  + O(\epsilon {\rm{,}}{\lambda ^3}),
\end{eqnarray} 
Invoking  eq.~(\ref{matching}), this can be recast to specify the far field boundary condition for the disturbance in the inner region, 
 \begin{eqnarray}
 \mathop {\lim }\limits_{\tilde r \to \infty } {\tilde h^{inner}}(\tilde r,\phi ) = \frac{\lambda }{4}{\tilde r^2}\cos 2\phi+ O(\epsilon {\rm{,}}{\lambda^2}).
 \end{eqnarray}
This matching also clarifies the meaning of taking the limit  $\tilde r\to\infty$. This implies exploring regions at distances much larger than the particle radius from the particle: 
\begin{eqnarray}
{{\tilde h}^{inner}} = \frac{{{h_{qp}}}}{{a{{\tilde r}^2}}}\cos 2\phi  + \frac{\lambda }{4}\left({{{\tilde r}^2} - \frac{1}{{{{\tilde r}^2}}}}\right)\cos 2\phi  + O({\lambda ^2}).
\end{eqnarray}
To construct a uniformly valid (uv) solution (in dimensional form) over the entire domain, we define 
\begin{eqnarray}
{h^{\rm{uv}}} = {R_c}{{
\hat h }^{outer}} + a{{\tilde h}^{inner}} - {R_c}\mathop {\lim }\limits_{\scriptstyle\lambda  \to 0\hfill\atop
\scriptstyle\tilde r~{\rm{fixed}}\hfill} {{\hat h }^{outer}},\label{uvs}
\end{eqnarray}
Consequently, the dimensional disturbance will be,
\begin{eqnarray}
\eta  = {h^{{\rm{uv}}}} - {R_c}{{\hat h}^{outer}} = \frac{{{h_{qp}}}}{{{{\tilde r}^2}}}\cos 2\phi  - \lambda \frac{a}{{4{{\tilde r}^2}}}\cos 2\phi  + O({\lambda ^2}).
\end{eqnarray}

This disturbance is a decaying function dependent only on the inner variable $\tilde {r}$. This implies that its value is identically zero in the outer region. Thus, the particle results in a ``local'' disturbance which fades over a length scale comparable to its radius $a$. Furthermore, in evaluating the area owing to the particle disturbance [eq.~(\ref{int-1})], there is no coupling  between $\tilde h^1$ and higher order term owing to throrthogonality of $\cos n \phi$.  Therefore, there is no term in the energy expression of order $\lambda^3$ and the capillary energy given in eq.~(\ref{energy-22}) is exact up to $O(\lambda^4,\epsilon)$.

\section{Electrostatic analogies}
Capillary interactions are often likened to electrostatic interactions \cite{Griffiths}, with the amendment that like capillary ``charges'' attract.  In the analogy, the height of the interface $h$  corresponds to the electrostatic potential $\psi$ external to the particle.  The potential field at the particle edge corresponds to constraints on the interface height at the boundary conditions at the contact line. The  surface charge density $\sigma_s$ is derived from potential gradients to impose these conditions. The electrostatic energy external to the particle corresponds to the capillary energy of the system.\\
\\
We evaluate the electrostatic energy $U$ for two situations to probe this analogy: \\
\\
1) We first solve for the potential $\psi$ and the electrostatic energy $U$ around a grounded disk.  It can be shown for this situation that the analogy with capillarity is valid, term by term.  We find a solution for the potential field that agrees with the inner solution for the height of the interface around a disk with a pinned, planar contact line.  Upon evaluating the energy over the domain, we find $U=0$ is analogous to the capillary energy for the disk with a pinned contact line. \\
\\
2) We solve for $U$ around a disk with a surface charge density in the form of a quadrupole to mimic the undulated, pinned boundary condition. However, we find that the simple analogy with capillarity fails, as the boundary conditions for a charged particle in an external field are not analogous to a particle with an undulated contact line on a curved interface. The requirements for continuous potentials at the particle surface, and for jump conditions on the normal electric field are not
present in the capillary problem. Associated with these conditions are finite potential fields inside
the particle  which are absent for the capillary problem.\\

\subsection{A grounded disk in an external potential}
A grounded disk with radius $a$ is placed in an applied external potential of the form:
\begin{eqnarray}
{\psi _{ext}} = {\psi _0}\;{r^2}\cos 2\phi.\label{ext}
\end{eqnarray}
The electric potentials inside and outside of the disk are solutions of the Laplace equation subject to $\psi (r = a) = 0$: 
\begin{align}
\psi ({\bf{r}}) = \left\{ \begin{array}{l}
\begin{array}{*{20}{c}}
{0,}&{r < a}
\end{array}\\
\begin{array}{*{20}{c}}
{{\psi _0}\left({{r^2} - \frac{{{a^4}}}{{{r^2}}}}\right)\cos 2\phi ,}&{r \ge a}
\end{array}
\end{array} \right.
\end{align}
The total stored electrical energy $U$ can be calculated in two ways:\\
\\
(i) Evaluation of the integral directly \cite{Griffiths}:\\
\begin{eqnarray}
U = \frac{1}{2}\mathop{{\int\!\!\!\!\!\int}\mkern-21mu \bigcirc}\limits_{I+P} 
 {\rho ({\bf{r}})\psi ({\bf{r}})} {\rm{d}}A,
\end{eqnarray}
where $I+P$ is the entire domain, ${\rm{d}}A$ is the area element, and $\rho({\bf{r}})$ is the charge distribution in the system. The sole charge is $\sigma_s$, the induced charge on the surface of the disk given as
\begin{eqnarray}
{\left. {{{\bf{e}}_r}\cdot \left[{ - \nabla \psi (r \ge a)}\right]} \right|_{r = a}} = \frac{{{\sigma _s}}}{{{\epsilon _0}}},
\end{eqnarray}
where ${\bf{e}}_r$ is the unit normal vector pointing away from the disk, and $\epsilon _0$ is the permittivity of the free space. $\sigma _s$ corresponding to this potential field is: 
\begin{eqnarray}
{\sigma _s} =  - 4{\epsilon _0}{\psi _0}a\cos 2\phi.
\end{eqnarray}
The integral can be recast and evaluated as follows:
\begin{align}
&U = \frac{1}{2}\mathop{{\int\!\!\!\!\!\int}\mkern-21mu \bigcirc}\limits_{I+P} 
 {\rho ({\bf{r}})\psi ({\bf{r}})} {\rm{d}}A = \frac{1}{2}\int_a^R {\int_0^{2\pi } {{\sigma _s}\delta (r - a)\psi ({\bf{r}})r{\rm{d}}\phi {\rm{d}}r } }\nonumber \\
&=  - 2{\epsilon _0}\psi _0^2a\int_a^R {\delta (r - a)\left({{r^3} - \frac{{{a^4}}}{r}}\right){\rm{d}}r\left( {\int_0^{2\pi } {{{\cos }^2}\phi {\rm{d}}\phi } } \right)}  = 0,
\end{align}
where $\delta$ is the Dirac delta function and $R$ is any arbitrary radial location from the center of the disk. The above integral is zero since the potential at the disk surface is zero.\\
\\
(ii) Calculation using Gauss's law to recast $U$ \cite{Griffiths},
\begin{align}
&U = \frac{1}{2}\mathop{{\int\!\!\!\!\!\int}\mkern-21mu \bigcirc}\limits_{I} 
 {\rho ({\bf{r}})\psi ({\bf{r}})} {\rm{d}}A =  - \frac{{{\epsilon _0}}}{2}\mathop{{\int\!\!\!\!\!\int}\mkern-21mu \bigcirc}\limits_{I} 
 {\psi {\nabla ^2}\psi {\rm{d}}A},
 \end{align}
 using
 \begin{align}
&\psi {\nabla ^2}\psi  = \nabla \cdot(\psi \nabla \psi ) - \nabla \psi \cdot\nabla \psi,
 \end{align}
 $U$ can be rewritten as
  \begin{align}
&U =  - \frac{{{\epsilon _0}}}{2}\oint\limits_{\partial I} {(\psi \nabla \psi )\cdot{\bf{n}}\;{\rm{d}}l}  + \frac{{{\epsilon _0}}}{2}\mathop{{\int\!\!\!\!\!\int}\mkern-21mu \bigcirc}\limits_{I} 
 {{{(\nabla \psi )}^2}{\rm{d}}A},\label{en-elc}
\end{align}
where $\bf{n}$ is the outward pointing unit normal on the contours  enclosing  $I$ and $P$ refers to the domain of the disk. To obtain a better insight into the induced vs. external field contributions to the energy, we decompose the electric potential outside the disk as:
\begin{eqnarray}
\psi  = {\psi _{ext}} + {\psi _{induced}}~~{\rm{for}}~r\ge a,\label{decomposition}
\end{eqnarray}
where ${\psi _{ext}}$ is given in eq.~(\ref{ext}) and ${\psi _{induced}}$ is
\begin{eqnarray}
{\psi _{induced}} =  - {\psi _0}\frac{{{a^4}}}{{{r^2}}}\cos 2\phi,~~{\rm{for}}~r\ge a. \label{induced}
\end{eqnarray}
In this case, the second integral in eq.~(\ref{en-elc}) can be evaluated,
\begin{eqnarray}
\mathop{{\int\!\!\!\!\!\int}\mkern-21mu \bigcirc}\limits_{I} 
 {{{(\nabla \psi )}^2}{\rm{d}}A}  = \mathop{{\int\!\!\!\!\!\int}\mkern-21mu \bigcirc}\limits_{I} 
 {{{(\nabla {\psi _{ext}})}^2}{\rm{d}}A}  + \mathop{{\int\!\!\!\!\!\int}\mkern-21mu \bigcirc}\limits_{I} 
 {{{(\nabla {\psi _{ind}})}^2}{\rm{d}}A}  + \mathop{{\int\!\!\!\!\!\int}\mkern-21mu \bigcirc}\limits_{I} 
 {2\nabla {\psi _{ext}}\cdot\nabla {\psi _{induced}}{\rm{d}}A},\label{decomp-psi}
\end{eqnarray}
On the other hand, the contour integral in eq.~(\ref{en-elc}) (first term on the right hand side) can be rewritten as: 
\begin{align}
&\oint\limits_{\partial I} {(\psi \nabla \psi )\cdot{\bf{n}}~{\rm{d}}l}  = \oint\limits_{r = a} { - {{\bf{e}}_r}\cdot(\psi \nabla \psi )a{\rm{d}}\phi }  + \oint\limits_{r = R} {{{\bf{e}}_r}\cdot(\psi \nabla \psi )R{\rm{d}}\phi }\nonumber \\
&=  - \int_0^{2\pi } {{{\left. {\psi \frac{{\partial \psi }}{{\partial r}}} \right|}_{r = a}}a{\rm{d}}\phi }  + \int_0^{2\pi } {{{\left. {\psi \frac{{\partial \psi }}{{\partial r}}} \right|}_{r = R}}R{\rm{d}}\phi }.
\end{align}
Since $\psi(r=a)=0$, the first integral is zero and   
\begin{eqnarray}
\oint\limits_{\partial I} {(\psi \nabla \psi )\cdot{\bf{n}}~{\rm{d}}l}  = \int_0^{2\pi } {{{\left. {\psi \frac{{\partial \psi }}{{\partial r}}} \right|}_{r = R}}R{\rm{d}}\phi },
\end{eqnarray}
Using eq.~(\ref{decomposition}), the contour integral becomes
\begin{align}
&\int_0^{2\pi } {{{\left. {\psi \frac{{\partial \psi }}{{\partial r}}} \right|}_{r = R}}R{\rm{d}}\phi }  = \int_0^{2\pi } {{{\left. {{\psi _{ext}}\frac{{\partial {\psi _{ext}}}}{{\partial r}}} \right|}_{r = R}}R{\rm{d}}\phi } \nonumber \\
&+\int_0^{2\pi } {{{\left. {({\psi _{induced}}\frac{{\partial {\psi _{ext}}}}{{\partial r}} + {\psi _{ext}}\frac{{\partial {\psi _{induced}}}}{{\partial r}})} \right|}_{r = R}}R{\rm{d}}\phi }  + \int_0^{2\pi } {{{\left. {{\psi _{induced}}\frac{{\partial {\psi _{induced}}}}{{\partial r}}} \right|}_{r = R}}R{\rm{d}}\phi }.\label{c-int}
\end{align}
The integrand of the second integral on the right hand side of the above is exactly zero. Furthermore, the integrand in the third contour integral on the right hand side of  eq.~(\ref{c-int}) goes as $\sim R^{-4}$ [see eq.~(\ref{induced})]. In addition, the third integral tends to zero when $R\to\infty$. Thus we have
\begin{align}
&\oint\limits_{\partial I} {(\psi \nabla \psi )\cdot{\bf{n}}{\rm{d}}l}  = \int_0^{2\pi } {{{\left. {\psi \frac{{\partial \psi }}{{\partial r}}} \right|}_{r = R}}R{\rm{d}}\phi }\label{d-int}\\
&= \int_0^{2\pi } {{{\left. {{\psi _{ext}}\frac{{\partial {\psi _{ext}}}}{{\partial r}}} \right|}_{r = R}}R{\rm{d}}\phi }  = \int_0^R {\int_0^{2\pi } {{{(\nabla {\psi _{ext}})}^2}r{\rm{d}}r{\rm{d}}\phi } }  = \mathop{{\int\!\!\!\!\!\int}\mkern-21mu \bigcirc}\limits_{I+P} 
 {{{(\nabla {\psi _{ext}})}^2}{\rm{d}}A}, \nonumber
\end{align}
where we have integrated by parts, used  Green's theorem, and noted that  ${\nabla ^2}{\psi _{ext}} = 0$. Finally, by substituting eqs.~(\ref{decomp-psi}) and (\ref{d-int}) into (\ref{en-elc}) we have:
\begin{align}
&U = \frac{{{\epsilon _0}}}{2}\left[ {\mathop{{\int\!\!\!\!\!\int}\mkern-21mu \bigcirc}\limits_{I} 
 {{{(\nabla {\psi _{induced}})}^2}{\rm{d}}A}  + \mathop{{\int\!\!\!\!\!\int}\mkern-21mu \bigcirc}\limits_{I} 
 {2\nabla {\psi _{ext}}\cdot\nabla {\psi _{induced}}{\rm{d}}A}  - [\mathop{{\int\!\!\!\!\!\int}\mkern-21mu \bigcirc}\limits_{I+P} 
 {{{(\nabla {\psi _{ext}})}^2}{\rm{d}}A}  - \mathop{{\int\!\!\!\!\!\int}\mkern-21mu \bigcirc}\limits_{I} 
 {{{(\nabla {\psi _{ext}})}^2}{\rm{d}}A} ]} \right]\nonumber\\
&= {\epsilon _0}\left[ {\mathop{{\int\!\!\!\!\!\int}\mkern-21mu \bigcirc}\limits_{I} 
 {\frac{{{{(\nabla {\psi _{induced}})}^2}}}{2}{\rm{d}}A}  + \mathop{{\int\!\!\!\!\!\int}\mkern-21mu \bigcirc}\limits_{I} 
 {\nabla {\psi _{ext}}\cdot\nabla {\psi _{induced}}{\rm{d}}A}  - \mathop{{\int\!\!\!\!\!\int}\mkern-21mu \bigcirc}\limits_P 
 {\frac{{{{(\nabla {\psi _{ext}})}^2}}}{2}{\rm{d}}A} } \right],\label{energy-1}
\end{align}
This expression for the electrostatic energy is exactly analogous, term-for-term, to the capillary energy we derived above, except for the constant (Pieranski's) term; the first two integrals on the right hand side are the disturbance terms and the third term is the area of the hole under the particle. \\
The integrals in eq.~(\ref{energy-1}) can be evaluated explicitly as,
\begin{align}
&\int_0^a {\int_0^{2\pi } {\frac{{{{(\nabla {\psi _{ext}})}^2}}}{2}r{\rm{d}}\phi {\rm{d}}r} }  = \pi \psi _0^2{a^4},\label{1aa}\\
&\int_a^R {\int_0^{2\pi } {\frac{{{{(\nabla {\psi _{induced}})}^2}}}{2}rd\phi {\rm{d}}r} }  = \pi \psi _0^2\left({{a^4} - \frac{{{a^8}}}{{{R^4}}}}\right),\label{2aa}\\
&\int_a^R {\int_0^{2\pi } {\nabla {\psi _{ext}}\cdot\nabla {\psi _{induced}}r{\rm{d}}\phi {\rm{d}}r} }  = 4\psi _0^2\int_a^R {\frac{{{a^4}}}{{{r^2}}}r{\rm{d}}r\int_0^{2\pi } {({{\cos }^2}\phi  - {{\sin }^2}\phi ){\rm{d}}\phi } }  = 0,\label{contested}
\end{align}
In the limit of $R\to\infty$, eqs.~(\ref{1aa}) and (\ref{2aa}) are identical and hence eq.~(\ref{energy-1}) requires that  $U=0$.\\
\subsection{A charged disk in an external potential: Handle with care}
A disk of radius $a$ in an applied external potential of the form:
\begin{eqnarray}
{\psi _{ext}} = {\psi _0}\;{r^2}\cos 2\phi.\label{ext}
\end{eqnarray}
subject to a quadrupolar potential on its edge $\psi (r = a) = q_p\cos 2\phi$. \label{ext-1}
We can show that the corresponding surface charge density $\sigma_s$ is 
\begin{eqnarray}
{\sigma _s} = 4\epsilon_0(\frac{q_p}{a}- {\psi _0}a)\cos 2\phi.
\end{eqnarray}
The electric potentials inside and outside the disk are solutions of the Laplace equation, subject to the boundary conditions: 
\begin{eqnarray}
&{\left. {{\psi ^{inside}}} \right|_{r = a}} = {\left. {{\psi ^{outside}}} \right|_{r = a}},\label{eq-pot}\\
&{\left. {{{\bf{e}}_r} \cdot (\nabla {\psi ^{inside}} - \nabla {\psi ^{outside}})} \right|_{r = a}} = {{{\sigma _s}} \mathord{\left/
 {\vphantom {{{\sigma _s}} {{\epsilon _0}}}} \right.
 \kern-\nulldelimiterspace} {{\epsilon _0}}},\label{maxwel}
\end{eqnarray}
and
\begin{eqnarray}
&\frac{{{\sigma _s}}}{{{\epsilon _0}}} = 4\left({\frac{q_p}{a}- {\psi _0}a}\right)\cos 2\phi.
\end{eqnarray}
Equation~(\ref{eq-pot}) requires continuity of the electric potential condition at the disk edge  and eq.~(\ref{maxwel}) is the Maxwell boundary condition. Note that eq.~(\ref{eq-pot}) requires that $\psi^{inside}$ be finite. This finite potential has no analogy in the capillary problem and this disconnection between the two problems will propagate throughout the calculation of $U$.\\
The corresponding solutions for the potentials are: 
\begin{align}
&{\psi ^{inside}} = q_p\frac{r^2}{a^2}\cos 2\phi
\end{align}
and
\begin{align}
&{\psi ^{outside}} = q_p\frac{a^2}{r^2}\cos 2\phi+{\psi _0}({r^2} - \frac{{{a^4}}}{{{r^2}}})\cos 2\phi,
\end{align}
where $\psi^{outside}$ corresponds to the height solution of the interface around a particle with a pinning boundary condition at the particle contact line in the capillary problem. Once again, the total electrical energy $U$ in this system can be calculated two ways:\\
\\
(i) Direct evaluation of the integral:
\begin{eqnarray}
U = \frac{1}{2}\mathop{{\int\!\!\!\!\!\int}\mkern-21mu \bigcirc}\limits_{I+P} 
 {\rho ({\bf{r}})\psi ({\bf{r}}){\rm{d}}A}.\label{en-r}
\end{eqnarray}
Since the sole charge is the charge on the surface of the disk
\begin{align}
&U = \frac{1}{2}\int_0^R {\int_0^{2\pi } {{\sigma _s}\delta (r - a)\psi ({\bf{r}})r{\rm{d}}\phi {\rm{d}}r} } \nonumber\\
&= 2{\epsilon _0}(\frac{{{q_p}}}{a} - {\psi _0}a)\int_0^R {\delta (r - a){q_p}\frac{{{r^2}}}{{{a^2}}}r{\rm{d}}r\int_0^{2\pi } {{{\cos }^2}2\phi ~{\rm{d}}\phi } }  = 2\pi {\epsilon _0}\left({q_p^2 - {\psi _0}{q_p}{a^2}}\right),
\end{align}
where $R$ is any arbitrary radial location from the center of the disk. \\
\\
(ii) Using Gauss's law to recast eq.~(\ref{en-r}):
\begin{align}
&
&U =  - \frac{{{\epsilon _0}}}{2}\oint\limits_{\partial (I+P)} {(\psi \nabla \psi )\cdot{\bf{n}}\;dl}  + \frac{{{\epsilon _0}}}{2}\mathop{{\int\!\!\!\!\!\int}\mkern-21mu \bigcirc}\limits_{I+P} 
 {{{(\nabla \psi )}^2}dA},\label{en-elc-1}
\end{align}
where the second integral (area integral) in eq.~(\ref{en-elc-1}) can be decomposed into the domains inside and outside of the disk:
\begin{eqnarray}
\mathop{{\int\!\!\!\!\!\int}\mkern-21mu \bigcirc}\limits_{I+P} 
 {{{(\nabla \psi )}^2}{\rm{d}}A}  = \mathop{{\int\!\!\!\!\!\int}\mkern-21mu \bigcirc}\limits_P 
 {{{(\nabla {\psi ^{inside}})}^2}{\rm{d}}A}  + \mathop{{\int\!\!\!\!\!\int}\mkern-21mu \bigcirc}\limits_{I} 
 {{{(\nabla {\psi ^{outside}})}^2}{\rm{d}}A},\label{decomp-v-1}
\end{eqnarray} 
where $P$ denotes the area under the disk. The first integral on the right hand side of this expression is finite since the electrical potential inside the disk is no longer zero and hence
\begin{eqnarray}
\mathop{{\int\!\!\!\!\!\int}\mkern-21mu \bigcirc}\limits_P 
 {{{(\nabla {\psi ^{inside}})}^2}{\rm{d}}A = \int_0^{2\pi } {({{\cos }^2}2\phi  + {{\sin }^2}2\phi ){\rm{d}}\phi \int_0^a {\frac{{4q_p^2}}{{{a^4}}}{r^3}{\rm{d}}r}  = 2\pi } } q_p^2,
\end{eqnarray}
 There is no analogy to this term in the capillary problem! The second integral, along with  the contour integral in eq.~(\ref{en-elc-1}), can be rearranged to be identical to the capillary energy [except for the constant term, see eq.~(\ref{energy-1})] derived above. \\
\indent The process of charging the particle (which generates the potential inside of the particle) differs from the process of undulating the contact line (which relies on wetting energies or pinning sites). Thus, the analogy with electrostatic energy must be treated with care, respecting  the differences in the physics of the two systems.
\section{Experimental observations}
In this section, we describe experiments designed to allow comparison to analysis and simulation. Please note that we summarize work from our own research group.  We refer the readers to the literature for more comprehensive reviews \cite{Kralchevsky,Botto}. Before embarking on this discussion, we summarize what we have learned above.\\

 Capillary interactions are remarkably strong: typical surface tensions are roughly $10-20~{{{{{k}}_{{B}}}{{T}}} \mathord{\left/
 {\vphantom {{{{{k}}_{{B}}}{{T}}} {{\rm{n}}{{\rm{m}}^{\rm{2}}}}}} \right.
 \kern-\nulldelimiterspace} {{\rm{n}}{{\rm{m}}^{\rm{2}}}}}$, so the elimination of even $1~n{\rm{m^2}}$ of surface area can lower the system's free energy significantly.  When particles attach to interfaces, they eliminate a patch of interface (lowering the capillary energy), and distort the interface around them to satisfy their wetting conditions or because of pinning of the contact line on the particle surface (with concomitant increase in interface area and thus capillary energy). The energy reduction associated with the eliminated patch traps the particles at the interface \cite{Pieranski}.  The energy cost associated with the distortions allows particles to interact and assemble \cite{Stamou}. When the distortions made by neighboring particles overlap, particles attract to minimize the interfacial area in their distortions. Closer to contact, rearrangement of the wetting configurations and associated solid-fluid wetting energies can also play a role.  These interactions can be used to assemble a broad range of materials, as they depend only on particle shape and wetting conditions at a given interface. \\
\\

Particles distort the interface above and below some planar reference state.  Regions that rise above the reference plane are ``positive'' distortions, and those that fall below it are ``negative'' distortions. There are well established analogies between interacting charged particles in electrostatics and interacting distortions on fluid interfaces, with the exception that ``like charges attract'' in capillarity, i.e. rise attracts rise, fall attracts fall, as, by orienting in this manner, the interfacial area of the distortions is reduced. Furthermore, above, we alluded to analogies (which must be handled with care) of a particle moving by capillarity in a curved interface and the interaction of charged particles (multipoles) in external electrostatic fields \cite{Marcello,Disk,Sphere}.\\

\indent Typically, particles move in creeping flow (i.e. the Reynolds number $Re = {{\rho Ua} \mathord{\left/
 {\vphantom {{\rho Ua} \mu }} \right.
 \kern-\nulldelimiterspace} \mu } \ll 1$, where $\rho$ is the fluid density, $\mu$ is a characteristics viscosity of the fluids near the interface and $U$ is the particle velocity), and hence move with negligible inertia, so the sum of forces on the particles is zero.  They also move so that the magnitude of viscous stresses compared to surface tension is negligible, (i.e. the Capillary number $Ca = {{\mu U} \mathord{\left/
 {\vphantom {{\mu U} \gamma }} \right.
 \kern-\nulldelimiterspace} \gamma } \ll 1$, where $\gamma$ is the interfacial tension).  In this limit, the interface shape is independent of particle velocity, and we can interpret the capillary interactions using capillary hydrostatics arguments. For deterministic particle migrations like those reviewed here, this implies that capillary forces are balanced by viscous drag.  From appropriate gradients of the total surface energy, capillary forces and torques can be calculated in the far field. In experiment, the equality between viscous drag and capillary forces allows us to infer the capillary interaction energies form the energy dissipated over the particle path.  Predicted forms can be compared to experiment including particle trajectories (translation, rotation) and preferred orientations with respect to each other. \\
\begin{figure} 
\centering
\includegraphics[width=0.6 \textwidth]{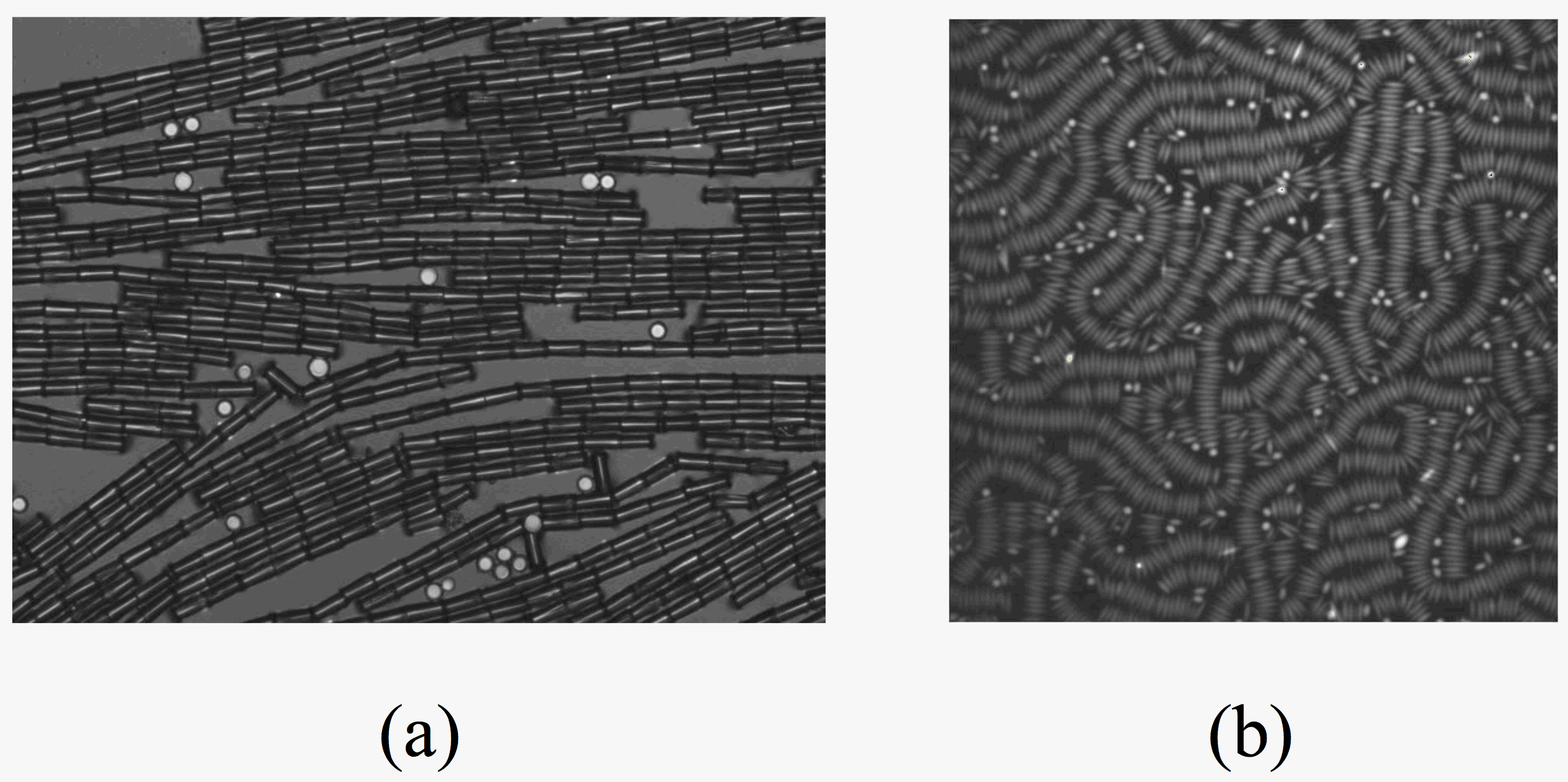}	     	    
 \caption{Assemblies formed by capillary interactions between elongated microparticles at air-water interfaces. (a) Cylinders chain in an end-to-end configuration to form  bamboo-like rigid assemblies.  In contrast, (b) ellipsoidal microparticles  assemble side-to-side to form flexible assemblies. (Micrograph in panel (b) is courtesy of Jan Vermant \cite{Vermant}).} 
\label{fig_10}
 \end{figure}
 
All particles excite a polar quadrupolar distortion in the far field.  These polar quadrupoles attract in mirror symmetric orientations -- with no preference for a particular orientation. Are all particles the same?  No.  Shape, patchy wetting, sharp edges, etc. all play a role in the near-field.  Electrostatic charge can create a number of interesting effects as well, which are beyond the scope of this discussion. We are particularly interested in the role of particle shape in guiding the formation of structure by capillary interactions on planar interfaces and in the role of curvature in guiding the formation of structures. Motivating images of microparticles assembled on planar interfaces are presented in fig.~\ref{fig_10}, in which cylindrical microparticles (roughly $20~\mu$m long) form rigid bamboo-like structures and ellipsoidal microparticles make flexible worm-like structures. These figures hint at the richness of these interactions in directing complex assemblies. Here, we discuss only particle pair interactions and particle-curvature interactions as well as what insights they can lend into the complex assemblies illustrated in these examples.\\
We seek to answer questions like:  Why do the cylinders assemble end-to-end?  Why do the ellipsoids assemble side-to-side?  Why is the chain of cylinders rigid, while the chain of ellipsoids is not?  Much of this material was published in a variety of papers, including refs.~\cite{Botto,Lorenzo,Eric,Lewandowski2010}.\\

\begin{figure} 
\centering
\includegraphics[width=0.5 \textwidth]{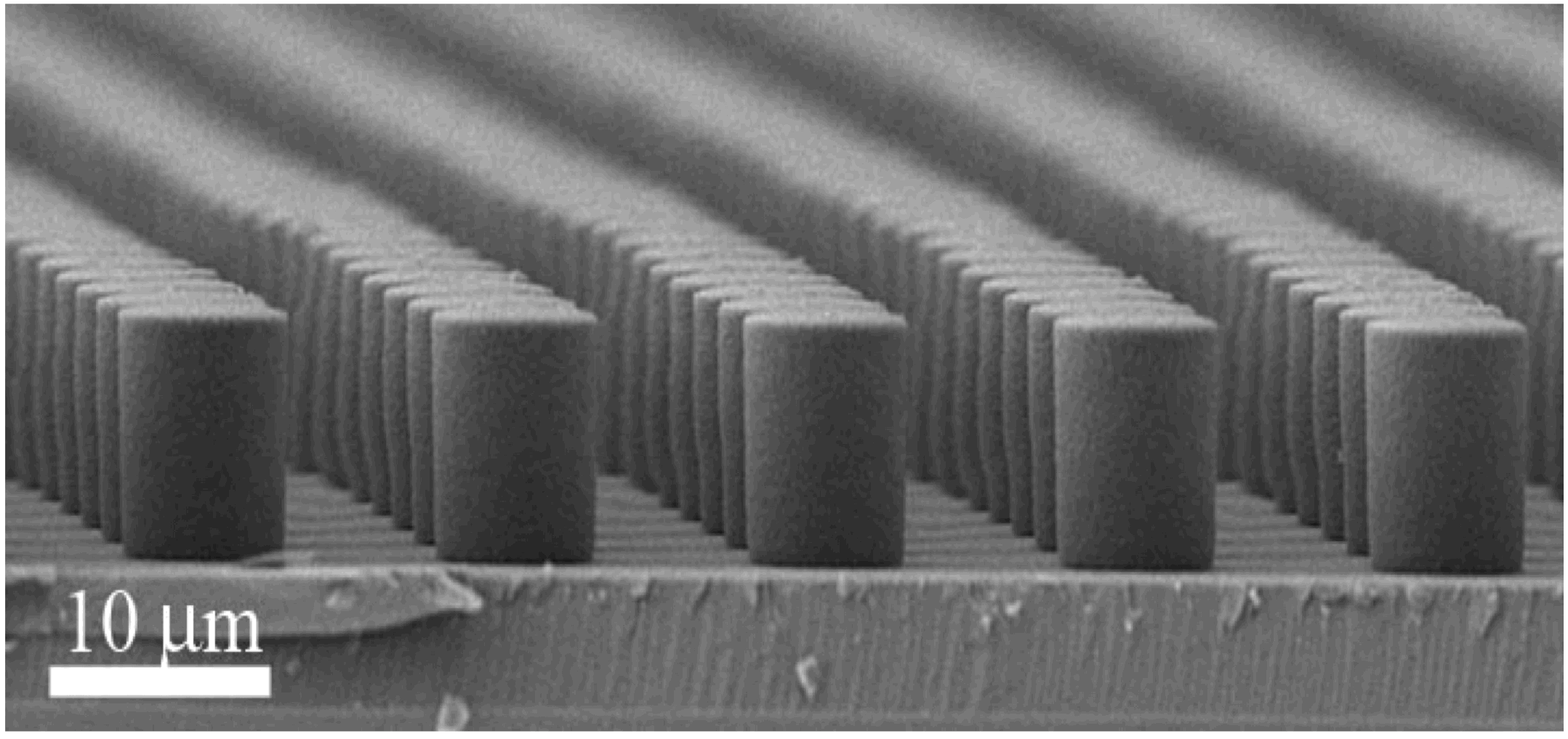}	     	    
 \caption{Microcylinders made by standard lithographic methods from the epoxy resin SU-8.} 
\label{fig_11}
 \end{figure}

In our experiments, we employ simple colloids like polystyrene microspheres.  We also use microparticles made from the epoxy resin SU-8 using standard lithographic techniques, like the microcylinders depicted in fig.~\ref{fig_11}. By gently sonicating this structure, the cylinders are liberated from the substrate and can be spread (or placed by spraying gently with compressed air) onto the air-water interface. These were exploited, for example, in the motivating figures above. In much of this work, we study the assembly of microcylinders that are $7-10~\mu$m in diameter, and roughly $20~\mu$m long.  An \emph{a priori} estimate of capillary interactions for these particles is $\gamma \pi {a^2} \sim {10^7}{k_B}T$.\\
 
\begin{figure} 
\centering
\includegraphics[width=0.6 \textwidth]{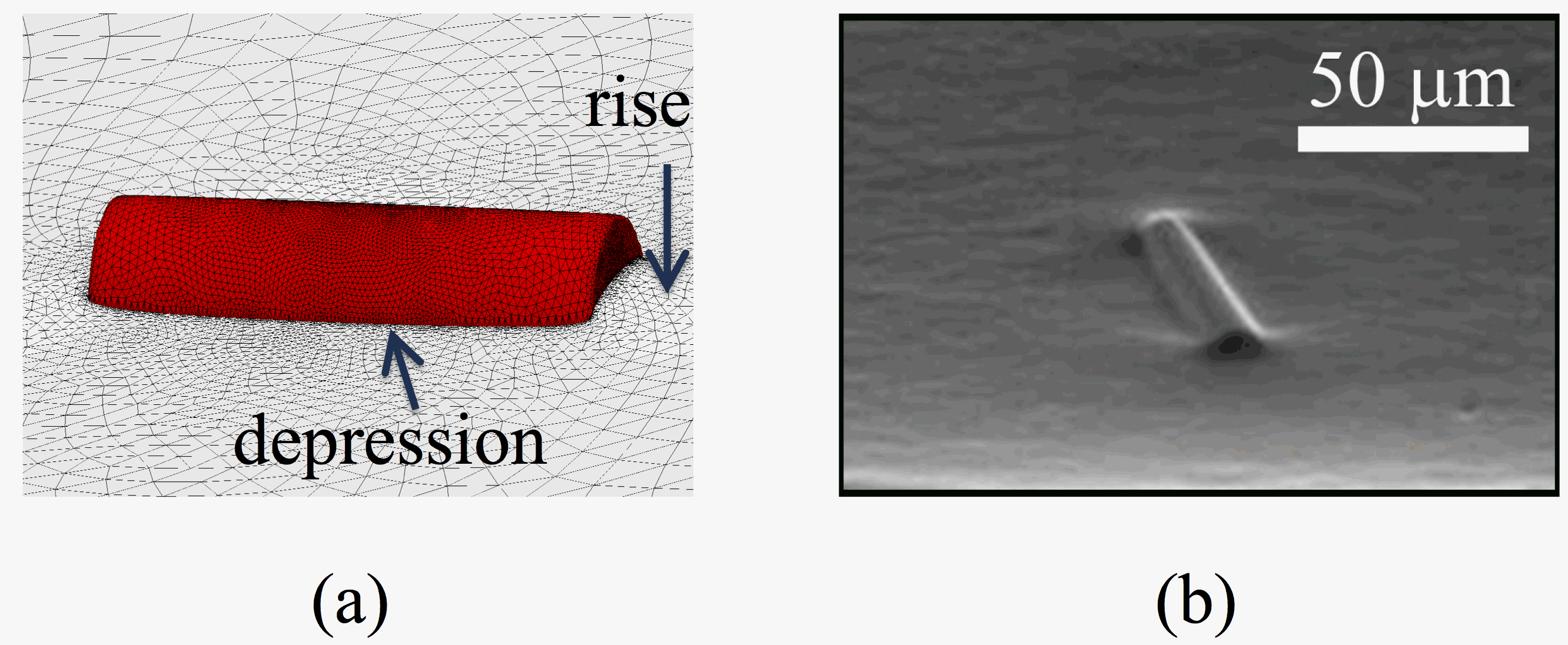}	     	    
 \caption{(a) The equilibrium wetting configuration for a microcylinder with equilibrium contact angle of $80^{\circ}$ obtained from simulation. (b) The shape of the interface around a microcylinder trapped in the air-aqueous  interface.  The image corresponds to an environmental SEM image of a SU-8 microparticle trapped in an air-gellan interface.} 
\label{fig_12}
 \end{figure}

When we initiate a study of a particle at the interface of isotropic fluids, simulation is typically used to find the minimum surface free energy configuration of the interface. See, for example, fig.~\ref{fig_12} which demonstrates interface shape around a microcylinder at an air-water interface (found using the open source code Surface Evolver \cite{S_evolver}).  Such equilibrium wetting configurations are simulated to give a bounding surface profile that guides our thinking.  We do not interpret this literally, however, as evidence is mounting that kinetic trapping of contact lines is the norm, as they pin at asperities, charged sites, or chemical patches \cite{Stamou,Manoharan,Furst1}. We compare these predicted profiles to experiment, e.g. using gellan to gel aqueous subphases, and imaging the wetting configuration by AFM. \\
  \begin{figure} 
\centering
\includegraphics[width=0.5 \textwidth]{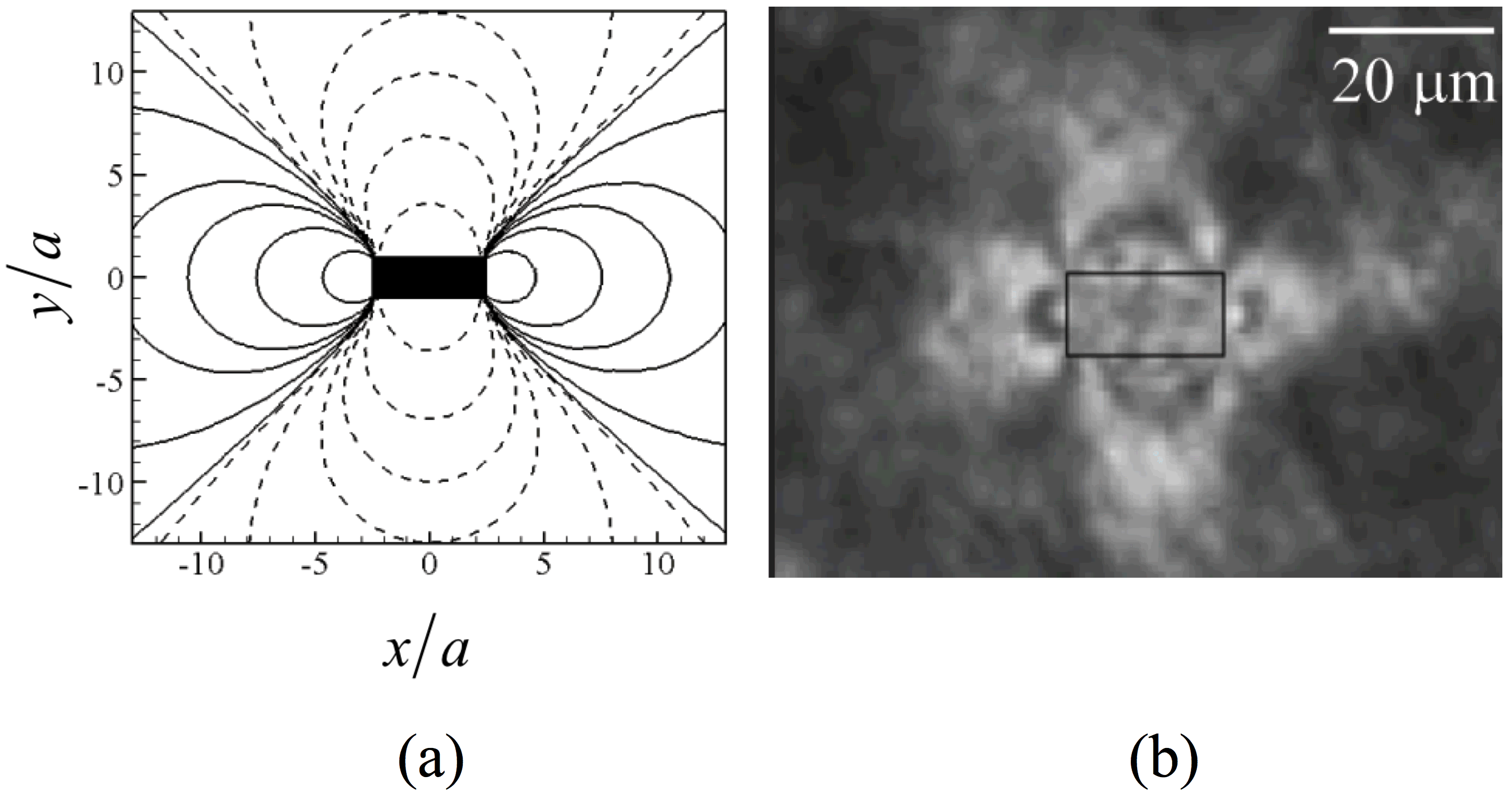}	     	    
 \caption{(a) Isoheight contours for the interface around a microcylinder at equilibrium obtained from simulation. (b) Interferometric images around a microcylinder trapped in the air-water interface.  The elongated quadrupolar structure of the distortion field is clearly evident.} 
\label{fig_13}
 \end{figure}
 
In experiment, we measure interfacial distortions made by particles using interferometry and compare these deformation fields both qualitatively and quantitatively to simulated predictions. For example, we extract multipoles contributing to this distortion field at various distances from the particle center of mass or change coordinate system as appropriate e.g. elliptical coordinates for elongated particles [see interferometry image in fig.~\ref{fig_13}(b)]. We also measure the interface profile using gel trapping methods and environmental SEM to discern near-field details of interface shape.  An example is shown below in fig.~\ref{fig_13}, in which the simulated isoheight contours around the microcylinder are compared to the interferometric images of the particle \cite{Lewandowski2010}. The isoheight contours agree excellently with a quadrupolar distortion in elliptical coordinates for ellipses of a particular eccentricity, i.e. those that corresponding to the smallest ellipse that circumscribes the rectangular projection of the cylinder. The simulated distortion field agrees to within better than $10\%$ at roughly three radii of contact and the agreement improves with distance from the particle.  Far from the particle the elliptical quadrupole decays to quadrupolar distortions in polar coordinates.  This is also confirmed in our experiments; see also fig.~\ref{fig_13}(a), in which interferograms are compared quantitatively to distortions corresponding to polar and elliptical quadrupolar distortions.\\
   \begin{figure} 
\centering
\includegraphics[width=0.60 \textwidth]{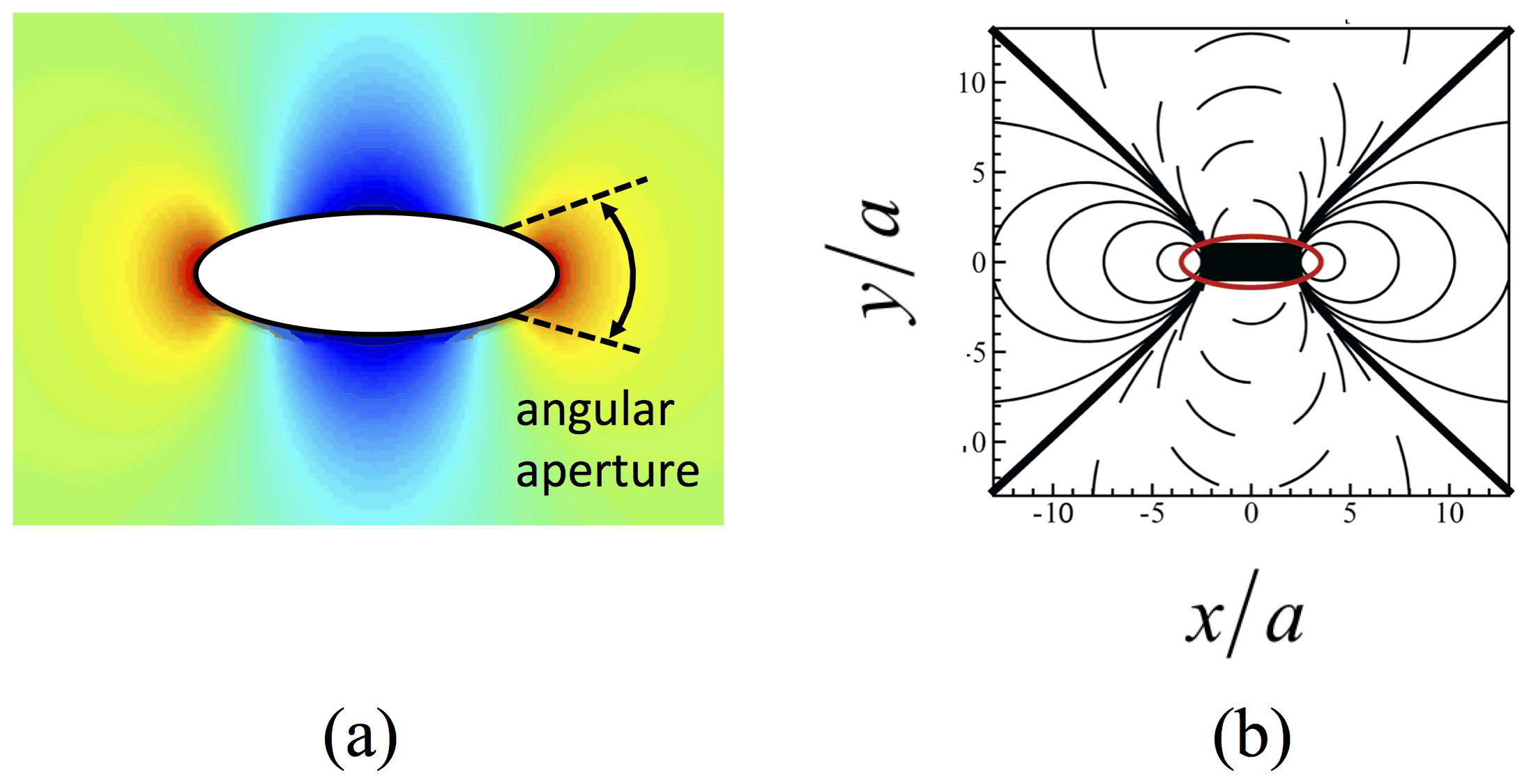}	     	    
 \caption{(a) The distribution of rise and fall around an elliptical quadrupole.  The distortions near the tips of this mode subtends a smaller angle.  As a result, excess area is concentrated in this region. (b) Isoheight contours for an elliptical quadrupole.} 
\label{fig_14}
 \end{figure}
 
A key feature of the elliptic quadrupole is that it concentrates interface distortion near the tip of the ellipse, whereas a polar quadrupole has a symmetric distribution of interface deformation, with regions of interface rise and depression evenly distributed over angular sectors having apertures of exactly ${{\pi  \mathord{\left/ {\vphantom {\pi  2}} \right. \kern-\nulldelimiterspace} 2}}$. For an elliptical quadrupole, the distortion at the tip has a smaller angular aperture than along the side, as can be seen in fig.~\ref{fig_14}. This asymmetry increases with particle aspect ratio. Owing to this asymmetry, the excess area is concentrated near the tips of the ellipse.  Thus, eliminating this area lowers the energy greatly, and tip-to-tip assembly is favored for elongated particles as they approach. We evaluate of the excess area created by two neighboring distortions as a function of their center-to-center distance and orientation \footnote{This is a recapitulation of the calculation performed in sec.~\ref{reflection}, where the distortions can be described in a coordinate system appropriate for elongated particles. Note that the derivation presented in sec.~\ref{reflection} for interacting polar quadrupoles is inspired by work by Stamou \emph{et al.} \cite{Stamou}. In that original work, the superposition approach was used to determine the interface shape.  We modified the analysis using the method of reflections to find surface shapes that obeys the boundary conditions on the particle and evaluated the associated capillary energy.}. We compare predicted forms to experiment by recording under an optical microscope the rotation and translation of pairs of interacting particles at an interface as they dimerize.\\
\begin{figure} 
\centering
\includegraphics[width=0.45 \textwidth]{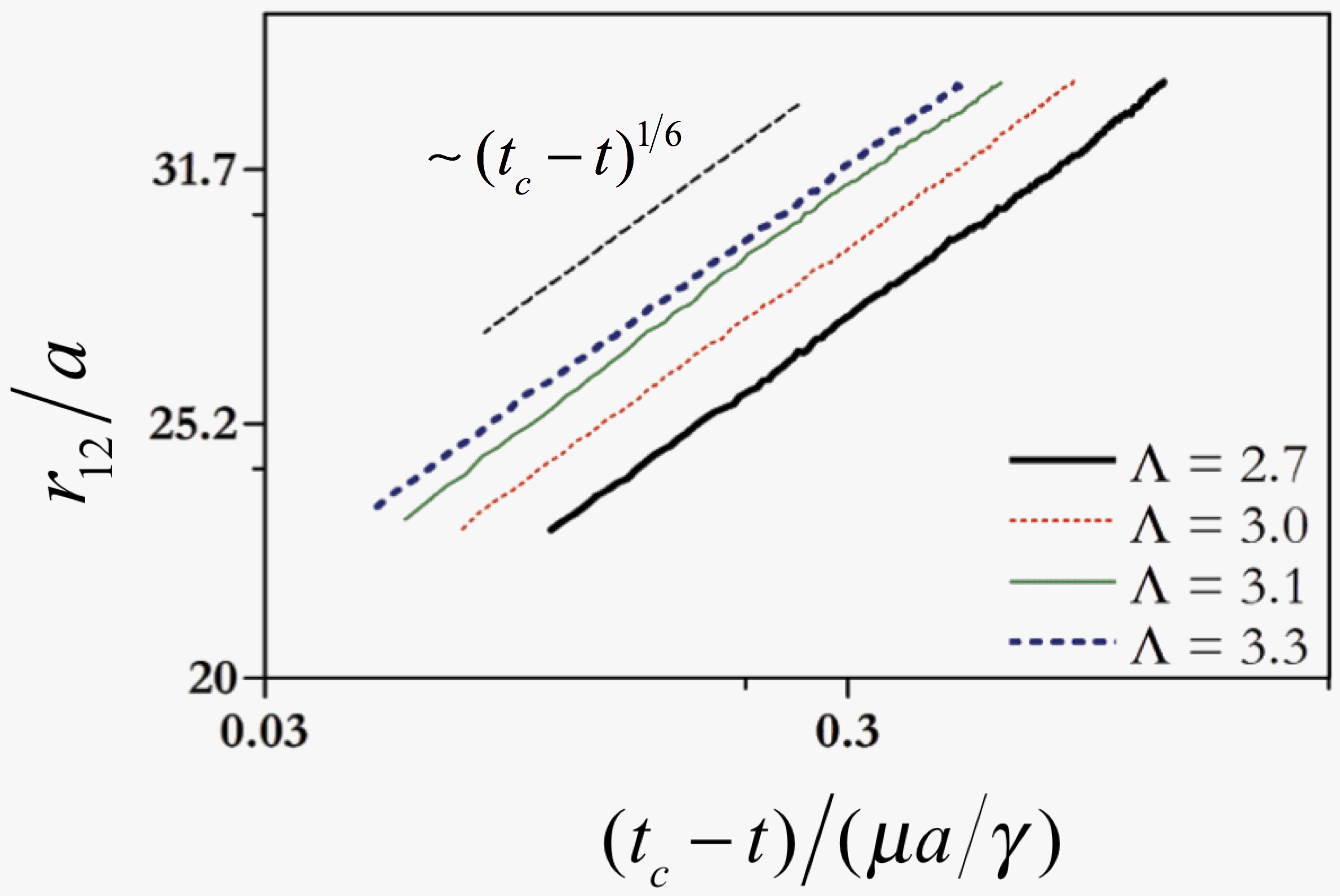}	     	    
 \caption{Power law dependence of separation distance versus time remaining to contact, where $t_c$ is the time in seconds when the particles attach end to end.  The trajectories in the far field obey the functional form predicted for interacting polar quadrupoles for various particle aspect ratios $\Lambda={L}/{2a}$. } 
\label{fig_15}
 \end{figure}
 
 In the far field, as pairs of cylinders approach, they obey the expected power law form for interacting polar quadrupoles ${r_{12}} \sim {({t_c} - t)^{1/6}}$ where $r_{12}$ is the center-to-center distance.  The power law can be found by equating viscous drag to the capillary force, which requires ${{{\rm{d}}{r_{12}}} \mathord{\left/ {\vphantom {{{\rm{d}}{r_{12}}} {{\rm{dt}}}}} \right. \kern-\nulldelimiterspace} {{\rm{dt}}}} \sim r_{12}^{ - 5}$. This is shown in fig.~\ref{fig_15} for particles of various aspect ratio $\Lambda  = {L}/{{2a}}$ (similar results were shown, prior to our findings, by Loudet and collaborators for ellipsoids \cite{Arjun}). The capillary energy of interaction can be estimated.  The capillary energy is balanced by viscous dissipation, i.e. ${C_D}\mu a\int_{r_{12}^I}^{r_{12}^{II}} {U{\rm{d}}{r_{12}}}  = {E_{cap}}(r_{12}^{II}) - {E_{cap}}(r_{12}^I)$.
\begin{figure} 
\centering
\includegraphics[width=0.75 \textwidth]{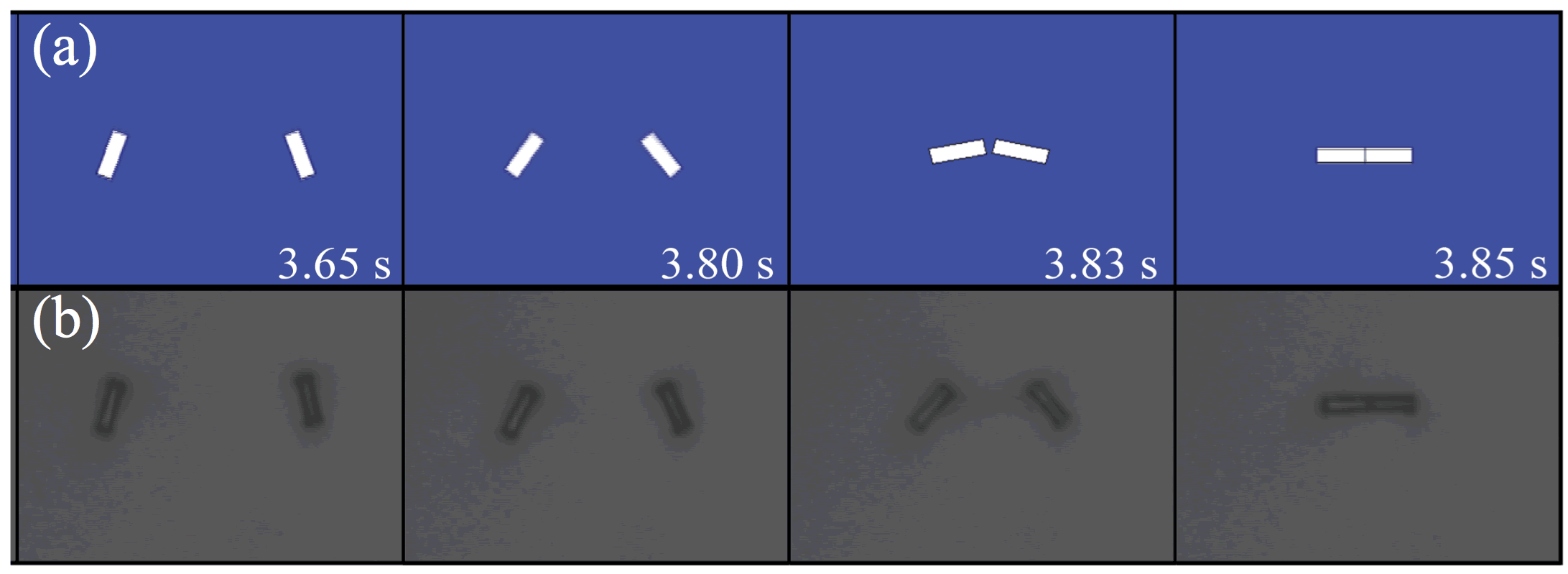}	     	    
 \caption{(a) Predicted trajectory based on interacting elliptical quadrupoles. The near-field torque enforces end-to-end alignment until contact. (b) Time-lapsed image of cylindrical microparticles rotating to assemble end-to-end. Key differences emerge, however, as elliptical quadrupoles minimize capillary energy by rotating near contact to side-to-side alignment, while cylinders form rigid chains.} 
\label{fig_17}
 \end{figure} 
Thus, by integrating the velocity of the particle over its path, we can estimate the energy dissipated in the far field, and find it to be $\left( {2.24 \pm 0.67} \right) \times {10^5}{k_B}T$.  Performing such integration all the way to contact, we find energies of $ \sim {10^7}{k_B}T$.\\

Experiments by us and others \cite{Vermant}, show that elongated microparticles in pair interaction rotate in the near-field, maintaining a mirror symmetric configuration, to bring the particle tips to contact (fig.~\ref{fig_17}). This is captured well by the analytical pair interaction potential, which predicts a near-field capillary torque that enforces this alignment that scales as $r_{12}^{-6}$, where $r_{12}$ is the center-to-center distance of the particles.  However, once elliptical quadrupoles attach tip-to-tip, they rotate in contact to side-to-side configurations. This is similar to the behavior of (weakly charged) ellipsoidal microparticles, and in marked contrast to cylindrical microparticles, which form rigid chains of end-to-end oriented cylinders. In fig.~\ref{fig_17}, we show a time-lapsed images of cylinders originally in nearly side-to-side rotating in the near-field to an end-to-end arrangement, and compare that motion to the predicted motion from analysis of interacting elliptical quadrupoles until contact.\\
\begin{figure} 
\centering
\includegraphics[width=0.6 \textwidth]{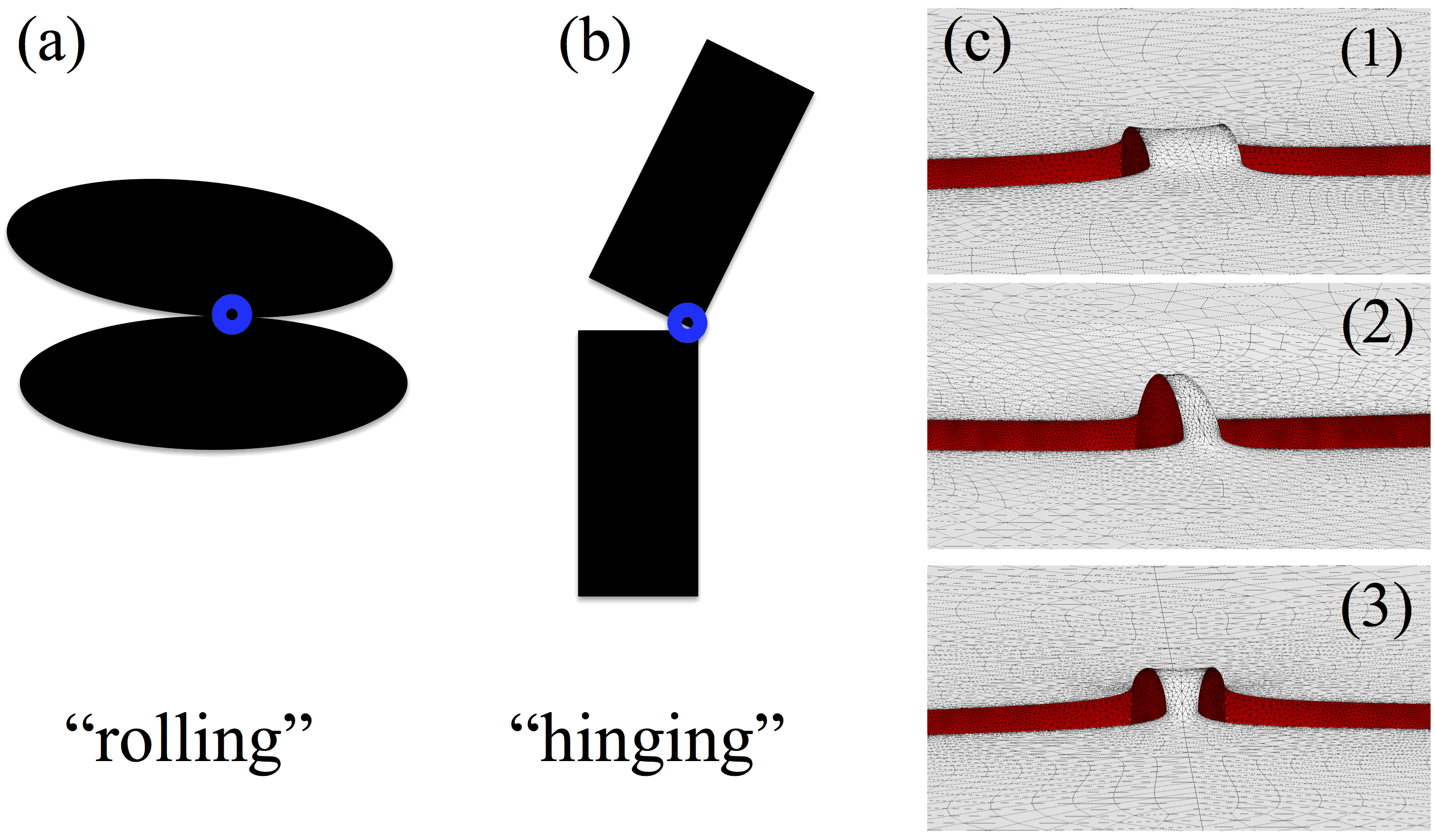}	     	    
 \caption{(a) Ellipsoids roll over each other close to contact and (b) cylindrical particles hinge around their sharp edge (c) numerically computed capillary bridge formed in gap between cylinders near contact. The bridge rearranges as the angle from end-to-end alignment is perturbed.  This rearranges the wetting configuration on the solid-liquid interfaces and dictates an energy landscape that differs strongly from that for interacting ellipsoids.} 
\label{fig_18}
 \end{figure}
 
Thereafter, for particles closer to contact, we must abandon the analysis in normal modes, and simply resort to simulation.  Here, we ask what cements end-to-end assembly for cylinders? The presence of sharp edges or facets \cite{Lorenzo} constrains cylinders to rotate in a hinging motion, whereas  ellipsoidal microparticles can roll over each other.  These differing steric constraints dictate distinct energy landscapes. In particular, in the gap within the hinged space between the cylinders, capillary bridges form with configurations strongly influenced by the wetting energy of the solid-liquid surfaces (see fig.~\ref{fig_18}).\\
\begin{figure} 
\centering
\includegraphics[width=0.65 \textwidth]{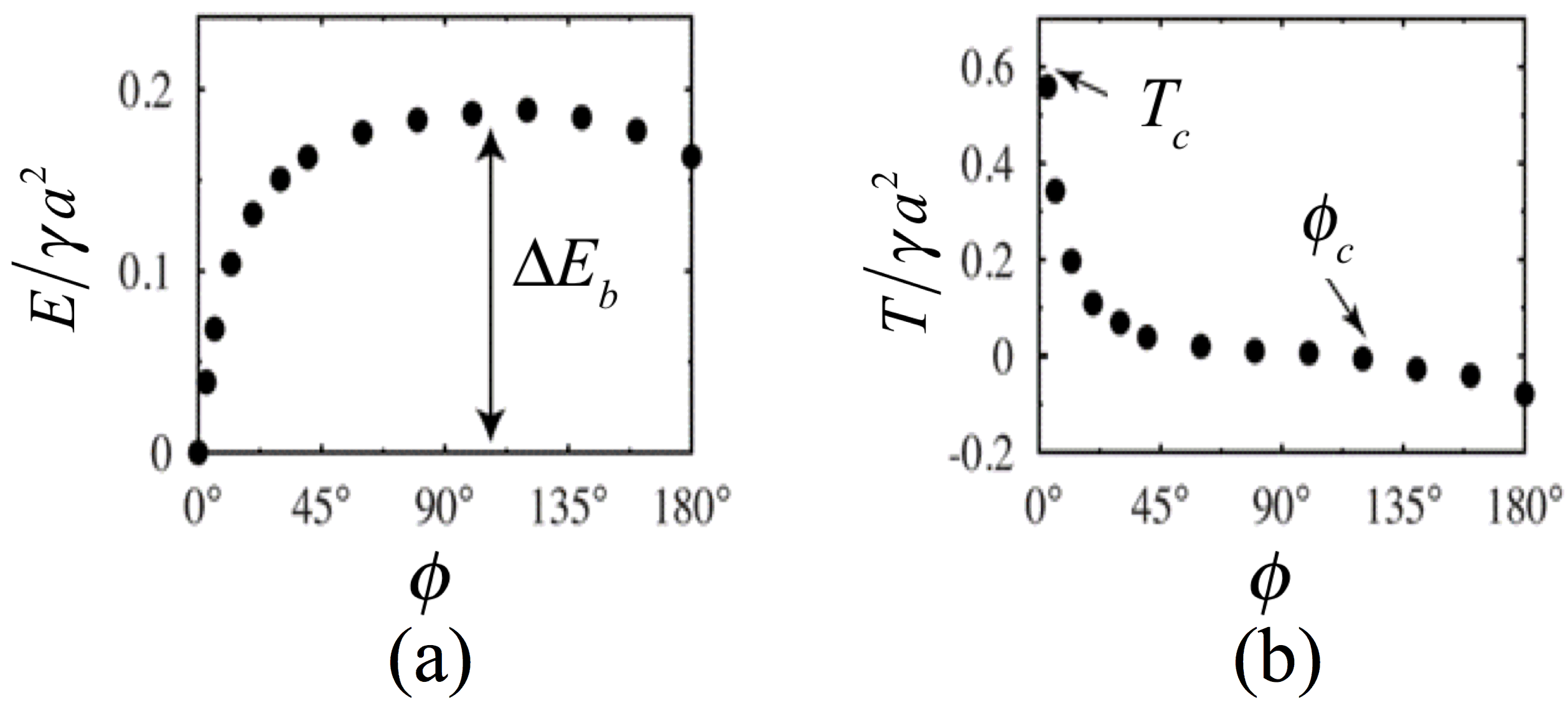}	     	    
 \caption{Ellipsoids rolling over each other close to contact. (a) Energy landscape for ellipsoids assembled tip-to-tip, rotating to an equilibrium side-to-side alignment. (b) Torque needed to rotate the ellipsoids from side-to-side alignment indicates an elastic regime.} 
\label{fig_19}
 \end{figure}
 
\begin{figure} 
\centering
\includegraphics[width=0.65 \textwidth]{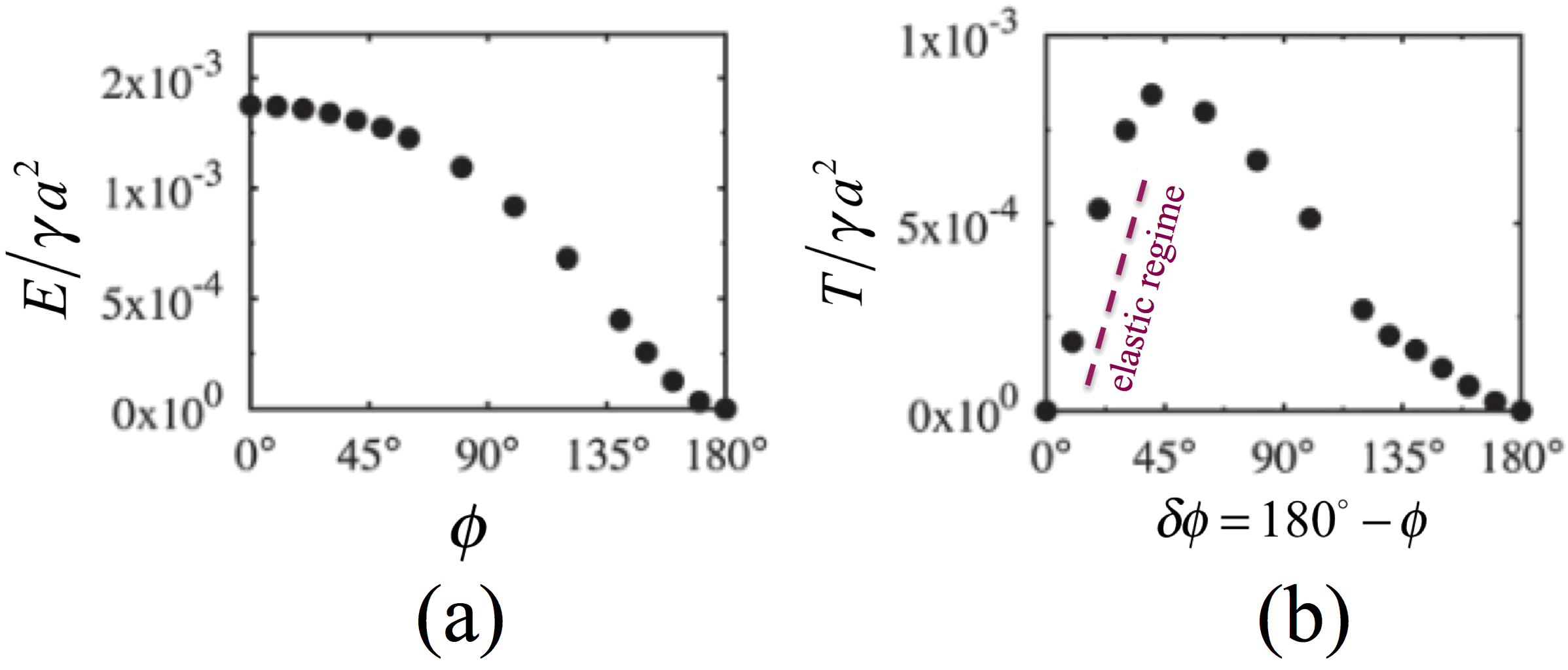}	     	    
 \caption{Cylinders hinging over sharp edges close to contact. (a) Energy landscape for cylinders assembled end-to-end, facing significant energy barrier $\Delta E_b$  between end-to-end  and side-to-side alignment. (b) Torque needed to rotate the cylinders from their equilibrium end-to-end alignment indicates a yield stress that renders cylinders rigid below that torque.} 
\label{fig_20}
 \end{figure}

\indent Simulations are performed for both ellipsoidal microparticles (fig.~\ref{fig_19}) and cylindrical microparticles (fig.~\ref{fig_20}) as they rotate from a tip-to -tip configuration to a side-to-side alignment. The reference energy for each graph is the equilibrium configuration, i.e. side-to-side for ellipsoids and end-to-end for microcylinders.  (The angle $\phi$ in these figures now represents the displacement from tip-to-tip alignment to a side-to-side configuration.  $\phi=0$ indicates end-to-end aligned objects; $\phi={{\pi  \mathord{\left/ {\vphantom {\pi  2}} \right. \kern-\nulldelimiterspace} 2}}$ indicates side-to-side aligned objects.)  We consider ellipsoids and spheres with identical wetting conditions and particle sizes.  For ellipsoids, the energy landscape is smooth and there is no energy barrier between the two states.  The equilibrium state with side-to-side oriented particles features a parabolic energy well.  If that well is perturbed, e.g. by rolling the ellipsoids to make a small distortion of the chain, the torque required to change their angle from the equilibrium angle in the plane of the interface is defined by the derivative of the energy well.  This is the torque shown on fig.~\ref{fig_19}(b).  It reveals that the ellipsoidal assembly is linearly elastic to small displacements from side-to-side alignment over a range of angles.\\
 
These simulations lend insight into the mechanics of the chains of ellipsoids.  The flexural rigidity $\kappa$ of the chain can be estimated. The chain curvature $C$ is given by the ratio of the infinitesimal change in angle $\delta \phi$ to the corresponding element of length. The curvature in correspondence to each bond is therefore $C={{{\delta \phi } \mathord{\left/ {\vphantom {{\delta \phi } {2R}}} \right. \kern-\nulldelimiterspace} {2R}}}$, where $2R$ is the distance between the bonds. The flexural rigidity is simply the proportionality constant between torque $T$ and $C$, i.e. $\kappa=e_1\gamma R^3$, where $e_1$ is a constant (which depends on aspect ratio and contact angle). For $\Lambda=3$ and $\theta_c=80^{\circ}$, our simulations give $e_1\sim 0.003$. For $\gamma  = 70~{{{\rm{m}}N} \mathord{\left/ {\vphantom {{{\rm{m}}N} {\rm{m}}}} \right. \kern-\nulldelimiterspace} {\rm{m}}}$ and $R=2.25~\mu$m, $\kappa$ is about $6\times10^5 k_BT \mu$m, which translates to $\sim 10^{3}-10^{4} k_BT$ of elastic energy for chain of ellipsoids having typical curvature $(100)^{-1}\mu$m. Thermal fluctuations alone are thus unable to bend chains formed by micro-ellipsoids. However, it should be emphasized that, in a compressed monolayer of microparticles, chain bending is caused by the contact forces between neighbouring chains, and not by thermal fluctuations. For an external torque of the order $10^5 k_BT$ (comparable to that supported by a rigid chain of cylinders in experiment (see ref.~\cite{Lorenzo} for detail), a chain of ellipsoids would exhibit a very high curvature.\\

The energy landscape (shown in fig.~\ref{fig_20}) for cylindrical microparticles differs from that of ellipsoids in several respects.  The hinging steric constraint imposed by the edges requires significant redistribution of liquid over the particle end faces as the assembly is distorted. This creates a large energy barrier to realignment, with end-to-end alignment being strongly preferred and side-to-side alignment being a local minimum, with a significantly higher energy than the observed end-to-end state.  The energy difference between end-to-end and side-to-side configurations is an order of magnitude larger for the cylinders than for the ellipsoids. Recall that this difference is solely due to the particle shape -- the radius and wetting conditions of the particles are similar. 
This energy landscape also has implications for the mechanics of chains; the torque to deform the chain is again the derivative of the energy with respect to angle. Chains of cylinders do not respond to bond bending as elastic elements, but as ÒbrittleÓ elements. Under a localized torque, they are predicted to remain rigid for torques lower than a critical torque $T_c$ and ``break'' for larger applied torques. For $a=2.25~\mu$m and $\gamma=70~{{{\rm{m}}N} \mathord{\left/ {\vphantom {{{\rm{m}}N} {\rm{m}}}} \right. \kern-\nulldelimiterspace} {\rm{m}}}$.\\

The strength of the capillary bond formed between dimerized particles can be measured by various means.  In our research team, we have studied these interactions for the cylindrical microparticles. We integrated a cylindrical particle made of nickel into a chain of SU-8 cylinders, and have used magnetic field gradients to rotate the chain of cylindrical particles at velocities for which viscous torques in excess of $10^5~k_BT$ were exerted.  The chain of cylindrical microparticles remained rigid, as this viscous torque was below the critical threshold to break the chain.  One could imagine, for weaker bonds, employing deformation using magnetic tweezers to compare to the predicted forms.\\

\indent In the prior discussions, we have focused on capillary attraction.  Under some circumstances, we can generate capillary repulsion, and use capillarity to define equilibrium particle separations.  We have explored this idea using particle roughness or contact line undulation with wavelengths small compared to the particle to generate repulsion only in the very near-field.
\section{Near-field repulsion}
\begin{figure} 
\centering
\includegraphics[width=0.75 \textwidth]{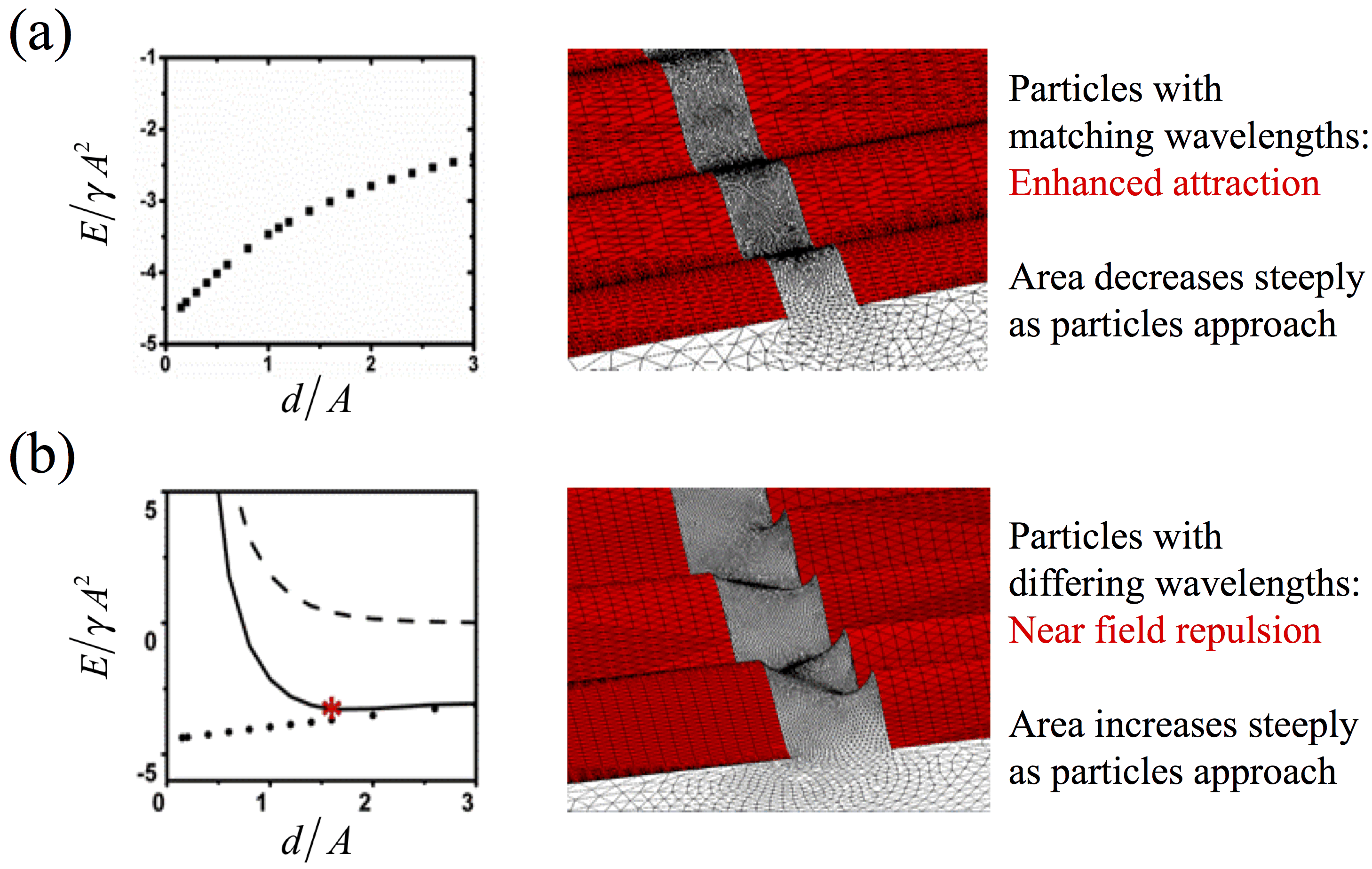}	     	    
 \caption{The concept of near-field capillary repulsion: Particles with polar quadrupolar interaction in the far field and near-field distortions owing to wavy contact lines. (a) Particles with matching near-field undulations attract until contact with enhanced capillary attraction. (b) Particles with mismatched near-field undulations attract in the far field and repel in near-field.} 
\label{fig_21}
 \end{figure} 
 Capillary bonds can be remarkably strong for microscale particles, with typical bond strengths of $\sim 10^5-10^7~k_BT$. Imparting repulsion to counter such attraction is a challenge, since the usual means for introducing repulsion, e.g. using charge to impart electrostatic repulsion, or by appending ligands to particles to impart steric repulsion, are too weak to overcome attractive interactions of this magnitude. In an analytical paper, Lucassen \cite{Lucassen} suggested an interesting approach: Use capillarity in the near-field to impart repulsion!\\
 
\indent Repulsive capillary interactions are associated with small scale contact line undulations. This concept can be understood by considering near-field capillary interactions between microparticles on interfaces in a highly simplified geometry, i.e. treating the particles as a pair of vertical parallel walls with pinned sinusoidal contact lines, connected by a fluid interface. The interface deforms over distances comparable to the wavelength; associated capillary energies decay over similar length scales. As the walls approach each other, the deformation fields interact; the interfacial area increases unless the sinusoidal contact lines are of identical amplitude and wavelength and are in phase.\\
\begin{figure} 
\centering
\includegraphics[width=0.75 \textwidth]{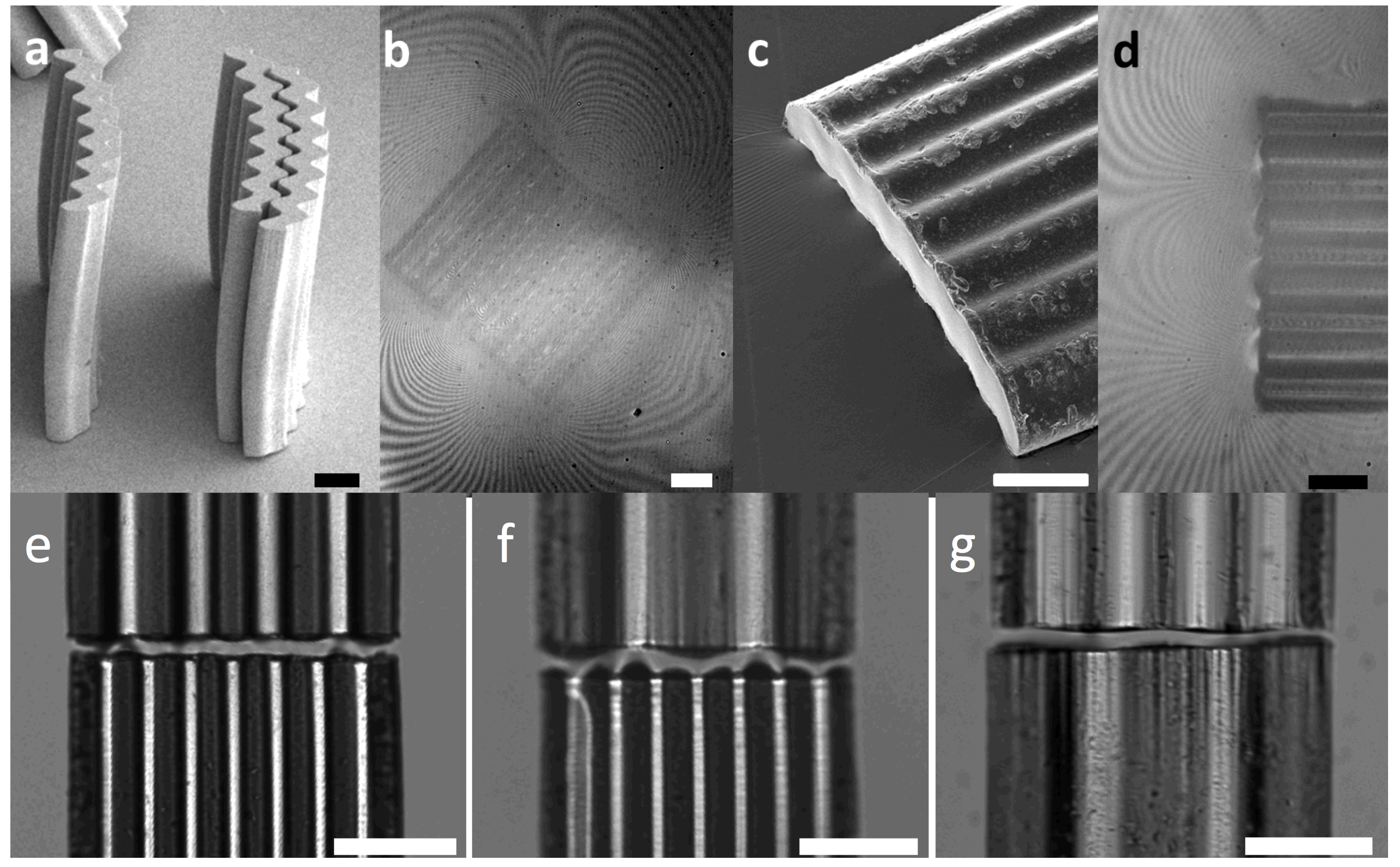}	     	    
 \caption{Experimental confirmation of capillary repulsion. (a) Undulated microparticles fabricated by lithographic methods. (b) Quadrupolar deformation around those microparticles at an air-water interface (c) PDMS replica of air-water interface showing contact line pinning on the undulated contact lines. (d) Near-field distortions owing to wavy contact line. For panels (a-d), scale bars: $50~\mu$m. (e-f) Assemblies of microparticles with mismatched wavelengths assembled to equilibrium separation distances. (e-f) Scale bars: $100~\mu$m.} 
\label{fig_22}
 \end{figure} 
 
This implies that undulated contact lines with wavelengths smaller compared to the particle alter pair interactions only in the near-field. If these features are identical, they strengthen the capillary bond and draw the particles into contact. If, however, they differ, they create repulsion owing to capillarity and allow particles to find equilibrium separation distances (see fig.~\ref{fig_21}). We have studied the energy of interaction of microparticles designed to impart a quadrupolar distortion in the far field, with contact line undulations having wavelength significantly smaller than the characteristic particle size. Using simulation and experiment, we show that identical microparticles with features in phase attract each other and microparticles with different wavelengths, under certain conditions, repel each other in the near-field, leading to a measurable equilibrium separation. We simulate the pair interaction between end-to-end aligned particles in close proximity as a function of separation distance $d$ between the particle end faces. Since the contact line is assumed to be pinned, the pair interaction energy $E$ is given by $E=\gamma S_{LV}$ where $\gamma$ is the interfacial tension and $S_{LV}$  is the area of the liquid-vapor interface. For the relatively small amplitude $A$ that we explore, the energy is approximately quadratic in wave amplitude $A$ for the corrugated particles. Therefore, we normalize the capillary energy by $\gamma A_2$, the capillary force by $\gamma A$, and distances by $A$. Figure \ref{fig_21}  compares interactions between a pair of corrugated particles with matching corrugations, which have enhanced attraction in the near-field. For  mismatched particles, e.g. particles with different wavelengths, particles attract in the far field owing to their quadrupole-quadrupole interaction, but repel in the near-field.\\

 We have performed experiments using lithographically-formed particles with wavy edges placed at interfaces (fig.~\ref{fig_22}).  We find that particles with matching contact lines approach until contact, but those with differing wavelengths have near-field capillary repulsion, allowing them to assemble at well-defined distances \cite{Lucassen, Lu}. This finding may be of broad and general interest, as particle roughness on scales far smaller than the particle dimensions can be modeled as such undulations.    
\section{Curved interface}
\begin{figure} 
\centering
\includegraphics[width=0.95 \textwidth]{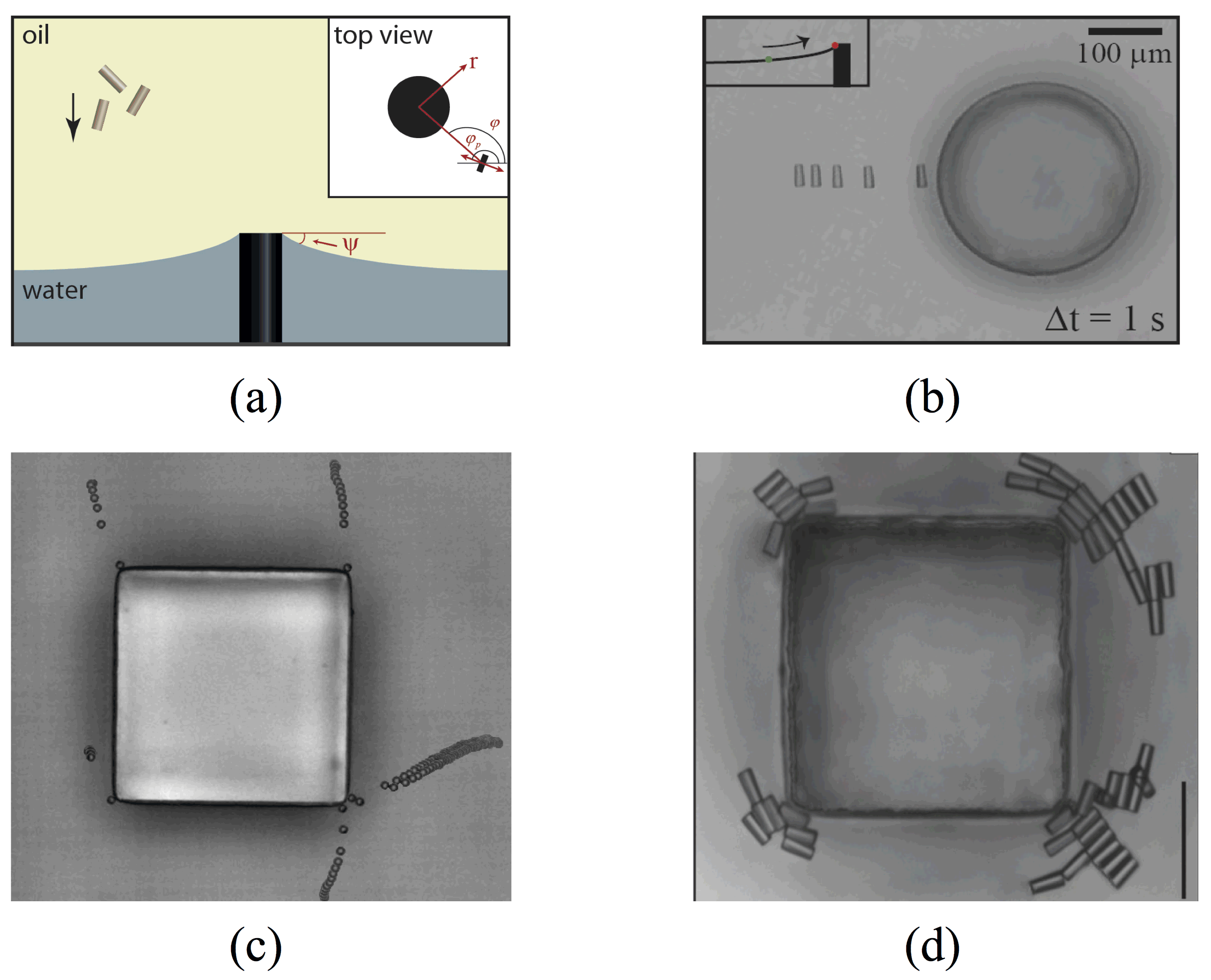}	     	    
 \caption{(a) Schematic of technique used to mold interface curvature.  A micropost with sharp edges is used to pin the interface at a shallow angle $\psi$.  Oil is placed gently atop this water layer.  The micropost can be circular (b) ellipsoidal (not shown), square (c) and (d), or other shape in cross section.  Particles placed in the oil settle through the oil, attach to the interface, migrate and assemble owing to curvature-particle and particle-particle interactions.} 
\label{fig_23}
 \end{figure}

 Fluid interfaces can be curved for many reasons, e.g. owing to confinement and volumetric constraints, the presence of pinning sites that distort the interface shape, or, as with a meniscus near a planar wall, because of the balance of surface energies and gravity. In all cases, we have developed theory and experiments that allows us to tailor interface shape to direct particles to given loci. Our particular interest is in designing fluid volumes with interface shapes that favor given configurations of particles, typically in situations where gravitational effects are negligible (i.e. confining fluid volumes have characteristic lengths smaller than the capillary length). We show particle trajectories and structures formed on curved interfaces in fig.~\ref{fig_23}. These interface curvatures are molded by pinning the fluid interfaces on microposts.\\
 
 The universal occurrence of the quadrupolar mode is an exciting fundamental finding that allows us to design curvature landscapes to direct assembly.  Key to this interface design are certain simple geometrical facts.  The height of the host surface in the absence of particles at any point satisfies the Young-Laplace equation, $2\gamma H_0=\Delta p$, where $H_0$ is the mean curvature, and $c_1$ and $c_2$  are the principal curvatures of the interface defined in terms of the principal radii of curvature $R_1$ and $R_2$. At any point on the interface, a polar coordinate $(r,\phi)$ can be defined tangent to the interface. Adopting the Monge representation, the host interface height can be expressed as ${h_{host}} = \frac{{{r^2}{H_0}}}{2} + \frac{{\Delta {c_0}}}{4}{r^2}\cos 2\phi$, where $\Delta c_0 = c_1-c_2$, and, by convention, $c_1>c_2$.  That is, the interface can be decomposed into a bowl-like $H_0$ and saddle-like $\Delta c_0$ curvatures. (We have discussed this surface decomposition in terms of a non-uniform asymptotic expansion and confirmed its validity in the small slope limit using a semi-analytic Green's function representation for the interface with numerically-determined coefficients for geometries that correspond to our experiments.  We discuss this representation in sec.~\ref{green}.) For small volumes of fluid, $H_0$ is typically constant.  However, in general, $\Delta c_0$ varies significantly, even in small confining volumes.  Notably, the deviatoric curvature field is described by a quadrupolar mode.  If a particle is located at $r=0$ in this curvature field, the particle sourced distortion ${h_{p}}\frac{{{a^2}}}{{{r^2}}}\cos 2(\phi-\alpha)$ will interact with this curved interface shape, where $\alpha$ is the angle made by the quadrupolar rise axis with respect to the first principal axis.   By orthogonality, this mode couples to the deviatoric curvature of the host interface, forcing regions of rise around the particle to align along the first principal axis, and regions of depression to align along the second principal axis.  Since this mode is ubiquitous, this effect is general, and will occur for all particles for which the associated capillary energy is greater than thermal energy.\\
 
In sec.~\ref{asymp}, we presented the theory to describe the curvature capillary energy.  In this theory, we assume small slopes, and the particles are assumed to be small compared to the principal radii of curvature of the interface.  We also assume that when a particle attaches to the interface, it rotates to lie in the tangent plane to the interface to eliminate dipolar deformations with respect to the tangent plane to the interface forbidden absent external torques. In this plane, the particle distorts the interface to impose a pinned quadrupolar mode of magnitude $h_p$ at its contact line.  In so doing, the particle induces a distortion that depends on the deviatoric curvature. The distortions created by the particle decays with distance from the particle.  Far from the particle, the interface recovers the host interface of the form given above. We then calculate the capillary energy associated with this distortion and find  ${E_{\rm{curvature}}} = {E_{{H_0}}} - \gamma \pi {a^2}\frac{{{h_{p}}\Delta {c_0}}}{2}\cos 2\alpha$, where we adopt the reference state of a particle attached to a planar interface. The first term ${E_{{H_0}}} =  - \gamma \pi {a^2}\frac{3}{4}{H_0}^2{a^2}$ is the dependence of the capillary energy on the mean curvature, assumed constant.   The second term is the energy associated with the deviatoric curvature $\Delta c_0$. Gradients of this energy are forces and torques.  Moreover, the host interface exerts a torque aligning the particle along the principal axes ($\alpha=0$), and a force that drives the particle to regions of deviatoric curvature. (This expression differs from that derived by others in the literature, who have adopted additional \emph{ad~hoc} closure conditions.)\\

This expression motivates us to form a host interface with a curvature field.  When particles attach to this interface, they will orient and migrate in this curvature field. Our laboratory has developed the use of lithographic methods to mold fluid interfaces for use in experiment that can be compared with theory.  We have studied the migration of particles with increasing complexity on curved fluid interfaces: Planar microdisks, microspheres, and microcylinders, trapped at oil-water interfaces.  For each shape, we assume pinned contact lines. These methods are described briefly here.   
 \subsection{Molding of the fluid interface}
  We commonly form a curved oil-water interface around a circular micropost of height $H_m$ and radius $R_m$. Far from the micropost is a confining ring, located several capillary lengths from the micropost center. This structure is filled with water so that the contact line is pinned at the top of the micropost and the slope of the interface is shallow ($\psi \sim 15^{\circ}-18^{\circ}$, as measured using a goniometer; see fig.~\ref{fig_23}). Oil (e.g. hexadecane) is gently poured on top of this water layer. The resulting interface shape close to the micropost is well approximated by ${h_0} = {H_m} - {R_m}\tan \psi \ln ({L}/{{{R_m}}})$, where $L$ is the radial distance measured from the center of the micropost. This interface shape (shown schematically) has zero mean curvature $H_0=0$ and position-dependent deviatoric curvature field $\Delta c_0(L)={2R_m \tan \psi}/{L^2}$. This expression is axisymmetric about the micropost.  The deviatoric curvature is greatest at the post and decays monotonically with distance from the post.  If particles follow trajectories of increasing deviatoric curvature, they should migrate radially toward the post along paths of decreasing $L$. Given this information, the predicted curvature capillary energy can be tested in many ways.  Note that torques aligning the particle along the principal axes are local, and particles rotate rapidly upon attachment to the interface to align their distortion along the principal axes $\alpha=0$. This is commented upon further, in sec.~\ref{curved-interface}.  We first summarize experiments performed to test the predicted force driving migration.\\
  
We have compared several aspects of particle trajectories to theoretical predictions.  For example,  
particles are predicted to migrate along contours defined by the deviatoric curvature gradient. This is confirmed for particles migrating around a circular micropost, which move in linear trajectories along radial paths toward the micropost, terminating at the post for isolated particles. We have also shown that particles can move along more complex paths on interfaces with more complex curvature fields, e,g., around microposts of elliptical or square cross section.\\

The energy dissipated along a particle path can be used to infer the capillary energy $\Delta E=E(L)-E(L_0)$. This can be graphed against the change in deviatoric curvature along the path $\Delta {c_0}(L) - \Delta {c_0}({L_0})$.  The graph is predicted to be linear.  Owing to their small size, the particles migrate in creeping flow. The energy dissipated can be inferred from experiment by integrating the viscous drag force over the particle path $\Delta E = \int_L^{{L_f}} {{F_{drag}}{\rm{d}}L}  = \mu a{C_d}({\theta _C},\Lambda )\int_L^{{L_f}} {U{\rm{d}}L}$ where $C_d$ is the drag coefficient. Note that this expression is valid only for particles far from neighbors and bounding walls. Power laws relating particle location with respect to the micropost center $L(t)$ and $t$ are implied by the equality of the viscous drag force to the capillary force. These power laws can be compared to the particle trajectories.  The force driving migration of a particle has the predicted form:
 \begin{eqnarray}
 {F_{\rm{curvature}}} =  - \frac{{d{E_{\rm{curvature}}}}}{{{\rm{d}}L}} = \gamma \pi {a^2}(\frac{{{h_{p}}}}{2})\frac{{{\rm{d}}\Delta {c_0}}}{{{\rm{d}}L}},
 \end{eqnarray}
where, around the circular micropost, ${{{\rm{d}}\Delta {c_0}} \mathord{\left/ {\vphantom {{{\rm{d}}\Delta {c_0}} {{\rm{d}}L}}} \right.
 \kern-\nulldelimiterspace} {{\rm{d}}L}} =  - {{4\tan \psi {R_m}} \mathord{\left/ {\vphantom {{4\tan \psi {R_m}} {{L^3}}}} \right.
 \kern-\nulldelimiterspace} {{L^3}}}$. Equating this force to the viscous drag and gathering constants, we can integrate along a particle path from an instantaneous position $L(t)$  to a final position $L(t_f)$; $- B\int_t^{{t_f}} {\rm{d}}t = \int_{L(t)}^{L({t_f})} {{L^3}{\rm{d}}L}$  where the constant $B = \frac{{2\pi \gamma a{h_{p}}{R_m}}}{{{C_d}\mu }}\tan \psi$. In this expression, analysis is terminated at a final value for $L(t_f)$, typically $\sim10$ radii from contact with the post, so that wall-particle interactions in the drag and capillary energy expressions are not required, and $t_f$ is the time at which this position is attained.   This implies a power law dependence between the particle's instantaneous position with respect to the center of the micropost and time remaining until contact, 
\begin{eqnarray}
L(t) = {\left[ {B({t_f} - t) + L^4{{({t_f})}}} \right]^{1/4}},\label{power}
\end{eqnarray}
When the particle position is graphed against time remaining to contact in log-log form, a line of slope ${{1 \mathord{\left/
 {\vphantom {1 4}} \right. \kern-\nulldelimiterspace} 4}}$ should result.
Below, we summarize key observations for disks, microspheres and microcylinders around circular microposts.  Thereafter, we discuss more complex curvature fields.  
\subsection{Observations of microdisk migration on curved interfaces around a circular micropost}
 \begin{figure} 
\centering
\includegraphics[width=0.95 \textwidth]{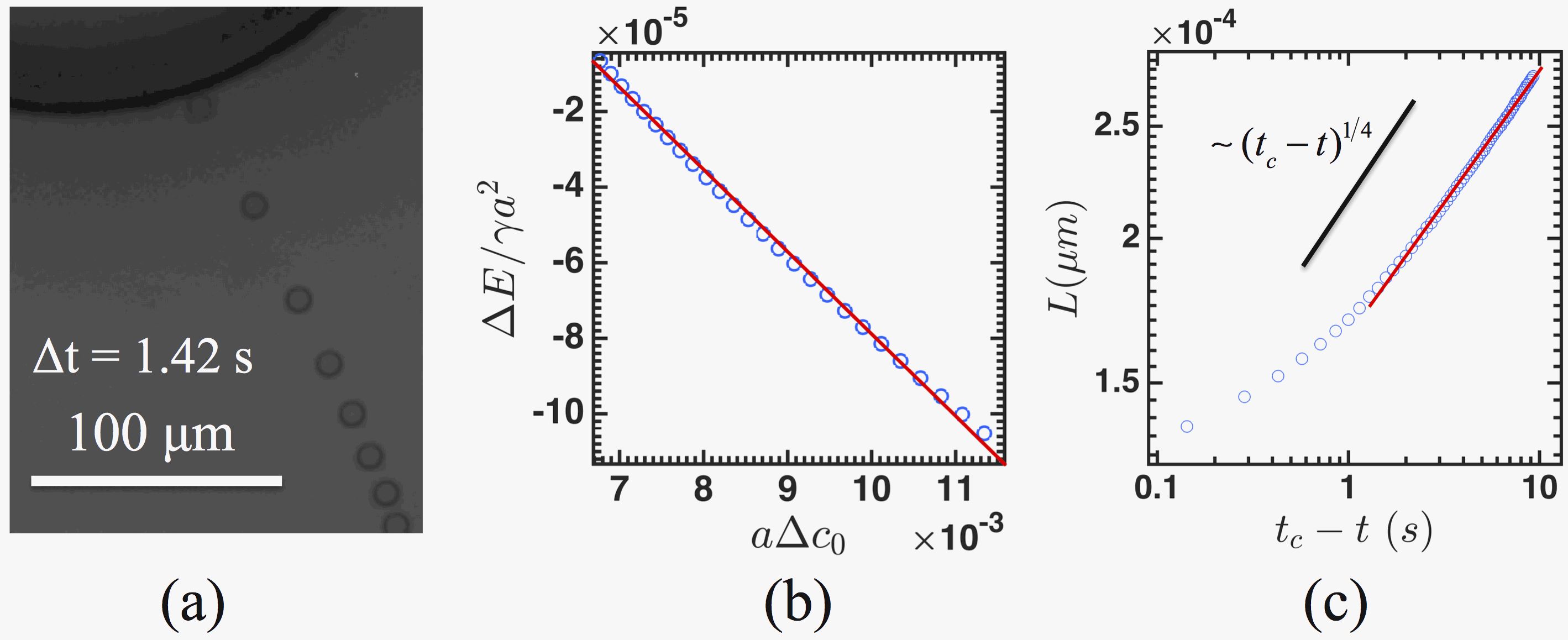}	     	    
 \caption{Capillary migration of microdisks on a curved interface. (a) Time-stamped trajectories of a microdisk around a  circular micropost. (b) Comparison of eq.~(\ref{energy-22}) (solid line) against those extracted from experiment. (c) Power law dependence of radial distance to the center of micropost $L$ versus time remaining to contact.  The trajectories in the region far from the post obey the functional form predicted (the slope of the solid line on experimental data is $0.23$). } 
\label{fig_24}
 \end{figure} 
Disks of radius $a=5~\mu$m and roughness $\sim 20-30~n$m in height (characterized by AFM) were fabricated using SU-8 epoxy resin and standard lithographic methods.  The deformation fields owing to the disks are so weak that they do not interact on planar interfaces unless they are within $\sim10$ radii of contact.  Disks are gently introduced to the upper (oil) phase, and allowed to sediment under gravity until they attach to the interface. Once attached, disks migrate along radial trajectories toward the circular micropost, deviating only if they are within $10-15$ radii of particles already attached to the post. Time-lapsed images reporting particle location at equal time increments clearly convey that the particles migrate more rapidly the closer they are to the micropost [fig.~\ref{fig_24}(a)]. \\
The capillary energy along these trajectories is inferred by determining the energy lost to viscous dissipation  $\Delta E = E({L_f}) - E(L)$. For the disks, we adopt Lamb's drag coefficient and find dissipation energies over paths in excess of $\Delta E > {10^5}{k_B}T$ that are indeed linear in $\Delta c_0$, as shown in fig.~\ref{fig_24}(b).  The slopes can be related to the amplitude of the quadrupolar distortion excited by the particle , which are inferred to fall between $25-30~n$m, of similar scale to the particle roughness.
We further test this prediction by testing for the power law dependency predicted in eq.~(\ref{power}).  The results, presented in fig.~\ref{fig_24}(c), indeed indicate that the predicted form for the curvature capillary energy is valid.\\
\subsection{Observations for Microsphere Migration on Curved Interfaces Around a Circular Micropost}
 \begin{figure} 
\centering
\includegraphics[width=0.95 \textwidth]{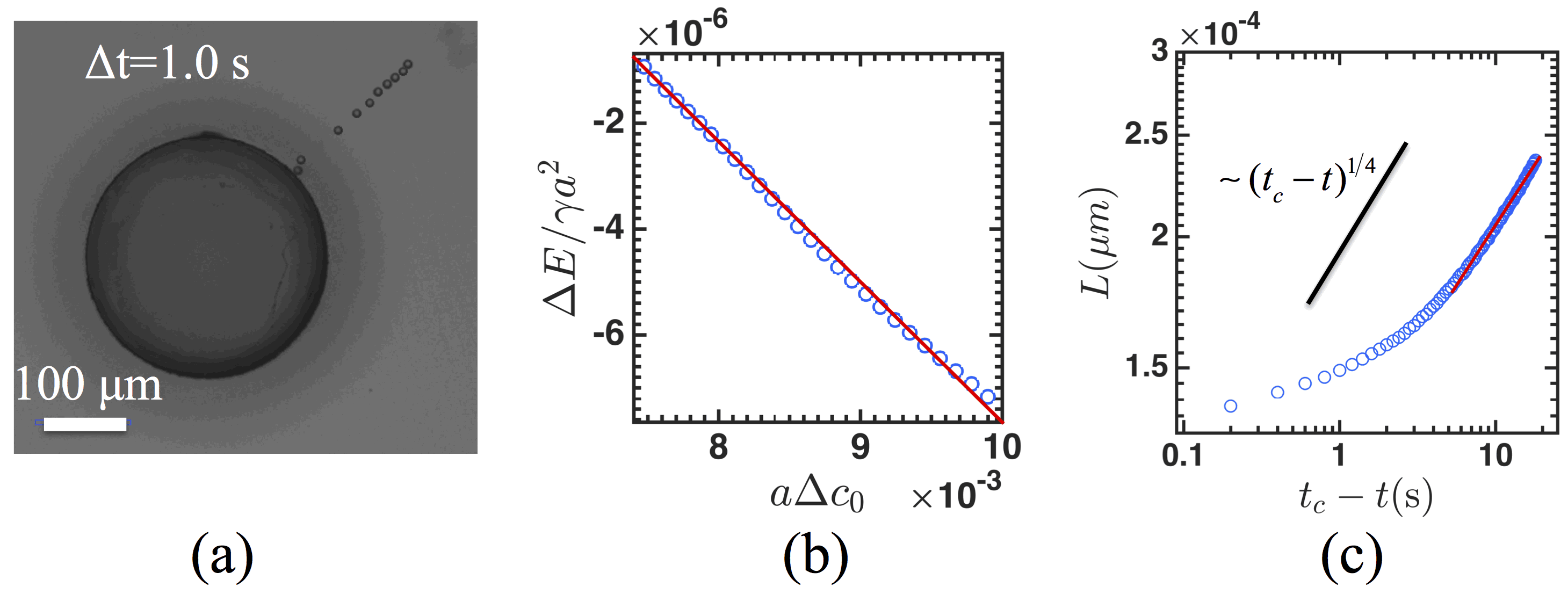}	     	    
 \caption{Capillary migration of a microsphere on a curved interface.  (a), (b) and (c): similar to Fig.~\ref{fig_24}, except that the particle is a $10~\mu m$ polystyrene sphere. Note that, in (a), the particle moves more rapidly in the regions of steep curvature.  The slope of the data in the power law region is $0.23$.} 
\label{fig_25}
 \end{figure} 
Polystyrene colloidal spheres (Polysciences, Inc.) with radius $a=5~\mu m$ were also studied on interfaces molded around microposts of circular cross section. The microspheres were introduced to the interface in a manner similar to the disks; they were suspended in hexadecane and gently introduced to the upper oil phase.  The particles settle under gravity, attach to the interface, and migrate.  They migrate uphill along radial paths until they contact the micropost, with deviations occurring only if the particles are close to neighboring particles or interacting with particles already attached to the micropost.  In a manner strikingly similar to the microdisks, the microspheres migrate along trajectories radial to the micropost [fig.~\ref{fig_25}(a)] with faster migrations in sites of high curvature.  They have capillary energies that are linear in the deviatoric curvature change along their paths [fig.~\ref{fig_25}(b)], with  $6 \times {10^3}{k_B}T < \Delta E < 5 \times {10^5}{k_B}T$.  Finally, they obey the power law implied by the equality of the curvature capillary force to the drag force [fig.~\ref{fig_25}(c)].
\subsection{Observations of microcylinders on curved interfaces around a circular micropost}
  \begin{figure} 
\centering
\includegraphics[width=0.95 \textwidth]{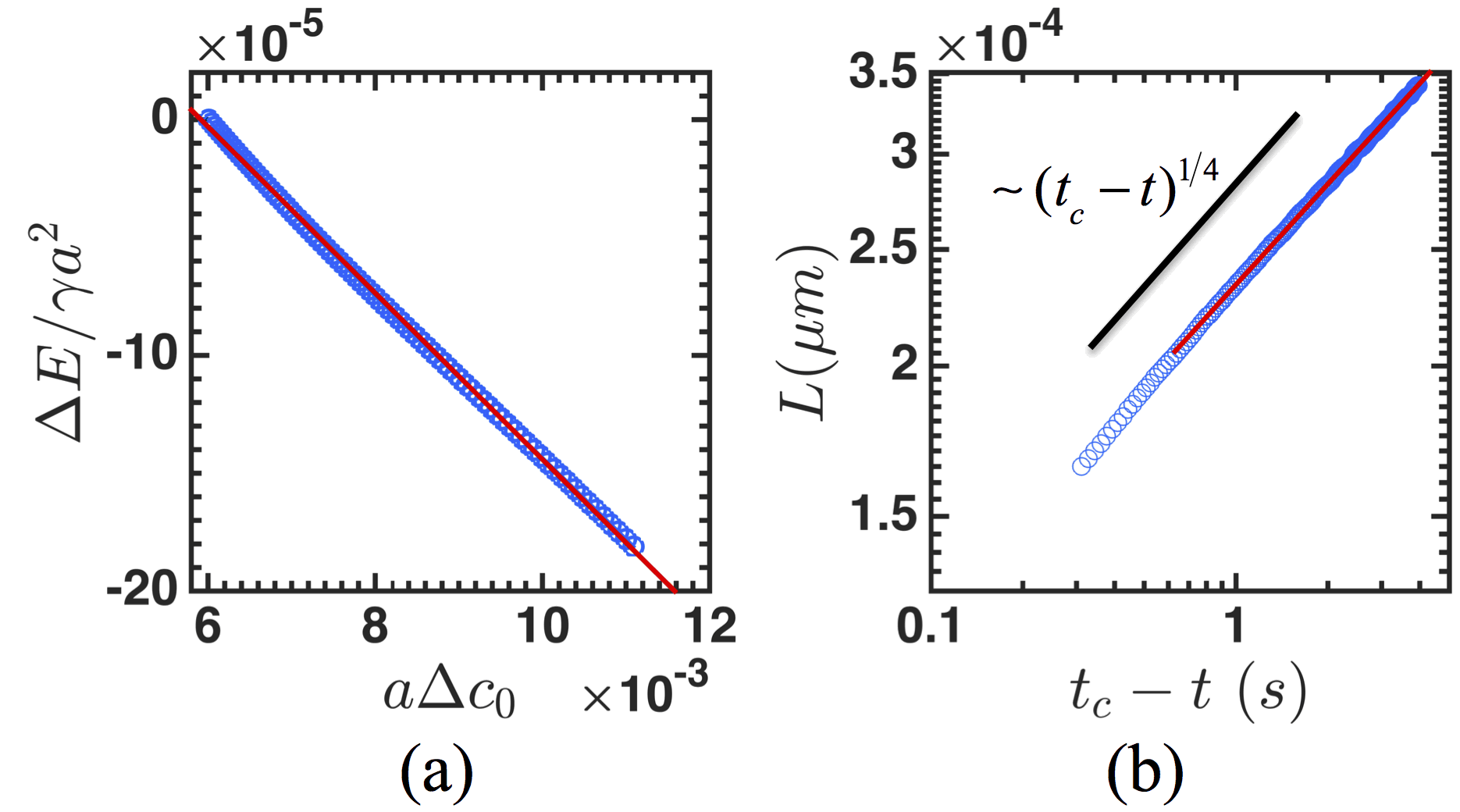}	     	    
 \caption{Capillary migration of a microcylinder on a curved interface.  (a), (b) and (c): similar to Fig.~\ref{fig_24}, except that the particle is an SU-8 cylinder. The slope of the data in the power law region is 0.24. Note that in (a), not only does the particle move more rapidly in regions of steep curvature, but the microcylinder maintains an alignment with its quadrupolar rise axis orientated along the rise axis of the host interface.} 
\label{fig_26}
 \end{figure} 
We study both alignment and migration of cylindrical microparticles on this interface. The microcylinders are fabricated by lithography from the epoxy resin SU-8 with radius $a = 5 \pm 1~\mu$m and length $L_p$, which varied weakly, so that aspect ratio $\Lambda  = {L_p}/2a$ fell within the range $2.0 \le \Lambda  \le 3.0$. The contact angle of water-SU-8-hexadecane is ${\theta _c} = {130^{\circ}}$ as determined by a sessile drop.\\
The cylindrical microparticles have key differences from the simple prior examples of spheres and disks. Owing to the highly anisotropic shape, the contact line is highly undulated and the microparticles create far greater distortions, on average, than do disks or spheres, which rely on random roughness or pinning events to distort the interface.  Furthermore, the anisotropic shape of the particle allows us to relate the orientation of the quadrupolar mode of the particle-sourced distortion to the particle geometry. This identification allows us to relate the particle orientation to the minimum energy state and to comment on the torque driving orientation along principal axes as well as on the force driving particle migration. The microcylinders are introduced into the oil superphase, allowing them to sediment through oil and to attach to the interface.  Immediately upon attaching to the interface, the cylinder rotates rapidly to orient along the principal axis. Thereafter, it moves along radial lines toward the micropost, as shown in fig.~\ref{fig_23}(b). At all locations on the interface, the torque driving rotation and alignment is proportional to the curvature capillary energy itself $-{\rm{d}}E/ {\rm{d}}L$.  Moreover, the curvature capillary force is repulsive until the particle sourced quadrupole is properly aligned.  Hence, particles remain fixed until they are properly aligned and then maintain alignment along the principal axes throughout the migration.  The migration again obeys the tests of the prediction: The energy is linear in the deviatoric curvature [see fig.~\ref{fig_26}(a)] and obeys the predicted power law form [as shown in fig.~\ref{fig_26}(b)] owing to the very large distortions sourced into the interface.  The energy dissipated along a trajectory is typically in excess of $10^7~k_BT$; this greater magnitude is in keeping with the larger magnitude of the quadrupolar distortion in the interface. 
 \subsection{Cylinders on interfaces with more complex curvature fields}
 Using microposts of differing cross section, we can impose deviatoric curvature fields of greater complexity.   For example, by pinning the oil-water interface on the edge of a micropost with elliptical cross section, we impose curvature fields that are strongest at the poles of the ellipse.  Once again, the curvature field can be described analytically, as the interface is well approximated by a monopole in elliptical coordinates.  Hence, we have closed-form solutions for the curvature field that can be compared in detail to the particle alignment and migration.  We do not go into the details here, but refer the reader to the original work \cite{Marcello}. 
If regions of local, steep curvature gradients are sites of attraction, corners should present Ôhot spotsÕ for migration and assembly.  We use square microposts, for which the interface curvature changes steeply at the corners. In fact, for ideal corners with infinitely sharp edges, the curvature diverges, similar to an electric field gradient near a pointed conductor.  In experiment, the finite radius of curvature of the corner $r_c$ limits this divergence. Microcylinders are strongly attracted to the corners.  Intriguingly and importantly, when particles accumulate in the region of the square micropost, they form complex structures guided by particle-particle and particle-curvature field interactions. We argue that the microcylinders, microdisks and microspheres migrate owing to similar physics -- they each create a distortion owing to their contact line configurations that has quadrupolar symmetry, and that this symmetry drives them to sites of elevated deviatoric curvature.  To further test this concept, we have recently studied microsphere migration in this setting \cite{Sphere}. We report in fig.~\ref{fig_23}(c) that isolated microspheres and dimerized particles indeed migrate to corners.
\subsection{Interface shape}
 \begin{figure} 
\centering
\includegraphics[width=0.75 \textwidth]{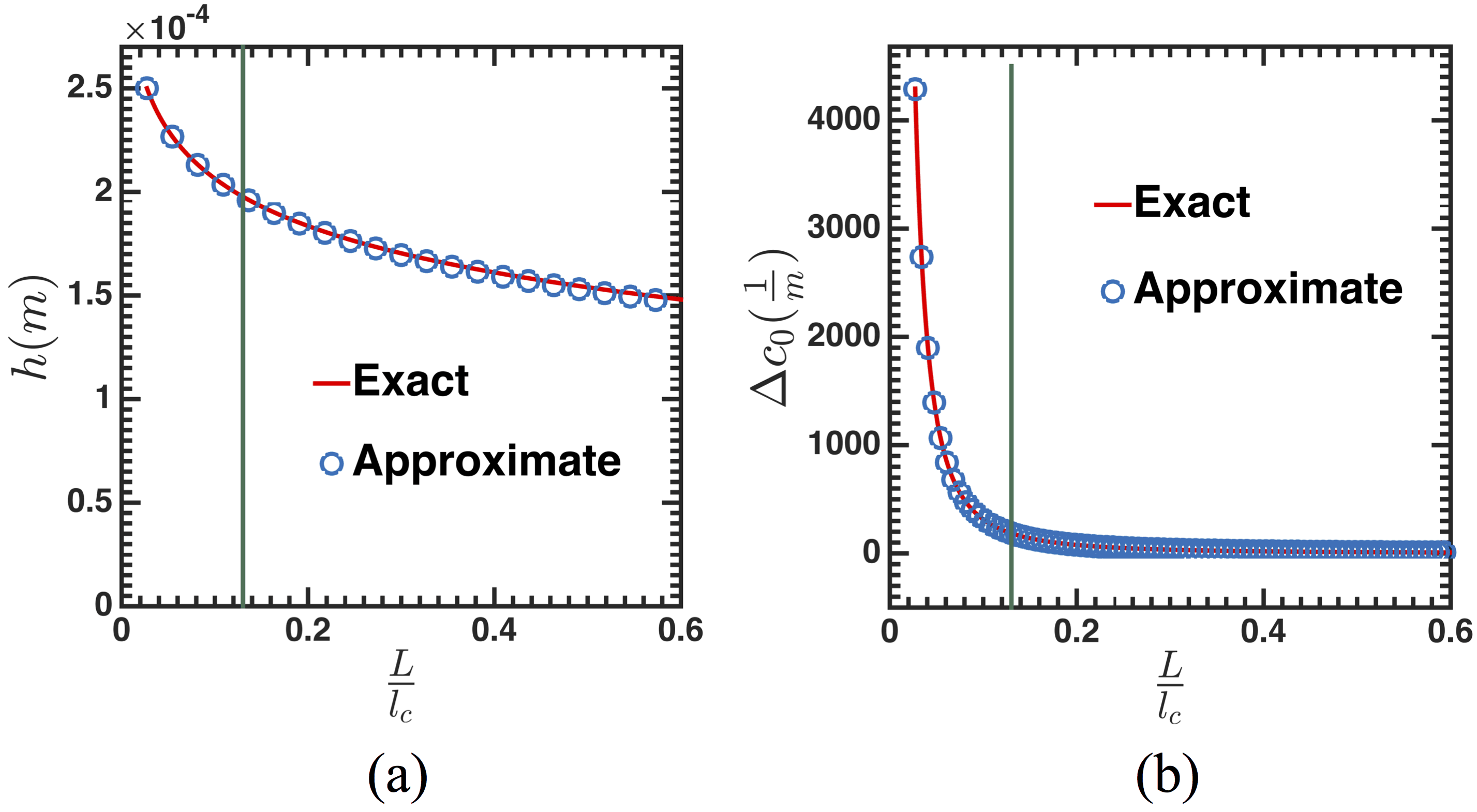}	     	    
 \caption{Comparison of the exact [eq.~(\ref{exact_h})] and approximate [eq.~(\ref{app_h})] solutions for (a) height and (b) deviatoric curvature. The vertical solid line defines the border of the region where we have seen particle migration.} 
\label{fig_27}
 \end{figure}
\begin{figure} 
\centering
\includegraphics[width=0.45 \textwidth]{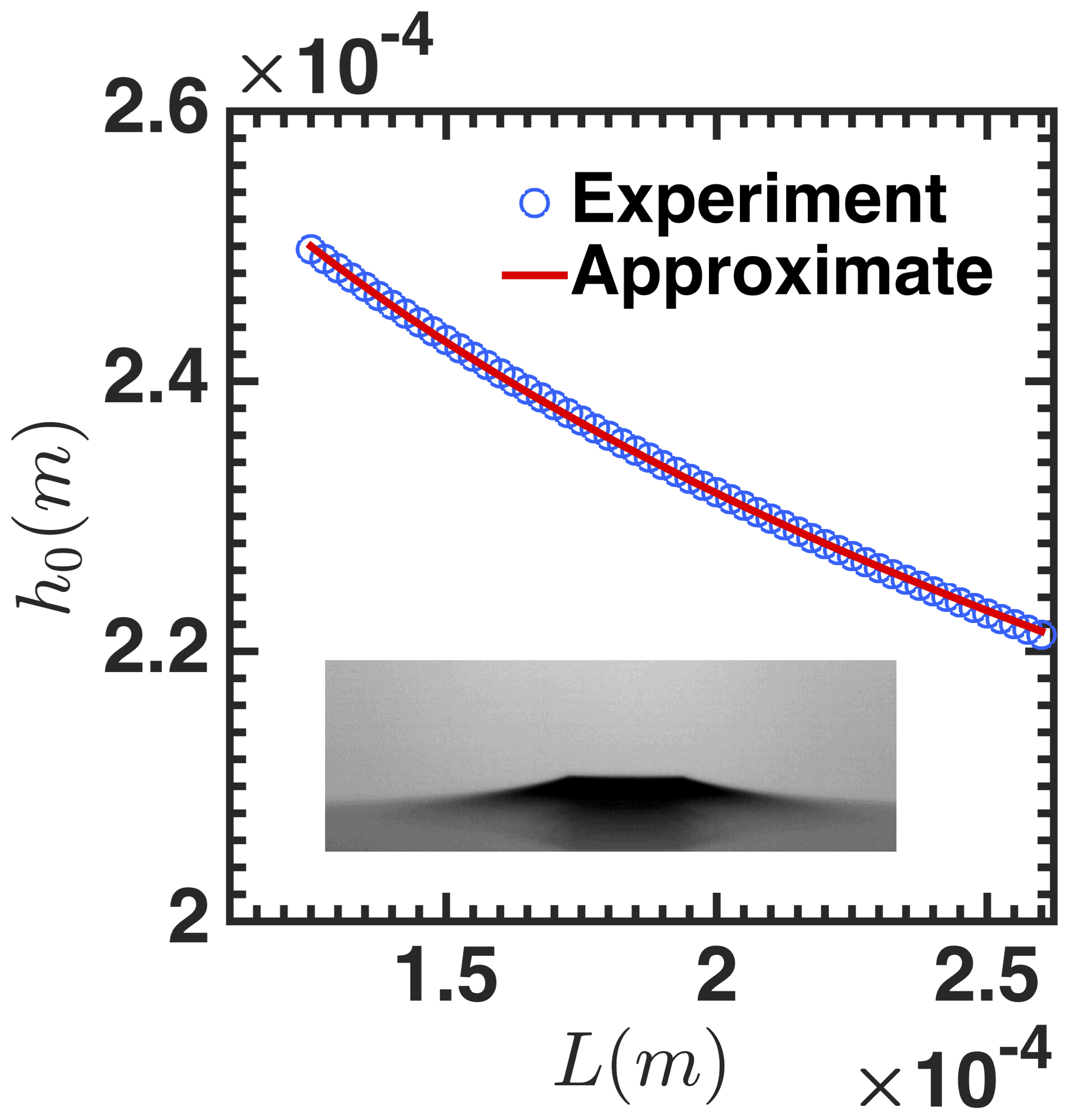}	     	    
 \caption{Comparison of the approximate solution [eq.~(\ref{app_h})] with the height profile observed in the experiment. The inset is the side view of the curved interface close to the micropost.} 
\label{fig_28}
 \end{figure}

In this section, we review what we know about the interface shape around the micropost and the shape of the interface locally around a particle. An expression for the shape of the interface around a circular micropost can be found by solving the linearized Young-Laplace equation:
\begin{eqnarray}
\gamma {\nabla ^2}{h_0} = \Delta p + \Delta \rho g{h_0},
\end{eqnarray}
with the solution in the following form
\begin{eqnarray}
{h_{host}} = A{I_0}\left( {\frac{L}{{{l_c}}}} \right) + B{K_0}\left( {\frac{L}{{{l_c}}}} \right) - \frac{{\Delta p}}{{\Delta \rho g}},\label{exact_h}
\end{eqnarray}
where $I_0$ and $K_0$ are modified Bessel function of the first and second kind, $A$ and $B$ are constants evaluated with the pinning boundary conditions on the post and the outer ring, $L$ is the radial distance from the center of the post, and ${l_c} = \sqrt {{\gamma  \mathord{\left/
 {\vphantom {\gamma  {\Delta \rho g}}} \right.
 \kern-\nulldelimiterspace} {\Delta \rho g}}}$ is the capillary length.\\
\indent In the region close to the post where ${L}/{l_c} \ll 1$, the above expression can be expanded further to yield the approximate logarithmic relation according to
\begin{eqnarray}
{h_{host}} \sim {H_m} - {R_m}\tan \psi \ln \left({\frac{L}{{{R_m}}}}\right).\label{app_h}
\end{eqnarray}
Figure~\ref{fig_27} illustrates the comparison of the approximate expression with the exact solution. Clearly, in the region close to the post where we have observed the particle migration, the approximate solution can accurately capture the height and the curvature field.
Figure~\ref{fig_28} demonstrates the comparison between the height profile obtained from experiment and the approximate expression in the region close to the post. 
 \subsection{Shape of the interface in the presence of a particle}\label{green}
   \begin{figure} 
\centering
\includegraphics[width=0.75 \textwidth]{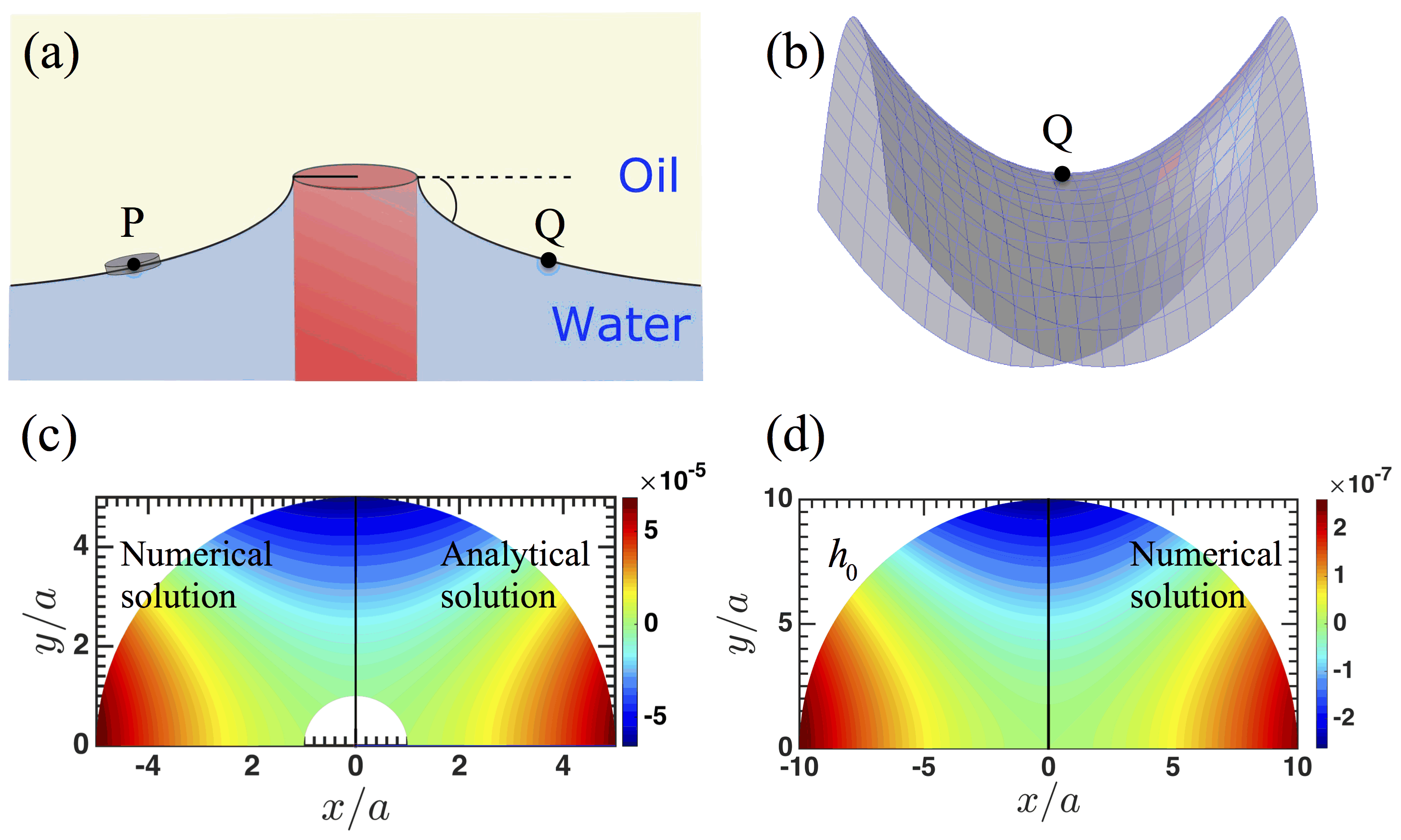}	     	    
 \caption{(a) Schematic of the host interface close to a micropost with circular cross section. (b) Side view of the host interface (saddle shape) at point $Q$ with respect to the tangent plane. (c) Comparison of height obtained from numerical (left plane) with the analytical solution (right plane) given in eq.~(\ref{uvs}) for a disk located at $L_0 =300 \mu$m with $h_{qp} \sim 4~n$m. (d) Comparison of the quadrupolar description found from numerical solution with the saddle shape $h_0=\frac{\Delta c_0}{4}r^2\cos2\phi$ at an arbitrary point $Q$ far from the particle.} 
\label{fig_29}
 \end{figure} 
 The expansion of the host interface around the center of mass of the particle in terms of local deviatoric curvature is key to our analysis to find the disturbance made by the particle in the interface. To confirm both the functional form for the host interface and the disturbance made by the particle, we find the shape of the interface for a particle placed on an interface using boundary integral formulations. This provides an independent verification of the asymptotic solution. Having assumed small slopes throughout the region, the height of the interface at any point on a plane can be obtained according to
\begin{align}
h({\bf{x}}) &= \int {\int {G({\bf{x}},{\bf{x}}')\rho ({\bf{x}}')L'{\rm{d}}L'{\rm{d}}\theta '} }  - {R_m}\int {{{\left. {h({\bf{x}}')\frac{{\partial G({\bf{x}},{\bf{x}}')}}{{\partial L'}}} \right|}_{r' = {R_m}}}{\rm{d}}\theta '} \\
 &+ {R_{ring}}\int {{{\left. {h({\bf{x}}')\frac{{\partial G({\bf{x}},{\bf{x}}')}}{{\partial L'}}} \right|}_{r' = {R_{ring}}}}{\rm{d}}\theta '},\label{bi-eq}\nonumber
\end{align}
where $G({\bf{x}},{\bf{x}}')$ is the Green's function of the Laplacian operator which satisfies
\begin{eqnarray}
{\nabla ^2}G({\bf{x}},{\bf{x}}') =  - \delta^2({\bf{x}} - {\bf{x}}'),
\end{eqnarray}
where $\delta^2({\bf{x}} - {\bf{x}}')$ is a two dimensional Dirac delta function.  The Green's function for this geometry is developed utilizing the fundamental solution to satisfy the Dirichlet boundary condition at the microcylinder and the outer confining ring. It is:
\begin{align}
&G\left( {L,L',\theta ,\theta '} \right) = \frac{1}{{2\pi }}\ln \left( {\frac{{{R_{ring}}}}{{{L_ > }}}} \right) + \frac{{\ln \left( {\frac{{{R_{ring}}}}{L}} \right)}}{{2\pi \ln \left( {\frac{{{R_{ring}}}}{{{R_m}}}} \right)}}\ln \left( {\frac{L}{{{R_{ring}}}}} \right)\\
&+ \sum\limits_{n = 1}^\infty  {\left[ {{{\left( {\frac{{{L_ < }}}{{{L_ > }}}} \right)}^n} - {{\left[ {{{\left( {\frac{{{R_{ring}}}}{{{R_m}}}} \right)}^n} - {{\left( {\frac{{{R_m}}}{{{R_{ring}}}}} \right)}^n}} \right]}^{ - 1}}\left\{ {\left[ {{{\left( {\frac{L}{{{R_m}{R_{ring}}}}} \right)}^n} - {{\left( {\frac{{{R_m}}}{{{R_{ring}}L}}} \right)}^n}} \right]{L^n}} \right.} \right.} \nonumber\\
&\left. {\left. { + \left[ {{{\left( {\frac{{{R_m}{R_{ring}}}}{L}} \right)}^n} - {{\left( {\frac{{{R_m}L}}{{{R_{ring}}}}} \right)}^n}} \right]{L^{ - n}}} \right\}} \right]\frac{{\cos n\left( {\theta  - \theta '} \right)}}{{2\pi n}},\nonumber
\end{align}
where $(L,\theta)$ is the polar coordinate located at the center of circular micropost and $L_{>}$ denotes the greatest value between $L$ and $L'$. To satisfy the boundary condition, we place $N$ singularities with unknown magnitudes on the contact line and thereafter we impose the boundary condition via eq.~(\ref{bi-eq}) to obtain the magnitudes of the singularities. Having determined the magnitude of the singularities, the height of interface at any point in the plane can be evaluated. Figure~\ref{fig_29} illustrates the comparison of the numerical solution and the asymptotic solution given previously.
\section{Conclusions}
In this work, we have addressed how inclusions made by particles at fluid interfaces can be used to direct particle assembly. In some sense, this is classical.  In the limit small slopes, the distortions are well-described by solutions to the Laplace equation, and analogies to electrostatic interactions are invoked to describe particle interactions.  However, the devil is in the details, and these analogies must be handled with care.  For example, the work to charge a particle with electrostatic charge differs significantly from the physicochemical processes associated with wetting that define contact lines and Òcapillary chargesÓ.   Furthermore, colloidal particle length scales are typically smaller than the radii of curvature of the interfaces, the length scale that defines the vessel size and other physicochemical length scales like the capillary length. The particles create distortions of about the size of their radii, and those distortions decay over similar distances.  Thus, the boundary value problems associated with understanding the distortions and their associated energies must be handled with care. Throughout the discussion, we allude to pinned contact lines. Contact lines can also be considered in an equilibrium state defined by solutions to Young's equation.  There is a growing body of work in the field of contact line pinning, even on simple spherical colloids, which is thought to be associated with asperities, discrete charges, or patchy wetting. There are interesting open issues in this field: What energies must be supplied to jump from one trapped state to another? Do they couple into the other phenomena we study here, like the curvature field that we employ to drive migration?\\ 

We and others are deeply interested in harnessing these effects to create hierarchical assemblies of particles, e.g. using curvature fields to guide structure formation. We are well on the way to this goal for colloids of characteristic size of tens of micrometers, which interact with nearly deterministic trajectories to form kinetically trapped structures essentially cemented by capillary interactions of order $10^7k_BT$. A more intriguing goal is to study these interactions in a weaker limit, where interactions are closer to thermal excitations, e.g. by decreasing particle size, weakening curvature gradients, and studying interfaces of lower tension. In such cases, reconfigurable assemblies could be formed.   These are just a few of the open questions and areas of ongoing research.  We opened the Chapter with allusions to other systems in which inclusions made by particles can be harnessed to guide assembly, e.g. particles on lipid bilayers and particles immersed in liquid crystals, which interact in the bulk via elasticity and via topological constraints. Both of these systems have intriguing similarities and differences from the simple isotropic fluid interfaces, and provide a rich landscape for experiment, theory,  and simulation as we build this toolkit to direct structure formation. \\ 
\newpage
\acknowledgments
This work is partially supported by NSF grants CBET-$1133267$ and CBET-$1066284$, GAANN P200A120246, and MRSEC grant DMR11- $20901$.  We thank Nate Bade and Ningwei Li for discussions.  We gratefully acknowledge permission granted by Soft Matter to present material originally published in that journal.


\begin{thebibliography}{0}%
\makeatletter
\providecommand \@ifxundefined [1]{%
 \@ifx{#1\undefined}
}%
\providecommand \@ifnum [1]{%
 \ifnum #1\expandafter \@firstoftwo
 \else \expandafter \@secondoftwo
 \fi
}%
\providecommand \@ifx [1]{%
 \ifx #1\expandafter \@firstoftwo
 \else \expandafter \@secondoftwo
 \fi
}%
\providecommand \natexlab [1]{#1}%
\providecommand \enquote  [1]{``#1''}%
\providecommand \bibnamefont  [1]{#1}%
\providecommand \bibfnamefont [1]{#1}%
\providecommand \citenamefont [1]{#1}%
\providecommand \href@noop [0]{\@secondoftwo}%
\providecommand \href [0]{\begingroup \@sanitize@url \@href}%
\providecommand \@href[1]{\@@startlink{#1}\@@href}%
\providecommand \@@href[1]{\endgroup#1\@@endlink}%
\providecommand \@sanitize@url [0]{\catcode `\\12\catcode `\$12\catcode
  `\&12\catcode `\#12\catcode `\^12\catcode `\_12\catcode `\%12\relax}%
\providecommand \@@startlink[1]{}%
\providecommand \@@endlink[0]{}%
\providecommand \url  [0]{\begingroup\@sanitize@url \@url }%
\providecommand \@url [1]{\endgroup\@href {#1}{\urlprefix }}%
\providecommand \urlprefix  [0]{URL }%
\providecommand \Eprint [0]{\href }%
\providecommand \doibase [0]{http://dx.doi.org/}%
\providecommand \selectlanguage [0]{\@gobble}%
\providecommand \bibinfo  [0]{\@secondoftwo}%
\providecommand \bibfield  [0]{\@secondoftwo}%
\providecommand \translation [1]{[#1]}%
\providecommand \BibitemOpen [0]{}%
\providecommand \bibitemStop [0]{}%
\providecommand \bibitemNoStop [0]{.\EOS\space}%
\providecommand \EOS [0]{\spacefactor3000\relax}%
\providecommand \BibitemShut  [1]{\csname bibitem#1\endcsname}%
\let\auto@bib@innerbib\@empty
\end{thebibliography}%


\begin{thebibliography}{29}%
\makeatletter
\providecommand \@ifxundefined [1]{%
 \@ifx{#1\undefined}
}%
\providecommand \@ifnum [1]{%
 \ifnum #1\expandafter \@firstoftwo
 \else \expandafter \@secondoftwo
 \fi
}%
\providecommand \@ifx [1]{%
 \ifx #1\expandafter \@firstoftwo
 \else \expandafter \@secondoftwo
 \fi
}%
\providecommand \natexlab [1]{#1}%
\providecommand \enquote  [1]{``#1''}%
\providecommand \bibnamefont  [1]{#1}%
\providecommand \bibfnamefont [1]{#1}%
\providecommand \citenamefont [1]{#1}%
\providecommand \href@noop [0]{\@secondoftwo}%
\providecommand \href [0]{\begingroup \@sanitize@url \@href}%
\providecommand \@href[1]{\@@startlink{#1}\@@href}%
\providecommand \@@href[1]{\endgroup#1\@@endlink}%
\providecommand \@sanitize@url [0]{\catcode `\\12\catcode `\$12\catcode
  `\&12\catcode `\#12\catcode `\^12\catcode `\_12\catcode `\%12\relax}%
\providecommand \@@startlink[1]{}%
\providecommand \@@endlink[0]{}%
\providecommand \url  [0]{\begingroup\@sanitize@url \@url }%
\providecommand \@url [1]{\endgroup\@href {#1}{\urlprefix }}%
\providecommand \urlprefix  [0]{URL }%
\providecommand \Eprint [0]{\href }%
\providecommand \doibase [0]{http://dx.doi.org/}%
\providecommand \selectlanguage [0]{\@gobble}%
\providecommand \bibinfo  [0]{\@secondoftwo}%
\providecommand \bibfield  [0]{\@secondoftwo}%
\providecommand \translation [1]{[#1]}%
\providecommand \BibitemOpen [0]{}%
\providecommand \bibitemStop [0]{}%
\providecommand \bibitemNoStop [0]{.\EOS\space}%
\providecommand \EOS [0]{\spacefactor3000\relax}%
\providecommand \BibitemShut  [1]{\csname bibitem#1\endcsname}%
\let\auto@bib@innerbib\@empty
\bibitem [{\citenamefont {Whitesides}\ and\ \citenamefont
  {Grzybowski}(2002)}]{Whitesides}%
  \BibitemOpen
  \bibfield  {author} {\bibinfo {author} {\bibfnamefont {G.~M.}\ \bibnamefont
  {Whitesides}}\ and\ \bibinfo {author} {\bibfnamefont {B.}~\bibnamefont
  {Grzybowski}},\ }\href {\doibase 10.1126/science.1070821} {\bibfield
  {journal} {\bibinfo  {journal} {Science}\ }\textbf {\bibinfo {volume}
  {295}},\ \bibinfo {pages} {2418} (\bibinfo {year} {2002})}\BibitemShut
  {NoStop}%
\bibitem [{\citenamefont {Pieranski}(1980)}]{Pieranski}%
  \BibitemOpen
  \bibfield  {author} {\bibinfo {author} {\bibfnamefont {P.}~\bibnamefont
  {Pieranski}},\ }\href {\doibase 10.1103/PhysRevLett.45.569} {\bibfield
  {journal} {\bibinfo  {journal} {Phys. Rev. Lett.}\ }\textbf {\bibinfo
  {volume} {45}},\ \bibinfo {pages} {569} (\bibinfo {year} {1980})}\BibitemShut
  {NoStop}%
\bibitem [{\citenamefont {Kralchevsky}\ and\ \citenamefont
  {Nagayama}(2000)}]{Kralchevsky}%
  \BibitemOpen
  \bibfield  {author} {\bibinfo {author} {\bibfnamefont {P.~A.}\ \bibnamefont
  {Kralchevsky}}\ and\ \bibinfo {author} {\bibfnamefont {K.}~\bibnamefont
  {Nagayama}},\ }\href {\doibase
  http://dx.doi.org/10.1016/S0001-8686(99)00016-0} {\bibfield  {journal}
  {\bibinfo  {journal} {Adv. Colloid Interface Sci.}\ }\textbf {\bibinfo
  {volume} {85}},\ \bibinfo {pages} {145} (\bibinfo {year} {2000})}\BibitemShut
  {NoStop}%
\bibitem [{\citenamefont {Stamou}\ \emph {et~al.}(2000)\citenamefont {Stamou},
  \citenamefont {Duschl},\ and\ \citenamefont {Johannsmann}}]{Stamou}%
  \BibitemOpen
  \bibfield  {author} {\bibinfo {author} {\bibfnamefont {D.}~\bibnamefont
  {Stamou}}, \bibinfo {author} {\bibfnamefont {C.}~\bibnamefont {Duschl}}, \
  and\ \bibinfo {author} {\bibfnamefont {D.}~\bibnamefont {Johannsmann}},\
  }\href {\doibase 10.1103/PhysRevE.62.5263} {\bibfield  {journal} {\bibinfo
  {journal} {Phys. Rev. E}\ }\textbf {\bibinfo {volume} {62}},\ \bibinfo
  {pages} {5263} (\bibinfo {year} {2000})}\BibitemShut {NoStop}%
\bibitem [{\citenamefont {Loudet}\ \emph {et~al.}(2005)\citenamefont {Loudet},
  \citenamefont {Alsayed}, \citenamefont {Zhang},\ and\ \citenamefont
  {Yodh}}]{Arjun}%
  \BibitemOpen
  \bibfield  {author} {\bibinfo {author} {\bibfnamefont {J.~C.}\ \bibnamefont
  {Loudet}}, \bibinfo {author} {\bibfnamefont {A.~M.}\ \bibnamefont {Alsayed}},
  \bibinfo {author} {\bibfnamefont {J.}~\bibnamefont {Zhang}}, \ and\ \bibinfo
  {author} {\bibfnamefont {A.~G.}\ \bibnamefont {Yodh}},\ }\href {\doibase
  10.1103/PhysRevLett.94.018301} {\bibfield  {journal} {\bibinfo  {journal}
  {Phys. Rev. Lett.}\ }\textbf {\bibinfo {volume} {94}},\ \bibinfo {pages}
  {018301} (\bibinfo {year} {2005})}\BibitemShut {NoStop}%
\bibitem [{\citenamefont {Botto}\ \emph
  {et~al.}(2012{\natexlab{a}})\citenamefont {Botto}, \citenamefont
  {Lewandowski},\ and\ \citenamefont {Cavallaro}}]{Botto}%
  \BibitemOpen
  \bibfield  {author} {\bibinfo {author} {\bibfnamefont {L.}~\bibnamefont
  {Botto}}, \bibinfo {author} {\bibfnamefont {E.~P.}\ \bibnamefont
  {Lewandowski}}, \ and\ \bibinfo {author} {\bibfnamefont {K.~J.}\ \bibnamefont
  {Cavallaro}, \bibfnamefont {M.and~Stebe}},\ }\href {\doibase
  10.1039/C2SM25929J} {\bibfield  {journal} {\bibinfo  {journal} {Soft Matter}\
  }\textbf {\bibinfo {volume} {8}},\ \bibinfo {pages} {9957} (\bibinfo {year}
  {2012}{\natexlab{a}})}\BibitemShut {NoStop}%
\bibitem [{\citenamefont {Botto}\ \emph
  {et~al.}(2012{\natexlab{b}})\citenamefont {Botto}, \citenamefont {Yao},
  \citenamefont {Leheny},\ and\ \citenamefont {Stebe}}]{Lorenzo}%
  \BibitemOpen
  \bibfield  {author} {\bibinfo {author} {\bibfnamefont {L.}~\bibnamefont
  {Botto}}, \bibinfo {author} {\bibfnamefont {L.}~\bibnamefont {Yao}}, \bibinfo
  {author} {\bibfnamefont {R.~L.}\ \bibnamefont {Leheny}}, \ and\ \bibinfo
  {author} {\bibfnamefont {K.~J.}\ \bibnamefont {Stebe}},\ }\href {\doibase
  10.1039/C2SM25211B} {\bibfield  {journal} {\bibinfo  {journal} {Soft Matter}\
  }\textbf {\bibinfo {volume} {8}},\ \bibinfo {pages} {4971} (\bibinfo {year}
  {2012}{\natexlab{b}})}\BibitemShut {NoStop}%
\bibitem [{\citenamefont {Cavallaro}\ \emph {et~al.}(2011)\citenamefont
  {Cavallaro}, \citenamefont {Botto}, \citenamefont {Lewandowski},
  \citenamefont {Wang},\ and\ \citenamefont {Stebe}}]{Marcello}%
  \BibitemOpen
  \bibfield  {author} {\bibinfo {author} {\bibfnamefont {M.}~\bibnamefont
  {Cavallaro}}, \bibinfo {author} {\bibfnamefont {L.}~\bibnamefont {Botto}},
  \bibinfo {author} {\bibfnamefont {E.~P.}\ \bibnamefont {Lewandowski}},
  \bibinfo {author} {\bibfnamefont {M.}~\bibnamefont {Wang}}, \ and\ \bibinfo
  {author} {\bibfnamefont {K.~J.}\ \bibnamefont {Stebe}},\ }\href {\doibase
  10.1073/pnas.1116344108} {\bibfield  {journal} {\bibinfo  {journal} {Proc.
  Natl. Acad. Sci.}\ }\textbf {\bibinfo {volume} {108}},\ \bibinfo {pages}
  {20923} (\bibinfo {year} {2011})}\BibitemShut {NoStop}%
\bibitem [{\citenamefont {Cavallaro}\ \emph {et~al.}(2013)\citenamefont
  {Cavallaro}, \citenamefont {Gharbi}, \citenamefont {Beller}, \citenamefont
  {{\v C}opar}, \citenamefont {Shi}, \citenamefont {Baumgart}, \citenamefont
  {Yang}, \citenamefont {Kamien},\ and\ \citenamefont {Stebe}}]{Cavallaro-LC}%
  \BibitemOpen
  \bibfield  {author} {\bibinfo {author} {\bibfnamefont {M.}~\bibnamefont
  {Cavallaro}}, \bibinfo {author} {\bibfnamefont {M.~A.}\ \bibnamefont
  {Gharbi}}, \bibinfo {author} {\bibfnamefont {D.~A.}\ \bibnamefont {Beller}},
  \bibinfo {author} {\bibfnamefont {S.}~\bibnamefont {{\v C}opar}}, \bibinfo
  {author} {\bibfnamefont {Z.}~\bibnamefont {Shi}}, \bibinfo {author}
  {\bibfnamefont {T.}~\bibnamefont {Baumgart}}, \bibinfo {author}
  {\bibfnamefont {S.}~\bibnamefont {Yang}}, \bibinfo {author} {\bibfnamefont
  {R.~D.}\ \bibnamefont {Kamien}}, \ and\ \bibinfo {author} {\bibfnamefont
  {K.~J.}\ \bibnamefont {Stebe}},\ }\href {\doibase 10.1073/pnas.1313551110}
  {\bibfield  {journal} {\bibinfo  {journal} {Proc. Natl. Acad. Sci.}\ }\textbf
  {\bibinfo {volume} {110}},\ \bibinfo {pages} {18804} (\bibinfo {year}
  {2013})}\BibitemShut {NoStop}%
\bibitem [{\citenamefont {Koltover}\ \emph {et~al.}(1999)\citenamefont
  {Koltover}, \citenamefont {R\"adler},\ and\ \citenamefont
  {Safinya}}]{Safinya}%
  \BibitemOpen
  \bibfield  {author} {\bibinfo {author} {\bibfnamefont {I.}~\bibnamefont
  {Koltover}}, \bibinfo {author} {\bibfnamefont {J.~O.}\ \bibnamefont
  {R\"adler}}, \ and\ \bibinfo {author} {\bibfnamefont {C.~R.}\ \bibnamefont
  {Safinya}},\ }\href {\doibase 10.1103/PhysRevLett.82.1991} {\bibfield
  {journal} {\bibinfo  {journal} {Phys. Rev. Lett.}\ }\textbf {\bibinfo
  {volume} {82}},\ \bibinfo {pages} {1991} (\bibinfo {year}
  {1999})}\BibitemShut {NoStop}%
\bibitem [{\citenamefont {Dan}\ \emph {et~al.}(1993)\citenamefont {Dan},
  \citenamefont {Pincus},\ and\ \citenamefont {Safran}}]{Pincus}%
  \BibitemOpen
  \bibfield  {author} {\bibinfo {author} {\bibfnamefont {N.}~\bibnamefont
  {Dan}}, \bibinfo {author} {\bibfnamefont {P.}~\bibnamefont {Pincus}}, \ and\
  \bibinfo {author} {\bibfnamefont {S.~A.}\ \bibnamefont {Safran}},\ }\href
  {\doibase 10.1021/la00035a005} {\bibfield  {journal} {\bibinfo  {journal}
  {Langmuir}\ }\textbf {\bibinfo {volume} {9}},\ \bibinfo {pages} {2768}
  (\bibinfo {year} {1993})}\BibitemShut {NoStop}%
\bibitem [{\citenamefont {Lipowsky}\ and\ \citenamefont
  {D{\"o}bereiner}(1998)}]{Lipowsky}%
  \BibitemOpen
  \bibfield  {author} {\bibinfo {author} {\bibfnamefont {R.}~\bibnamefont
  {Lipowsky}}\ and\ \bibinfo {author} {\bibfnamefont {H.-G.}\ \bibnamefont
  {D{\"o}bereiner}},\ }\href {http://stacks.iop.org/0295-5075/43/i=2/a=219}
  {\bibfield  {journal} {\bibinfo  {journal} {Eur. Phys. Lett.}\ }\textbf
  {\bibinfo {volume} {43}},\ \bibinfo {pages} {219} (\bibinfo {year}
  {1998})}\BibitemShut {NoStop}%
\bibitem [{\citenamefont {Yu}\ and\ \citenamefont {Granick}(2009)}]{Granick}%
  \BibitemOpen
  \bibfield  {author} {\bibinfo {author} {\bibfnamefont {Y.}~\bibnamefont
  {Yu}}\ and\ \bibinfo {author} {\bibfnamefont {S.}~\bibnamefont {Granick}},\
  }\href {\doibase 10.1021/ja905900h} {\bibfield  {journal} {\bibinfo
  {journal} {J. Am. Chem. Soc.}\ }\textbf {\bibinfo {volume} {131}},\ \bibinfo
  {pages} {14158} (\bibinfo {year} {2009})}\BibitemShut {NoStop}%
\bibitem [{\citenamefont {Nayfeh}(2000)}]{Nayfeh}%
  \BibitemOpen
  \bibfield  {author} {\bibinfo {author} {\bibfnamefont {A.~H.}\ \bibnamefont
  {Nayfeh}},\ }\href@noop {} {\emph {\bibinfo {title} {Perturbation Methods}}}\
  (\bibinfo  {publisher} {Wiley},\ \bibinfo {address} {New Yrok},\ \bibinfo
  {year} {2000})\BibitemShut {NoStop}%
\bibitem [{\citenamefont {Young}(1805)}]{Young}%
  \BibitemOpen
  \bibfield  {author} {\bibinfo {author} {\bibfnamefont {T.}~\bibnamefont
  {Young}},\ }\href {\doibase 10.2307/107159} {\bibfield  {journal} {\bibinfo
  {journal} {Phil. Trans. R. Soc. Lond.}\ }\textbf {\bibinfo {volume} {95}},\
  \bibinfo {pages} {65} (\bibinfo {year} {1805})}\BibitemShut {NoStop}%
\bibitem [{\citenamefont {Kaz}\ \emph {et~al.}(2012)\citenamefont {Kaz},
  \citenamefont {McGorty}, \citenamefont {Mani}, \citenamefont {Brenner},\ and\
  \citenamefont {Manoharan}}]{Manoharan}%
  \BibitemOpen
  \bibfield  {author} {\bibinfo {author} {\bibfnamefont {D.~M.}\ \bibnamefont
  {Kaz}}, \bibinfo {author} {\bibfnamefont {R.}~\bibnamefont {McGorty}},
  \bibinfo {author} {\bibfnamefont {M.}~\bibnamefont {Mani}}, \bibinfo {author}
  {\bibfnamefont {M.~P.}\ \bibnamefont {Brenner}}, \ and\ \bibinfo {author}
  {\bibfnamefont {V.~N.}\ \bibnamefont {Manoharan}},\ }\href
  {http://dx.doi.org/10.1038/nmat3190} {\bibfield  {journal} {\bibinfo
  {journal} {Nat. Mater.}\ }\textbf {\bibinfo {volume} {11}},\ \bibinfo {pages}
  {138} (\bibinfo {year} {2012})}\BibitemShut {NoStop}%
\bibitem [{\citenamefont {Park}\ and\ \citenamefont {Furst}(2011)}]{Furst1}%
  \BibitemOpen
  \bibfield  {author} {\bibinfo {author} {\bibfnamefont {B.~J.}\ \bibnamefont
  {Park}}\ and\ \bibinfo {author} {\bibfnamefont {E.~M.}\ \bibnamefont
  {Furst}},\ }\href {\doibase 10.1039/C1SM00005E} {\bibfield  {journal}
  {\bibinfo  {journal} {Soft Matter}\ }\textbf {\bibinfo {volume} {7}},\
  \bibinfo {pages} {7676} (\bibinfo {year} {2011})}\BibitemShut {NoStop}%
\bibitem [{\citenamefont {Lewandowski}\ \emph {et~al.}(2010)\citenamefont
  {Lewandowski}, \citenamefont {M.}, \citenamefont {Botto}, \citenamefont
  {Bernate}, \citenamefont {Garbin},\ and\ \citenamefont
  {Stebe}}]{Lewandowski2010}%
  \BibitemOpen
  \bibfield  {author} {\bibinfo {author} {\bibfnamefont {E.~P.}\ \bibnamefont
  {Lewandowski}}, \bibinfo {author} {\bibfnamefont {C.}~\bibnamefont {M.}},
  \bibinfo {author} {\bibfnamefont {L.}~\bibnamefont {Botto}}, \bibinfo
  {author} {\bibfnamefont {J.~C.}\ \bibnamefont {Bernate}}, \bibinfo {author}
  {\bibfnamefont {V.}~\bibnamefont {Garbin}}, \ and\ \bibinfo {author}
  {\bibfnamefont {K.~J.}\ \bibnamefont {Stebe}},\ }\href {\doibase
  10.1021/la1012632} {\bibfield  {journal} {\bibinfo  {journal} {Langmuir}\
  }\textbf {\bibinfo {volume} {26}},\ \bibinfo {pages} {15142} (\bibinfo {year}
  {2010})}\BibitemShut {NoStop}%
\bibitem [{\citenamefont {Lucassen}(1992)}]{Lucassen}%
  \BibitemOpen
  \bibfield  {author} {\bibinfo {author} {\bibfnamefont {J.}~\bibnamefont
  {Lucassen}},\ }\href {\doibase
  http://dx.doi.org/10.1016/0166-6622(92)80268-7} {\bibfield  {journal}
  {\bibinfo  {journal} {Colloids Surf.}\ }\textbf {\bibinfo {volume} {65}},\
  \bibinfo {pages} {131} (\bibinfo {year} {1992})}\BibitemShut {NoStop}%
\bibitem [{\citenamefont {Yao}\ \emph {et~al.}(2013)\citenamefont {Yao},
  \citenamefont {Botto}, \citenamefont {Cavallaro}, \citenamefont {Bleier},
  \citenamefont {Garbin},\ and\ \citenamefont {Stebe}}]{Lu}%
  \BibitemOpen
  \bibfield  {author} {\bibinfo {author} {\bibfnamefont {L.}~\bibnamefont
  {Yao}}, \bibinfo {author} {\bibfnamefont {L.}~\bibnamefont {Botto}}, \bibinfo
  {author} {\bibfnamefont {M.}~\bibnamefont {Cavallaro}}, \bibinfo {author}
  {\bibfnamefont {B.~J.}\ \bibnamefont {Bleier}}, \bibinfo {author}
  {\bibfnamefont {V.}~\bibnamefont {Garbin}}, \ and\ \bibinfo {author}
  {\bibfnamefont {K.~J.}\ \bibnamefont {Stebe}},\ }\href {\doibase
  10.1039/C2SM27020J} {\bibfield  {journal} {\bibinfo  {journal} {Soft Matter}\
  }\textbf {\bibinfo {volume} {9}},\ \bibinfo {pages} {779} (\bibinfo {year}
  {2013})}\BibitemShut {NoStop}%
\bibitem [{\citenamefont {Yao}\ \emph {et~al.}(2015)\citenamefont {Yao},
  \citenamefont {Sharifi-Mood}, \citenamefont {Liu},\ and\ \citenamefont
  {Stebe}}]{Disk}%
  \BibitemOpen
  \bibfield  {author} {\bibinfo {author} {\bibfnamefont {L.}~\bibnamefont
  {Yao}}, \bibinfo {author} {\bibfnamefont {N.}~\bibnamefont {Sharifi-Mood}},
  \bibinfo {author} {\bibfnamefont {I.~B.}\ \bibnamefont {Liu}}, \ and\
  \bibinfo {author} {\bibfnamefont {K.~J.}\ \bibnamefont {Stebe}},\ }\href
  {\doibase http://dx.doi.org/10.1016/j.jcis.2014.12.070} {\bibfield  {journal}
  {\bibinfo  {journal} {J. Colloid Interface Sci.}\ }\textbf {\bibinfo {volume}
  {449}},\ \bibinfo {pages} {436} (\bibinfo {year} {2015})}\BibitemShut
  {NoStop}%
\bibitem [{\citenamefont {Lewandowski}\ \emph {et~al.}(2008)\citenamefont
  {Lewandowski}, \citenamefont {Bernate}, \citenamefont {Searson},\ and\
  \citenamefont {Stebe}}]{Eric}%
  \BibitemOpen
  \bibfield  {author} {\bibinfo {author} {\bibfnamefont {E.~P.}\ \bibnamefont
  {Lewandowski}}, \bibinfo {author} {\bibfnamefont {J.~A.}\ \bibnamefont
  {Bernate}}, \bibinfo {author} {\bibfnamefont {P.~C.}\ \bibnamefont
  {Searson}}, \ and\ \bibinfo {author} {\bibfnamefont {K.~J.}\ \bibnamefont
  {Stebe}},\ }\href {\doibase 10.1021/la801167h} {\bibfield  {journal}
  {\bibinfo  {journal} {Langmuir}\ }\textbf {\bibinfo {volume} {24}},\ \bibinfo
  {pages} {9302} (\bibinfo {year} {2008})}\BibitemShut {NoStop}%
\bibitem [{\citenamefont {G.~B.~Arfken}\ and\ \citenamefont
  {Harris}(2005)}]{Arfken}%
  \BibitemOpen
  \bibfield  {author} {\bibinfo {author} {\bibfnamefont {H.~J.~W.}\
  \bibnamefont {G.~B.~Arfken}}\ and\ \bibinfo {author} {\bibfnamefont {F.~E.}\
  \bibnamefont {Harris}},\ }\href@noop {} {\emph {\bibinfo {title}
  {Mathematical Methods for Physicists}}}\ (\bibinfo  {publisher} {Academic
  Press},\ \bibinfo {address} {Waltham, Massachusett, USA},\ \bibinfo {year}
  {2005})\BibitemShut {NoStop}%
\bibitem [{\citenamefont {Danov}\ \emph {et~al.}(2005)\citenamefont {Danov},
  \citenamefont {Kralchevsky}, \citenamefont {Naydenov},\ and\ \citenamefont
  {Brenn}}]{Danov_bipolar}%
  \BibitemOpen
  \bibfield  {author} {\bibinfo {author} {\bibfnamefont {K.~D.}\ \bibnamefont
  {Danov}}, \bibinfo {author} {\bibfnamefont {P.~A.}\ \bibnamefont
  {Kralchevsky}}, \bibinfo {author} {\bibfnamefont {B.~N.}\ \bibnamefont
  {Naydenov}}, \ and\ \bibinfo {author} {\bibfnamefont {G.}~\bibnamefont
  {Brenn}},\ }\href {\doibase http://dx.doi.org/10.1016/j.jcis.2005.01.079}
  {\bibfield  {journal} {\bibinfo  {journal} {J. Colloid Interface Sci.}\
  }\textbf {\bibinfo {volume} {287}},\ \bibinfo {pages} {121 } (\bibinfo {year}
  {2005})}\BibitemShut {NoStop}%
\bibitem [{\citenamefont {Griffiths}(1999)}]{Griffiths}%
  \BibitemOpen
  \bibfield  {author} {\bibinfo {author} {\bibfnamefont {D.~J.}\ \bibnamefont
  {Griffiths}},\ }\href@noop {} {\emph {\bibinfo {title} {Introduction to
  Electrodynamics}}},\ \bibinfo {edition} {3rd}\ ed.\ (\bibinfo  {publisher}
  {Prentice Hall},\ \bibinfo {address} {Upper Saddle River, New Jersey, USA},\
  \bibinfo {year} {1999})\BibitemShut {NoStop}%
\bibitem [{\citenamefont {Sharifi-Mood}\ \emph {et~al.}(2015)\citenamefont
  {Sharifi-Mood}, \citenamefont {Liu},\ and\ \citenamefont {Stebe}}]{Sphere}%
  \BibitemOpen
  \bibfield  {author} {\bibinfo {author} {\bibfnamefont {N.}~\bibnamefont
  {Sharifi-Mood}}, \bibinfo {author} {\bibfnamefont {I.~B.}\ \bibnamefont
  {Liu}}, \ and\ \bibinfo {author} {\bibfnamefont {K.~J.}\ \bibnamefont
  {Stebe}},\ }\href {\doibase 10.1039/C5SM00310E} {\bibfield  {journal}
  {\bibinfo  {journal} {Soft Matter}\ }\textbf {\bibinfo {volume} {11}},\
  \bibinfo {pages} {6768} (\bibinfo {year} {2015})}\BibitemShut {NoStop}%
\bibitem [{\citenamefont {Basavaraj}\ \emph {et~al.}(2006)\citenamefont
  {Basavaraj}, \citenamefont {Fuller}, \citenamefont {Fransaer},\ and\
  \citenamefont {Vermant}}]{Vermant}%
  \BibitemOpen
  \bibfield  {author} {\bibinfo {author} {\bibfnamefont {M.~G.}\ \bibnamefont
  {Basavaraj}}, \bibinfo {author} {\bibfnamefont {G.~G.}\ \bibnamefont
  {Fuller}}, \bibinfo {author} {\bibfnamefont {J.}~\bibnamefont {Fransaer}}, \
  and\ \bibinfo {author} {\bibfnamefont {J.}~\bibnamefont {Vermant}},\ }\href
  {\doibase 10.1021/la060465o} {\bibfield  {journal} {\bibinfo  {journal}
  {Langmuir}\ }\textbf {\bibinfo {volume} {22}},\ \bibinfo {pages} {6605}
  (\bibinfo {year} {2006})}\BibitemShut {NoStop}%
\bibitem [{\citenamefont {Brakke}(1992)}]{S_evolver}%
  \BibitemOpen
  \bibfield  {author} {\bibinfo {author} {\bibfnamefont {K.~A.}\ \bibnamefont
  {Brakke}},\ }\href {\doibase 10.1080/10586458.1992.10504253} {\bibfield
  {journal} {\bibinfo  {journal} {Exp. Math.}\ }\textbf {\bibinfo {volume}
  {1}},\ \bibinfo {pages} {141} (\bibinfo {year} {1992})}\BibitemShut {NoStop}%
\bibitem [{Note1()}]{Note1}%
  \BibitemOpen
  \bibinfo {note} {This is a recapitulation of the calculation performed in
  sec.~\ref {reflection}, where the distortions can be described in a
  coordinate system appropriate for elongated particles. Note that the
  derivation presented in sec.~\ref {reflection} for interacting polar
  quadrupoles is inspired by work by Stamou \protect \emph {et al.} \cite
  {Stamou}. In that original work, the superposition approach was used to
  determine the interface shape. We modified the analysis using the method of
  reflections to find surface shapes that obeys the boundary conditions on the
  particle and evaluated the associated capillary energy.}\BibitemShut {Stop}%
\end{thebibliography}
%
\end{document}